\documentclass[reqno,11pt]{amsart}
\usepackage{amsmath, amsthm, amscd,amsfonts, amssymb,hyperref,floatrow}
\usepackage{graphicx, color,dsfont,xcolor}

\numberwithin{equation}{section}

\newtheorem{theorem}{Theorem}[section]

\newtheorem{rem}[theorem]{Remark}

\def\tr{\mbox{tr}}

\renewcommand{\tilde}{\widetilde}          
\DeclareMathSymbol{\leqslant}{\mathalpha}{AMSa}{"36} 
\DeclareMathSymbol{\geqslant}{\mathalpha}{AMSa}{"3E} 
\DeclareMathSymbol{\eset}{\mathalpha}{AMSb}{"3F}     
\renewcommand{\leq}{\;\leqslant\;}                   
\renewcommand{\geq}{\;\geqslant\;}                   

\newcommand{\R}{\mathbb{R}}

\newcommand{\N}{\mathbb{N}}

\newcommand \be  {\begin{equation*}}
\newcommand \bea {\begin{eqnarray} \nonumber }
\newcommand \ee  {\end{equation*}}

\newcommand \ba  {\begin{align}}
\newcommand \ea  {\end{align}}

\definecolor{remi}{rgb}{0,0,0}

\begin{document}

\title{Eigenvector dynamics: general theory and some applications \\}

\author{Romain Allez$^{1,2}$ and Jean-Philippe Bouchaud$^1$}
\address{$^1$ Capital~Fund~Management, 6--8 boulevard Haussmann, 75\,009 Paris, France}
\address{$^2$ Universit{\'e} Paris-Dauphine, Ceremade, 75\,016 Paris, France.}
\date{\today}
\begin{abstract}
We propose a general framework to study the stability of the subspace spanned by $P$ consecutive eigenvectors of a generic symmetric matrix ${\bf H}_0$, when a small perturbation is added. This problem is relevant in various contexts, including quantum dissipation (${\bf H}_0$ is then the Hamiltonian) 
and financial risk control (in which case ${\bf H}_0$ is the assets return covariance matrix). We argue that the problem can be formulated in terms of 
the singular values of an overlap matrix, that allows one to define an overlap distance. We specialize our results for the case of a Gaussian Orthogonal ${\bf H}_0$, for which 
the full spectrum of 
singular values can be explicitly computed. We also consider the case when ${\bf H}_0$ is a covariance matrix and illustrate the usefulness of our results using financial data. The special case where the top eigenvalue is much larger than all the other ones can be investigated in full detail. In particular, the dynamics of the angle made by the top eigenvector and its true direction defines an interesting new class of random processes.  
\end{abstract}

\maketitle
\tableofcontents

\section{Introduction}

Random Matrix Theory (RMT) is extraordinary powerful at describing the eigenvalues statistics of large random, or pseudo-random, matrices \cite{Verdu,Handbook,AGZ,Mehta}. Eigenvalue 
densities, two-point correlation functions, level spacing distributions, etc. can be characterized with exquisite details. The ``dynamics'' of 
these eigenvalues, i.e. the way these eigenvalues evolve when the initial matrix ${\bf H}_0$ is perturbed by some small matrix $\varepsilon {\bf P}$, is also well understood \cite{Simon}. 
The knowledge of the corresponding eigenvectors is comparatively much poorer (but see \cite{Wilkinson2}). One reason is that many RMT results concern 
rotationally invariant matrix ensembles, 
such that by definition the statistics of eigenvectors is featureless. Still, as we will show below, some interesting results can be derived for the
dynamics of these eigenvectors. Let us give two examples for which this question is highly relevant.

One problem where the evolution of eigenvectors is important is Quantum Dissipation \cite{Wilkinson} (see also the related recent strand of the literature
on Quantum ``Fidelity'' \cite{Fidelity}). As the parameters of the Hamiltonian ${\bf H}_t = {\bf H}_0 + 
\varepsilon {\bf P}_t$ of a system evolve with time $t$, the average energy changes as well. One term corresponds to the average (reversible) change 
of the Hamiltonian that leads to a shift of the energy levels (the eigenvalues). But if the external perturbation is not infinitely slow, some 
transitions between energy levels will take place, leading to a dissipative (irreversible) term in the evolution equation of the average energy 
of the system. The adiabaticity condition which ensures that no transition takes place amounts to comparing the speed of change of the perturbation 
$\varepsilon {\bf P}_t$ with a quantity proportional to the typical spacing between energy levels. For systems involving a very large number $N$ of 
degrees of freedom, the average level spacing of the $N \times N$ Hamiltonian ${\bf H}$ goes to zero as $N^{-1}$. For $N \to \infty$, any finite speed of change therefore
corresponds to the ``fast'' limit, where a large number of transitions between states is expected. In fact, if the quantum system is in state 
$|\phi_i^0 \rangle$ at time $t=0$, that corresponds to the $i$th eigenvector of ${\bf H}_0$, the probability to jump to the $j$th eigenvector of ${\bf H}_1$,  
$|\phi_j^1 \rangle$, at time $t=1$ is given by $|\langle \phi_j^1 | \phi_i^0 \rangle|^2$, where we use the bra-ket notation for vectors and scalar
products. The way energy is absorbed by the system will therefore be determined by the perturbation-induced distortion of the eigenvectors. More precisely, 
if $|\phi_i^0 \rangle$ is different from $|\phi_i^1 \rangle$, some transitions must take place in the non-adiabatic limit, that involve
all the states $j$ which have a significant overlap with the initial state. 

Another very relevant situation is Quantitative Finance, where the covariance matrix ${\bf C}$ between the returns of $N$ assets (for example stocks) plays a
major role in risk control and portfolio construction \cite{Review}. More precisely, the risk of a portfolio that invests $w_\alpha$ in asset $\alpha$ is given by 
${\mathcal R}^2 = \sum_{\alpha\beta} w_\alpha {\bf C}_{\alpha\beta} w_\beta$. Constructing low risk portfolios requires the knowledge of the $n$ 
largest eigenvalues of ${\bf C}$ ($n$ is often chosen empirically, keeping only the statistically meaningful eigenvalues that lie outside the Marchenko-Pastur sea, see 
\cite{bookBP} for details), 
$\lambda_1 \geq \dots \geq \lambda_n$ and their corresponding eigenvectors $|\phi_1 \rangle, \dots, |\phi_n \rangle$. The top eigenvalues and 
eigenvectors represent the most risky directions in a financial context. A portfolio such that the vector of weights $| w  \rangle$ 
has zero overlap with the first $n$ eigenvectors of ${\bf C}$ has a risk that is bounded from above by $\lambda_{n+1}$. The problem with this idea is that
it relies on the assumption that the covariance matrix ${\bf C}$ is perfectly known and {\it constant in time}. The observation of a sufficiently long time series of past returns would thus allow one, in such a stable world, to determine ${\bf C}$ and to immunize the portfolio against risky investment modes. 

Unfortunately, this idea is thwarted by two (inter-related) predicaments: a) time series are always of finite length, and lead to substantial 
``noise'' in empirical estimates of ${\bf C}$ \cite{Review} and b) the world is clearly not stationary and there is no guarantee that the covariance matrix 
corresponding to the pre-crisis period 2000-2007 is the same as the one corresponding to the period 2008-2011. For one thing, some companies disappear 
and others are created in the course of time. But even restricted to companies that exist throughout the whole period, it is by no means granted that
the correlation between stock returns do not evolve in time. This is why it is common practice in the financial industry to restrict the period
used to determine the covariance matrix to windows of a few years into the recent past. This leads to the measurement noise problem alluded above. 
Now, if the ``future'' large eigenvectors do not coincide with the past ones, a supposedly low risk portfolio will in fact be exposed to large 
risks directions in the future. Denoting as $|\phi_i^0 \rangle$ the past eigenvectors and $|\phi_j^1 \rangle$ the future ones, the total risk of
the portfolio $| w  \rangle = |\phi_i^0 \rangle $ can be defined as $\sum_{j=1}^n \lambda_j^1 \langle \phi_j^1 |\phi_i^0 \rangle^2$. Therefore, 
as for the quantum dissipation problem, the statistics of the overlaps $\langle \phi_j^1 | \phi_i^0 \rangle$ is a crucial piece of information.  

In practice, one computes the empirical covariance matrix ${\bf E}$ using past stock returns, which is defined as: $${\bf E}_{ij} = \frac{1}{T} \sum_{t=1}^{T} r_i^t r_j^t\,,$$ where $T$ is the length of the period on which the measurement is done and $r_i^t$ is the return of stock number $i$ on day $t$. 
If the true covariance matrix ${\bf C}$ exists and is stable in this period, the empirical matrix ${\bf E}$ can be seen as a perturbation of ${\bf C}$, since one can write ${\bf E}= {\bf C}+ {\bf \mathcal{E}}$ where ${\bf \mathcal{E}}$ is a matrix whose elements are of order $1/\sqrt{T}$ 
(by the central limit theorem) to be considered small as $T$ is usually quite large. In this sense, the problem falls in the more general context introduced above. 

The paper is organized as follows. In the next section \ref{stat_tools}, we introduce the main statistical tools and objects studied in different contexts in the following sections 
and we also briefly recall standard perturbation theory.  In the next two sections, we turn to two explicit illustrations, first in the context of matrices in the 
Gaussian Orthogonal ensemble (GOE), 
and then in the context of covariance matrices.
More precisely, in section \ref{GOE}, we study the stability of the eigenvectors for a matrix ${\bf H}_0$ in the GOE by computing the ``overlap distance'' between 
the perturbed space and the non-perturbed space in the limit of large matrices ${\bf H}_0$ when the 
perturbation matrix $\bf P$ is also in the GOE. Furthermore, we are able to compute the full spectrum of the overlap matrix in this limit, 
which gives a precise idea of the perturbation induced distortion for the eigenvectors. In section \ref{section_cov_matrix}, we go through the same steps in the context of covariance matrices. 
We study the link between the 
{\it population} eigenvectors (the eigenvectors of the {\it true} covariance matrix) and the {\it sample} eigenvectors (the eigenvectors of the empirical covariance matrix).
Then, in section \ref{isolated_top_eigenvalue}, we analyse more precisely the case of a population covariance matrix with an isolated top eigenvalue much larger than the 
other ones. 
We measure the empirical covariance matrix with an exponential moving average estimator and characterize the temporal evolution of the angle made by the top eigenvector 
and its true direction which defines an interesting new class of random processes.      
Finally, in section \ref{empirical_results}, we apply our ideas to the analysis of financial market correlations. Our purpose here is to study whether 
correlations between stock returns evolve or not. In particular, is there a constant in time correlation matrix (population correlation matrix)? Do the economical 
sectors (eigenvectors of the correlation matrix) evolve or not ? We find that there is indeed a genuine evolution of the correlation matrix of stocks returns for different markets in the U.S, 
in Europe and in Japan, a result that confirms recent studies (see e.g. \cite{Ingve,pa_leverage,Munnix}). We also give a partial description of this temporal evolution.   

\section{Perturbation theory and Statistical tools}\label{stat_tools}
In the whole paper, we will mainly be interested in the eigenvectors of a matrix ${{\bf H}}_1$ that can be written as 
\begin{equation*}
{\bf H}_1 = {\bf H}_0 + \varepsilon {\bf P}
\end{equation*} 
where ${\bf H}_0$ and $\bf P$ are two $N \times N$ symmetric matrices and $\varepsilon$ a small (positive) parameter. The matrix ${\bf H}_0$ is the 
{\it true signal} which is perturbed by the adding of the small term $\varepsilon {\bf P}$. The matrix ${\bf H}_1$ will be referred as the {\it perturbed} matrix. 
The eigenvalues of the matrix ${\bf H}_i, i=0,1$ will be denoted as $\lambda_1^i \geq\lambda_2^i\geq\dots\geq\lambda_N^i$ and the corresponding eigenvectors 
$|\phi_1^i\rangle,\dots,|\phi_N^i\rangle$.

Our aim is to describe the relation between the perturbed eigenvectors $|\phi_i^1\rangle$ and 
the non-perturbed eigenvectors $|\phi_i^0\rangle$ when the parameter $\varepsilon$ tends to $0$. 

When trying to follow the evolution of a given eigenvector $|\phi_i \rangle$ when the small perturbation $\varepsilon \bf P$ is added, one immediately faces a problem when
eigenvalues ``collide''. It is well known that true collisions (degeneracies) are non generic; the collisions are in fact avoided and levels do not
cross. However, upon the pseudo-collision of $\lambda_i$ and $\lambda_{i+1}$ (say), the eigenvectors $|\phi_i \rangle$ and $|\phi_{i+1} \rangle$ 
strongly hybridize (this phenomenon was observed for example in \cite[Fig. 1]{vallejos}). 
The eigenvector $|\phi_i^0\rangle$ will in fact hybridize with all the eigenvectors associated to eigenvalues $\lambda_j^0,j\neq i$ that are close to 
$\lambda_i^0$ and will therefore be strongly affected by the adding of the perturbation if the level spacing is too small. Indeed, using standard perturbation theory to second order in 
$\varepsilon$, the
perturbed eigenvectors can be expressed in terms of the initial eigenvectors, for small $\varepsilon$, as:
\begin{align}\label{perturbationtheory}
|\phi^1_i\rangle &= {\left(1- \frac{\varepsilon^2}{2} \sum_{j \neq i} \left(\frac{P_{ij}}{\lambda_i^0 - \lambda_j^0} \right)^2 \right)} |\phi_i^0 \rangle 
+ \varepsilon \sum_{j \neq i} \frac{P_{ij}}{\lambda_i^0 - \lambda_j^0} |\phi_j^0\rangle \\ 
&+  \varepsilon^2 \sum_{j \neq i} \frac{1}{\lambda_i^0 - \lambda_j^0} \left( \sum_{\ell \neq i} \frac{P_{j\ell} P_{\ell i}}{\lambda_i^0 - \lambda_\ell^0} - 
\frac{P_{ii}P_{ij}}{\lambda_i^0 - \lambda_j^0} \right) | \phi_j^0 \rangle \notag
\end{align}
where $P_{ij} \equiv  \langle \phi_j^0|{\bf P}|\phi_i^0 \rangle$. 
The denominators $\lambda_i^0 - \lambda_j^0$ remind us that eigenvectors are strongly affected by eigenvalue pseudo-collisions, as alluded to above.
We mention in passing that perturbation theory to second order in $\varepsilon$ for the eigenvalues gives 
\begin{equation}\label{pt_eigenvalues}
\lambda_i^1= \lambda_i^0 + \varepsilon P_{ii} + \varepsilon^2 \sum_{j\neq i} \frac{P_{ij}^2}{\lambda_i^0-\lambda_j^0}\,.
\end{equation}
It is important to mention at this point that equations \eqref{perturbationtheory} and \eqref{pt_eigenvalues} are {\it a priori} only valid in the regime where the entries of the perturbation 
matrix $\varepsilon {\bf P}$ are small compared to the level spacing of the non-perturbed matrix ${\bf H}_0$. 
This condition ensures that the asymptotic correction terms appearing in \eqref{perturbationtheory} and \eqref{pt_eigenvalues} are small compared to the leading term 
of order $1$ corresponding to the non-perturbed system. 

The idea is then to study the stability of a whole subspace $V_0$ spanned by 
$2p+1$ several consecutive eigenvalues: $\{|\phi_{k-p}^0\rangle, \dots |\phi_{k}^0\rangle, \dots, |\phi_{k+p}^0\rangle\}$. Motivated by the 
above examples, we ask the following question: how should one choose $q \geq p$ such that the subspace $V_1$ of dimension $2q+1$ spanned by the set
$\{|\phi_{k-q}^1\rangle, \dots |\phi_{k}^1\rangle, \dots, |\phi_{k+q}^1\rangle\}$ has a significant overlap with the initial subspace? In order
to answer this question, we consider the $(2q+1) \times (2p+1)$ rectangular matrix of overlaps ${\bf G}$ with entries: 
\begin{equation}
G_{ij} := \langle \phi_i^1 | \phi_j^0 \rangle\,.
\end{equation}

The $(2p+1)$ non zero singular values $1 \geq s_1 \geq s_2 \geq \dots \geq s_{2p+1} \geq 0$ of ${\bf G}$ give full information about the overlap between the two spaces. 
For example, the largest singular value $s_1$ indicates that there is a certain linear combination of the $(2q+1)$ perturbed eigenvectors that has a scalar 
product $s_1$ with a certain linear combination of the $(2p+1)$ unperturbed eigenvectors. If $s_{2p+1}=1$, then the initial subspace is entirely spanned by the
perturbed subspace. If on the contrary $s_1 \ll 1$, it means that the initial and perturbed eigenspace are nearly orthogonal to one another since even the 
largest possible overlap between any linear combination of the original and perturbed eigenvectors is very small. A good 
measure of what can be called an overlap distance $D(V_0,V_1)$ between the two spaces $V_0$ and $V_1$ 
is provided by the average of the logarithm of the singular values:
\begin{equation}\label{def_D}
D(V_0,V_1) :=-\frac{\sum_i \ln s_i}{2p+1},
\end{equation}
but alternative measures, such as $1-\sum_i s_i/(2p+1)$, can be considered as well. Since the singular values $s$ are obtained as the square-root of the eigenvalues 
of the matrix ${\bf G}^\dagger{\bf G}$, one has $D \equiv - \ln \det {\bf G}^\dagger{\bf G}/2P$, where we henceforth introduce for convenience the notations $P=2p+1$, $Q=2q+1$. 
The overlap distance $D$ was originally studied for $P=Q$ in \cite{Anderson}, see e.g. \cite{Ullmo,vallejos}, where a fundamental effect observed in many body systems, 
called the {\it Anderson Orthogonality catastrophe} (AOC) is introduced. Anderson \cite{Anderson} addressed the ground state of a finite system consisting of $P$ 
noninteracting electrons. Upon the introduction of a finite rank perturbation matrix $\varepsilon {\bf P}$, this ground state gets modified. 
It is then shown that the overlap between the original and the modified $P$-electron ground state, which 
is in fact exactly given by our overlap distance $D(V_0,V_1)$ between the two subspaces $V_0$ and $V_1$ (with $P=Q$), is proportional to a
negative power of $P$, and vanishes in the thermodynamic $P\rightarrow +\infty$ limit, hence the catastrophe. 
We will see that our idea of introducing a rectangular $Q\times P$ overlap matrix $\bf G$ enables to avoid this orthogonality catastrophe. 
Our objects introduced here will also allow us to revisit the 
AOC in the case of square matrices $\bf G$ showing that it occurs for the random matrix model studied in section \ref{GOE} (AOC for this RM model is also studied in \cite{vallejos}). 
In \cite{Ullmo,vallejos}, the AOC is also investigated 
through random matrix models as in our paper. The main difference with \cite{Ullmo} is that we consider here full rank perturbation instead of a localized perturbation of rank $1$, for which one can do explicit computations 
(and so treat the non-perturbative regime).

As an interesting benchmark, consider the case when two subspaces $W_0$ and $W_1$ respectively of dimensions $P$ and $Q$ 
are constructed using randomly chosen orthonormal vectors in a space of dimension $N$. In this case, one expects accidental overlaps, such that the $s_i$ are in fact non zero, and therefore $D(W_0,W_1)$ is finite. This distance can be calculated exactly using Random Matrix Theory tools in the limit $N,P,Q \to \infty$, with $\alpha=P/N$ and $\beta=Q/N$ held fixed. 
The result is \cite{RSVD}:
\begin{equation*}
D_{RMT}(W_0,W_1) = - \int_0^1 {\rm d} s \ln(s) \frac{\sqrt{(s^2-\gamma_-)_+ (\gamma_+-s^2)_+}}{\beta\pi s (1-s^2)}
\end{equation*}
where $\gamma_\pm = \alpha+\beta-2\alpha\beta \pm 2\sqrt{\alpha\beta(1-\alpha)(1-\beta)}$. In other words, in that limit, the full density of singular 
values is known; all singular values are within the interval $[\sqrt{\gamma_-},\sqrt{\gamma_+}]$. This provides a benchmark to test whether the two eigenspaces are accidentally close 
($D \approx D_{RMT}$), or if they are genuinely similar ($D \ll D_{RMT}$).

Endowed with the above formalism, we can now proceed to compute $D(V_0,V_1)$ in the case where the perturbation is small. Indeed equation \eqref{perturbationtheory} 
allows us to obtain the overlap matrix ${\bf G}$. 
Keeping only the relevant terms to order $\varepsilon^2$, we find:\footnote{see \cite{Wilkinson2} for similar calculations.}
\begin{equation}\label{expr_G}
G_{ij} = \begin{cases} 
1 - \frac{\varepsilon^2}{2} \sum_{\ell \neq i} \left(\frac{P_{i\ell}}{\lambda_i^0 - \lambda_\ell^0} \right)^2 & \text{if $i=j$},\\
 \frac{\varepsilon \,P_{ij}}{\lambda_i^0 - \lambda_j^0}  + \frac{\varepsilon^2}{\lambda_i^0 - \lambda_j^0} \left( \sum_{\ell \neq i} \frac{P_{j\ell} P_{\ell i}}{\lambda_i^0 - \lambda_\ell^0} - 
\frac{P_{ii}P_{ij}}{\lambda_i^0 - \lambda_j^0} \right) &  \text{if $i \neq j$}. 
\end{cases} 
\end{equation}
Using \eqref{expr_G}, we can also compute the matrix ${\bf G^\dagger G}$ to second order in $\varepsilon$, we obtain for $i\not = j$:
\begin{equation}\label{expr_GGnd}
(G^\dagger G)_{ij} = - \varepsilon^2 \sum_{\ell \not \in \{k-q;\dots;k+q\}} \frac{P_{\ell i}P_{\ell j}}{(\lambda_i^0-\lambda_\ell^0)(\lambda_j^0-\lambda_\ell^0)}\,,
\end{equation}
and, for $i=j$:
\begin{equation}\label{expr_GGod}
(G^\dagger G)_{ii} = 1 - \varepsilon^2 \sum_{j\not \in \{k-q;\dots;k+q\}} \left(\frac{P_{ij}}{\lambda_i^0-\lambda_j^0}\right)^2.
\end{equation}
It is then easy to derive the central result of our study: to second order in $\varepsilon$, the distance $D(V_0,V_1)$ between the initial and perturbed eigenspaces is:
\begin{equation}\label{expr_D_dimension_finie}
D(V_0,V_1) =  \frac{\varepsilon^2}{2P} \sum_{i=k-p}^{k+p} \sum_{j\notin \{k-q, \dots, k+q\}} \left(\frac{P_{ij}}{\lambda_j^0 - \lambda_i^0} \right)^2.
\end{equation}
The matrices ${\bf G}$ and ${\bf G^\dagger G}$ are both close to the identity matrix as they should.
Let us define the matrix ${\bf \Sigma}$ by $$ {\bf \Sigma} = \frac{1}{\varepsilon^2} \left( {\bf I} -  {\bf G^\dagger G}\right)\,.$$ 
One can note that the matrix $\bf \Sigma$ is positive definite and that its matrix elements are of order $1$ when $\varepsilon$ goes to $0$.

\section{Eigenvector stability in the GOE ensemble}\label{GOE}

We will now define a random matrix model for which we will apply the results of the previous section. 
Let ${\bf H}_0$ be a random matrix of the Gaussian Orthogonal Ensemble (GOE), i.e. a matrix of size $N\times N$ with gaussian entries randomly chosen 
with the probability measure on the space of real symmetric matrices  
\begin{equation*}
P({\rm d}H_0) = \exp(-\frac{N}{2\sigma^2} \tr(H_0^2)) \, {\rm d}H_0\,.
\end{equation*}
This definition implies that the matrix $\bf H_0$ is symmetric with independent Gaussian entries above the diagonal with variance $\sigma^2/N$ on the diagonal 
and $\sigma^2/2N$ off diagonal. 

The perturbation matrix is similarly defined as a random matrix of the GOE, independent of ${\bf H}_0$ with the same variance profile for the entries. 


We then define the ${\it perturbed}$ matrix ${\bf H}_1$ as before: 
\begin{equation}
{\bf H}_1 = {\bf H}_0 + \varepsilon {\bf P}.
\end{equation}  
It is very well known that the density of 
${\bf H}_0$-eigenvalues 
$\rho_N(\lambda):=1/N \sum_{i=1}^N \delta_{\lambda_i}$ tends in the large $N$ limit to the Wigner semi-circle law
\begin{equation}\label{semicircle}
\rho({\rm d}\lambda) \equiv\frac{1}{2\pi} \sqrt{4\sigma^2-\lambda^2} {\rm d}\lambda\,.
\end{equation}
For simplicity, we take $\sigma^2=2$ in the following.

\begin{rem}
Here the choice of a GOE random matrix for $\bf H_0$ is made to get an explicit expression for the density of states 
in the limit of large matrices. But our theory developed in the following would apply for sequences of matrices $({\bf H_0}({N}))_N$ such that the density of states 
converges to a general (not necessarily the semi-circle density) continuous density $\rho(\lambda) {\rm d}\lambda$. 
Moreover, the sequence $({\bf H_0}(N))_N$ can be supposed deterministic or random. 
The matrix $\bf H_0$ can be seen as 
the {\it{true signal}} to which a small noisy perturbation ${\bf P}_t$ is added.     
\end{rem}

In this whole current section, $\overline{\cdot\cdot\cdot}$ denotes an averaging over the random matrix ${\bf P}$ \footnote{There is no need in averaging over 
the random matrix ${\bf H}_0$ for the following results to be valid.}. 
 
\subsection{Distance between subspaces of perturbed and non-perturbed eigenvectors} \label{Distance_eigenspaces}

We consider the subspace $V_0$ of initial eigenvectors corresponding to all the eigenvalues $\lambda$ contained in a certain finite interval $[a,b]$ included 
in the Wigner sea $[-2,2]$. 
We want to compute the mean overlap distance 
$\overline{D}(V_0,V_1)$ between $V_0$ and the subspace $V_1$ spanned by 
the perturbed eigenvectors of ${\bf H}_1$, corresponding to all eigenvalues contained in $[a-\delta,b+\delta]$, where $\delta$ is a positive parameter. 

Using formula \eqref{expr_D_dimension_finie}, which is valid if the entries of the perturbation matrix $\varepsilon \bf P$ (of order $\varepsilon N^{-1/2}$) are much smaller than the mean level spacing 
of the matrix ${\bf H}_0$, of order $(N\rho(\lambda))^{-1}$, we can write for $\varepsilon \ll N^{-1/2}$:
\begin{equation}\label{comp_D_GOE}
\overline{D}(V_0,V_1)  = \frac{\varepsilon^2}{2P} \sum_{\lambda_i^0\in [a;b]} \sum_{\lambda_j^0 \not \in[a-\delta;b+\delta]} \frac{1}{(\lambda_j^0 - \lambda_i^0)^2} .
\end{equation}

It is easily seen that Eq. \eqref{comp_D_GOE} becomes, in the large $N$ limit: 
\begin{equation}\label{D_GOE}
\overline{D}(V_0,V_1)  =  \frac{\varepsilon^2}{2\int_a^b \rho(\lambda) {\rm d}\lambda} \int_a^b {\rm d}\lambda \int_{[-2;2]\setminus[a-\delta;b+\delta]} {\rm d}\lambda' \,\, \frac{\rho(\lambda)\rho(\lambda')}{(\lambda - \lambda')^2}\,,
\end{equation}
where $\rho$ is the Wigner semicircle density \eqref{semicircle}. 

Formula \eqref{D_GOE} is a priori only rigorously valid in the perturbative regime where $\varepsilon \ll {N}^{-1/2}$. We argue that 
in fact it remains valid in a wider regime where $\varepsilon \ll 1$.
Indeed although perturbation theory for the eigenvectors fails for ${\bf H}_0$ eigenvalues that are at distance of 
order of the mean level spacing of ${\bf H}_0$, it remains valid in the limit $\varepsilon \ll 1$ 
for eigenvalues at distance large compared to the order of the perturbation entries $\varepsilon N^{-1/2}$ and in particular for 
two eigenvalues lying respectively in the two well separated intervals $[a;b]$ and 
$[a-\delta;b+\delta]$ for which this distance is larger than $\delta$ (which indeed is $\gg \varepsilon  N^{-1/2}$).  We see that every terms appearing 
in \eqref{perturbationtheory} corresponding to overlap between 
eigenvectors associated to eigenvalues that are at distance smaller than $\delta$ disappear in 
formulas \eqref{expr_D_dimension_finie} (and also in \eqref{expr_GGnd}, \eqref{expr_GGod}). Therefore, we expect 
\eqref{D_GOE}, as well as our results below, to remain valid in the regime $N^{-1/2} \ll \varepsilon \ll 1$, provided the computed distance $\overline{D}(V_0,V_1)$ itself remains much 
smaller\footnote{For this condition to be valid, $\delta$ has to be fixed independent of $\varepsilon$, or at least such that $\varepsilon^2 |\ln(\delta)|\ll 1 $.  } 
than unity.  
We checked formula \eqref{D_GOE} using numerical simulations, with very good agreement for different values of $a,b,\delta,N, \varepsilon$. 
In those numerical tests we chose the parameters $N,\varepsilon,\delta$ so as to approach the regime $N^{-1/2} \ll \varepsilon\ll 1$ (for example, $N=4000,\varepsilon=0.1,\delta=0.5$).

We will now write $\overline{D}(a,b;\delta,\varepsilon)$ instead of $\overline{D}(V_0,V_1)$. 

%

It is interesting to study the above expression in the double limit $\delta \to 0$ and $\Delta = b-a \to 0$. One finds:
\begin{equation}\label{cases}
\frac{1}{\varepsilon^2} \overline{D}(a,a+\Delta;\delta,\varepsilon) \approx \begin{cases} 
\frac{\rho(a) \ln(\Delta/\delta)}{\Delta} & \text{if $\delta \ll \Delta\ll 1$},\\
\frac{\rho(a)}{\delta}  &  \text{if $\Delta \ll \delta\ll 1$}. 
\end{cases} 
\end{equation}
This last expression shows that when the width $\Delta$ of interval $[a,b]$ tends to zero, the corresponding eigenvectors are scattered in a region of 
width $\delta$ much larger than $\Delta$ itself as soon as $\varepsilon \gg \sqrt{\Delta}$. It also shows that for fixed $\Delta$, the distance $\overline{D}$ diverges logarithmically
when $\delta \to 0$. This is a consequence of the pseudo-collisions that occur between eigenvalues close to the boundaries of the interval $[a,b]$. 
When $\delta > 0$, these pseudo-collisions are avoided and $D$ remains finite. 

When $\delta = 0$, we can do a more precise analysis of the right hand side of \eqref{comp_D_GOE}. One can show, for large $N$, that the following result holds 
at least in the regime $\varepsilon \ll N^{-1/2}$:
\begin{equation}\label{delta=0}
\frac{1}{\varepsilon^2} \overline{D}(a,b;\delta=0,\varepsilon) \approx \ln N  \,\, \frac{\rho(a)^2+\rho(b)^2}{2\int_a^b \rho(\lambda) d\lambda } + A(a,b)
\end{equation}
where $A(a,b)$ is a constant independent of $N$ that can be explicitly computed, and involves the well known two-point function $g(r)$ that describes the level-level correlations in the GOE 
(see Appendix A for the details of this computation). 
The $\ln N$ term can be guessed from the logarithmic behavior of $D$ when $\delta \to 0$, since one indeed expects the divergence to be cut-off when $\delta$ becomes of the order of the 
level spacing, i.e. $\delta \sim (N \rho)^{-1}$. 
Eq. \eqref{delta=0} is precisely the Anderson orthogonality catastrophe as first introduced in \cite{Anderson} in the case of finite rank perturbation matrices. 
We recover here exactly the result of \cite{vallejos} (see their Eq. (31)) by taking $a=-2,b=0$ in our Eq. \eqref{delta=0}. 
\footnote{The authors of \cite{vallejos} expect deviations 
in \eqref{delta=0} when the parameter $x:=\varepsilon\sqrt{N}$ (which has to be $\ll 1$ for \eqref{delta=0} to be fully valid) is increased. However their numerical 
results (presented in Fig. 2 of \cite{vallejos}) show that the discrepancies are only noticeable for $x$ close to $1$. 
In addition, the authors of \cite{vallejos} explain that the failure of \eqref{delta=0} for not small enough $x$ is due 
to the first-order perturbation theory estimate that breaks down when used for levels in the vicinity of the edges $a,b$ of the initial interval. This problem was avoided previously by the use 
of rectangular matrices with $Q>P$ and the introduction of the $\delta$ margin at the edges $a$ and $b$.} 


As a side remark, we note that Eq. \eqref{cases} predicts that a fraction $\propto \delta^{-1}$ of the original eigenspace gets shoved away at distances larger than $\delta$ 
(in eigenvalue space). In the context of the non adiabatic evolution of a quantum system \cite{Wilkinson}, this implies that the energy of the system makes jump with a power-law distribution of sizes, such that all moments of order $q \geq 1$ 
diverge. This means that under an extreme non-adiabatic process, the energy is not diffusive but rather performs a ``Cauchy flight'' (i.e. a L\'evy flight with a tail exponent equal to $2$), see \cite{Wilkinson2}. When the
perturbation varies on a finite time $\tau$, one expects the tails of the jump process to be truncated beyond $\delta \sim \hbar/\tau$ \cite{Wilkinson}.

\subsection{Full distribution of the singular values of the overlap matrix}\label{full_distrib}

To order $\varepsilon^2$, the distance $D$ computed in the previous subsection is proportional to the mean position of the
singular values. One can actually be much more precise and compute, for $P,Q \to \infty$, the {\it full distribution of all 
singular values}, giving an indication of their scatter around the mean position $\langle s \rangle$.
The computation of the density of states (DOS) can be straightforwardly performed using free random matrices techniques.

Using Eq. \eqref{expr_GGnd} and  \eqref{expr_GGod}, it is easy to compute the entries of the matrix ${\bf \Sigma}$ in the perturbative regime $\varepsilon \sqrt{N}\ll 1$:
\footnote{We skip the subscript $0$ on the eigenvalues $\lambda_i$s.} 
\begin{equation}\label{expr_Sigma}
\Sigma_{ij} =  \sum_{\ell \not \in \{k-q;\dots;k+q\}} \frac{P_{\ell i}P_{\ell j}}{(\lambda_i-\lambda_\ell)(\lambda_j-\lambda_\ell)}
\end{equation}

Denote, for each $\ell \not \in \{k-q;\dots;k+q\}$ by ${\bf v}_{\ell}$ the random Gaussian vectors of $\R^P$
\begin{equation*}
{\bf v}_{\ell} = \left( \frac{P_{\ell,k-p}}{\lambda_{k-p}-\lambda_\ell},  \frac{P_{\ell,k-p+1}}{\lambda_{k-p+1}-\lambda_\ell} 
,  \hdots , \frac{P_{\ell, k+p}}{\lambda_{k+p} -\lambda_\ell} \right)^\dagger\,.
\end{equation*}


It is easily seen that in fact (changing to the equivalent notation for the summation on $\ell$ in term of $a,b$)  
 \begin{equation*}
\Sigma =  \sum_{ \ell : \lambda_\ell \not\in [a-\delta;b+\delta]} {\bf v}_\ell {\bf v}_{\ell}^\dagger 
\end{equation*}

This matrix $ {\bf v}_\ell {\bf v}_{\ell}^\dagger$ is clearly the matrix of a projector on ${\bf v}_\ell $ and has only one non-zero eigenvalue which is equal to
$$\sigma(\lambda_\ell)= || {\bf v}_\ell ||_2^2 = \sum_{j \in \{k-p;\dots;k+p\}} \left(\frac{P_{\ell j}}{\lambda_j-\lambda_\ell}\right)^2$$
which can be approximated in the limit of large matrices ($P\to \infty$) by 
$$\sigma(\lambda_\ell) \to  \int_a^b d\lambda \frac{\rho(\lambda)}{(\lambda-\lambda_\ell)^2}.$$
%
%
The resolvent $Z^\ell(z)\equiv \frac1P \tr((z-{\bf v}_\ell {\bf v}_{\ell}^\dagger)^{-1}) $ of the matrix ${\bf v}_\ell {\bf v}_{\ell}^\dagger$ is equal to:
\footnote{Resolvents are usually denoted by the letter $G$, but we do not want to confuse the reader with the overlap matrix $G$ of 
which we compute the singular value spectrum.}
\begin{equation*}
Z^\ell(z) = \frac1P \left(\frac{1}{z-\sigma(\lambda_\ell)} + \frac{P-1}{z}\right)\,.
\end{equation*}
The Blue function, which by definition is the functional inverse of the resolvent $ B^\ell(Z^\ell(z))=z$, can be computed to first order in $1/P$:
\begin{equation*}
B^\ell(z) = \frac1z +\frac1P \frac{\sigma(\lambda_\ell)}{1-\sigma(\lambda_\ell) z}
\end{equation*}
Finally, the Red function, defined as $ R^\ell(z) \equiv  B^\ell(z) - \frac1z$, is given by:
\begin{equation*} 
R^\ell(z) =\frac1P \frac{\sigma(\lambda_\ell)}{1-\sigma(\lambda_\ell) z}
\end{equation*}

\begin{figure}[h!btp] 
		\center
		\includegraphics[scale=0.8]{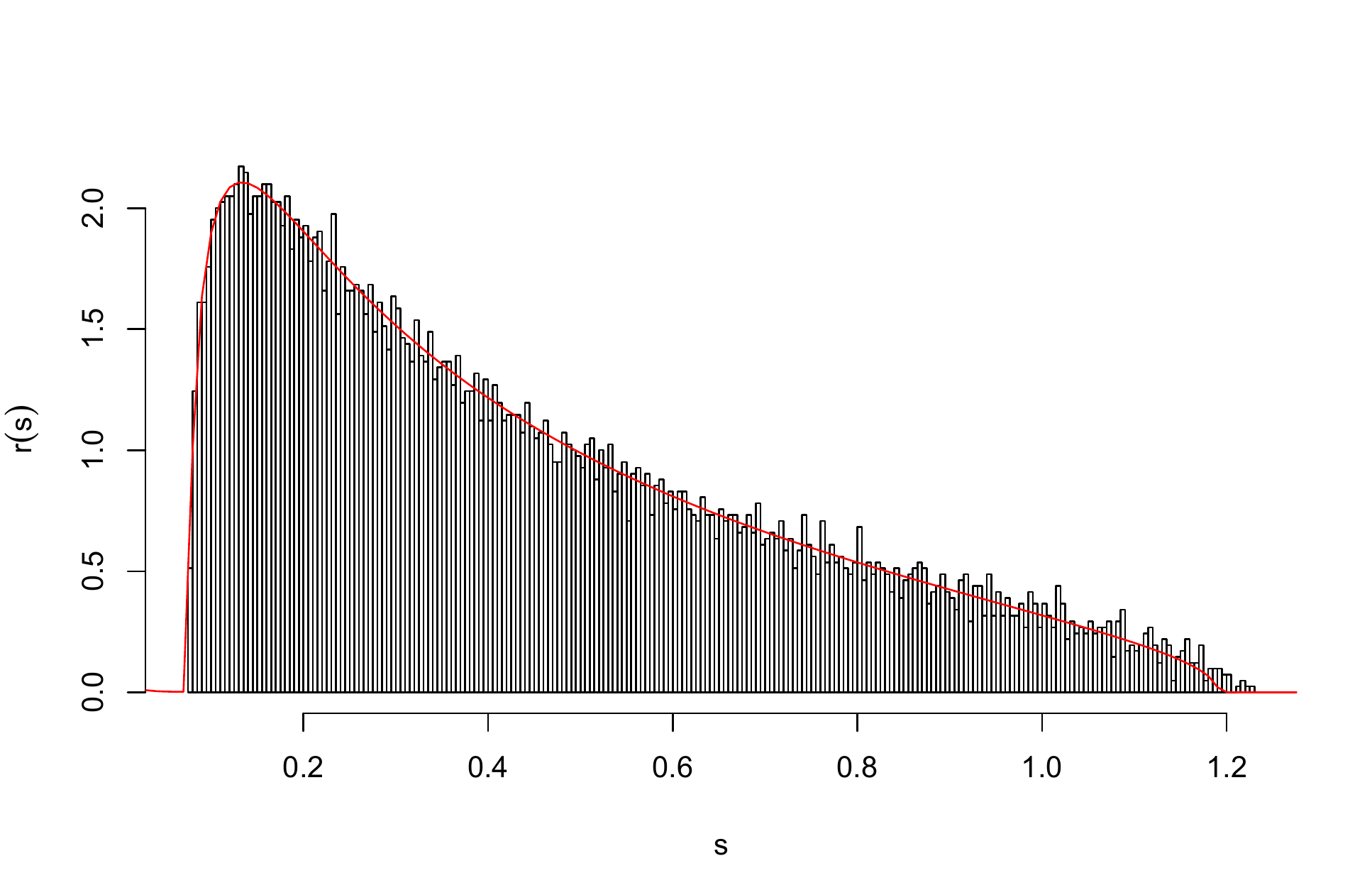}
                \caption{\small{The histogram represents a numerical simulation of the density of states of the matrix $\bf \Sigma$ (computed with $15$ independent samples).
                The red curve is the theoretical corresponding density for $r(s)$ obtained by solving numerically \eqref{eq_resolvent}.   
                For this figure, we chose $a=0,b=0.5,\delta=0.5$. We chose the parameters $N$ and $\varepsilon$ so as to approach the "less perturbative" regime 
                where $1/\sqrt{N}\ll \varepsilon\ll 1$ for this figure as 
                $N=4000$ and $\varepsilon=0.1$. }}      \label{fig_weak_pertub}
\end{figure}

The trick, coming from the theory of free matrices, is to use the additive property of the Red function (also called R-transform) for the asymptotically 
free matrices ${\bf v}_\ell {\bf v}_{\ell}^\dagger$.
Essentially, the R-transform of the matrix ${\bf \Sigma}$ can be computed as the sum of the R-transforms of the matrices ${\bf v}_\ell {\bf v}_{\ell}^\dagger$: 
\begin{equation*}
R(z) =  \sum_{\ell \not \in \{k-q;\dots;k+q\}} R^\ell(z) =  \frac1P\sum_{\ell \not \in \{k-q;\dots;k+q\}} \frac{\sigma(\lambda_\ell)}{1-\sigma(\lambda_\ell) z}
\end{equation*}
Finally, the Blue function of ${\bf \Sigma}$ is:
\begin{align*}
B(z) = \frac1z +  \frac1P\sum_{\ell \not \in \{k-q;\dots;k+q\}} \frac{\sigma(\lambda_\ell)}{1-\sigma(\lambda_\ell) z}
\end{align*}
which can be approximated in the limit of large $P$ as:
\begin{align}\label{eq_BlueFunction}
B(z) = \frac1z + \frac{1}{N_a^b} \int_{[-2;2]\setminus[a-\delta;b+\delta]} d\lambda \frac{ \rho(\lambda)\sigma(\lambda)}{1-\sigma(\lambda) z}.
\end{align}
where we note here and below $N_a^b := \int_a^b\rho(\lambda) d\lambda$. Rewriting equation \eqref{eq_BlueFunction} in terms of the resolvent gives our central result:
\begin{equation}\label{eq_resolvent}
\boxed{z = \frac{1}{Z(z)} + \frac{1}{N_a^b} \int_{[-2;2]\setminus[a-\delta;b+\delta]} d\lambda \frac{ \rho(\lambda)\sigma(\lambda)}{1-\sigma(\lambda) Z(z)}\,.}
\end{equation}
%
%
Equation \eqref{eq_resolvent} characterizes the density of states of the matrix $\bf \Sigma$ in the limit of large dimension. 
We ran numerical simulations to test the validity of Eq. \eqref{eq_resolvent} in the regime $1/\sqrt{N} \ll \varepsilon \ll 1$, see Fig. \ref{fig_weak_pertub}. 
The agreement is excellent. It would be interesting to run this numerical test for very large values of $N$ (here we took $N=4000$) so as to fully reach the regime $1/\sqrt{N} \ll \varepsilon \ll 1$. However, this becomes numerically demanding, and we leave this study for future work.

We now want to extract 
the qualitative informations 
about the distribution of all singular values of the matrix ${\bf G}$ from this equation. 
In particular, we will show in the next subsection that the density of singular values has a compact support for which we characterize the left and right edges. We also 
study the shape of this distribution in the two asymptotic regimes $\Delta\ll\delta$ and $\delta \ll \Delta \ll 1$.

\subsection{Qualitative properties of the spectrum of ${\bf \Sigma}$}\label{qual_prop_spectrum}

\subsubsection{Right and Left edges}
 
The relation between the resolvent $Z$ and the density of states $r(s)$ of the matrix ${\bf \Sigma}$ is 
$\lim_{\omega\to 0} \Im Z(s-i\omega)= \pi r(s), s \in \R$. Note that one should not confuse the density of 
states $\rho(\lambda)$ of the original matrix ${\bf H}_0$ with the density of eigenvalues $r(s)$ of ${\bf \Sigma}$.

From equation \eqref{eq_resolvent}, we can derive a system of equations for the real ($g(s)$) and imaginary ($r(s)$) parts of $Z(s)$, for
$\omega \to 0$:
\begin{align}\label{eq_g}
 s&= \frac{g(s)}{g(s)^2+\pi^2 r(s)^2} + \frac{1}{N_a^b} \int\limits_{[-2;2]\setminus[a-\delta;b+\delta]}dx\frac{\rho(x)\sigma(x)(1-\sigma(x)g(s))}{(1-\sigma(x)g(s))^2+\sigma(x)^2
\pi^2 r(s)^2}\,,\\
0&=  r(s)\left(\frac{-1}{g(s)^2+\pi^2 r(s)^2} + \frac{1}{N_a^b} \int\limits_{[-2;2]\setminus[a-\delta;b+\delta]}dx\frac{\rho(x)\sigma(x)^2}{(1-\sigma(x)g(s))^2+\sigma(x)^2
\pi^2 r(s)^2}\right)\,. \label{eq_r=0} 
\end{align}

The second equation \eqref{eq_r=0} always admits the solution $r=0$. Plugging $r=0$ into the first equation gives:
\begin{equation}\label{eq_gwithr=0}
s = \frac{1}{g(s)} +  \frac{1}{N_a^b} \int\limits_{[-2;2]\setminus[a-\delta;b+\delta]}dx\frac{\rho(x)\sigma(x)}{1-\sigma(x)g(s)}
\end{equation}
Equation \eqref{eq_gwithr=0} implies the asymptotic relation $g(s) \sim_{s \to \infty} 1/s$ and therefore large positive values of $s$ correspond to small values of $g(s)$. Set 
\begin{equation} \label{m_0}
m_0\equiv \max_{x\in[-2;2]\setminus[a-\delta;b+\delta]} \sigma(x),
\end{equation} 
the Right Hand Side (RHS) of the above equation is well defined provided $g(s) \in (0;1/m_0)$ . However, when $g(s) \to 0^+$ or when $g(s) \to (1/m_0)^-$, the RHS tends to $+\infty$. Thus, on the 
interval $g(s) \in (0;1/m_0)$, the RHS must reach a minimum which corresponds to the right edge of the density of states. 
The point $\bar{g} \in (0;1/m_0)$ for which this minimum is reached verifies: 
\begin{equation}\label{eq_barg}
-\frac{1}{\bar{g}^2} + \frac{1}{N_a^b}  \int\limits_{[-2;2]\setminus[a-\delta;b+\delta]}dx\frac{\rho(x)\sigma(x)^2}{(1-\sigma(x)\bar{g})^2} = 0\,,
\end{equation}
and we can compute the right edge of the spectrum $s_{max}$ from:
\begin{equation}\label{smax}
s_{max} = \frac{1}{\bar{g}} +  \frac{1}{N_a^b} \int\limits_{[-2;2]\setminus[a-\delta;b+\delta]}dx\frac{\rho(x)\sigma(x)}{1-\sigma(x)\bar{g}}.
\end{equation}

We can now turn to the left edge of the spectrum. Equation \eqref{eq_gwithr=0} implies also the asymptotic relation $g(s) \to  -\infty$ when $s \to 0$ 
and therefore small positive values of $s$
correspond to large negative values of $g(s)$. The RHS of equation \eqref{eq_gwithr=0} is well defined for negative values of $g(s)$; it goes 
to $0^-$ for very large and negative values of $g(s)$, and goes to $-\infty$ for $g(s)=0^-$, so it has a positive maximum somewhere in between. 
The value of this maximum corresponds to the left edge of the density of states and can be computed numerically like for the right edge. 
The point $\tilde{g} \in(-\infty;0)$ for which this maximum is reached verifies the same equation as $\bar{g}$ above, and the left edge  $s_{min}$ is now
given by:
\begin{equation}\label{smin}
s_{min} = \frac{1}{\tilde{g}} +  \frac{1}{N_a^b} \int\limits_{[-2;2]\setminus[a-\delta;b+\delta]}dx\frac{\rho(x)\sigma(x)}{1-\sigma(x)\tilde{g}}.
\end{equation}

\subsubsection{Small fluctuations regime $\Delta \ll \delta$}

We first consider the case where $\Delta \equiv b-a \ll \delta$, corresponding to $P \ll Q$, in particular the dimension of the perturbed subspace is 
much larger than the dimension of the unperturbed space and so the perturbed space almost surely spans the unperturbed subspace. 
We therefore expect small fluctuations in this regime. Equation \eqref{eq_resolvent} can 
be solved explicitly in this case. It is in fact possible to perform an asymptotic expansion in $\sigma(x)$, 
which is very small compared to $1$ for all $x\in[-2;2]\setminus[a-\delta;b+\delta]$ and then to solve equation \eqref{eq_resolvent}.  

More precisely, in this regime, we have for all $x\in[-2;2]\setminus[a-\delta;b+\delta]$:
\begin{equation*}
\sigma(x) \approx \frac{\rho(a)}{(x-a)^2} \Delta\,.  
\end{equation*}
We plug this approximation in equation \eqref{eq_resolvent} to obtain 
\begin{align*}
z &= \frac{1}{Z(z)} + \frac{\Delta \times \rho(a)}{N_a^b} \int\limits_{[-2;2]\setminus[a-\delta;b+\delta]} dx \frac{\rho(x)}{(x-a)^2 - \Delta \times \rho(a) Z(z)}\\
& \approx \frac{1}{Z(z)} + \int\limits_{[-2;2]\setminus[a-\delta;b+\delta]} dx \frac{\rho(x)}{(x-a)^2} +  \Delta \times \rho(a) Z(z) \int\limits_{[-2;2]\setminus[a-\delta;b+\delta]} dx  \frac{\rho(x)}{(x-a)^4}\,.
\end{align*}
Now setting $A\equiv \int_{[-2;2]\setminus[a-\delta;b+\delta]} dx \frac{\rho(x)}{(x-a)^2} $ and $B \equiv \Delta \times \rho(a) \int_{[-2;2]\setminus[a-\delta;b+\delta]} dx  
\frac{\rho(x)}{(x-a)^4}$, we see 
that $Z(z)$ is solution of the polynomial equation of degree two:
\begin{equation}\label{eq_polynome_G}
B Z(z)^2 + (A-z) Z(z) +1 = 0. 
\end{equation}
For $z=s \in \R$, this equation has solutions with non-zero imaginary part only if $s\in[A-2\sqrt{B};A+2\sqrt{B}]$, which are given by
$$Z(z) = \frac{-A+s\pm i \sqrt{4B-(A-s^2)}}{2B}.$$ 
Using the relation $\lim_{\omega\to 0} \Im Z(s-i\omega)= \pi r(s)$ for $s \in \R$, we find that $r(s)$ in this regime is given by
the semi-circle law
\begin{equation}\label{eq_density_r_small_fluct}
r(s) = \frac{1}{2B\pi} \sqrt{4B-(A-s)^2}, \quad A-2\sqrt{B}< s <A+2\sqrt{B} .
\end{equation}
This result is consistent with \eqref{D_GOE} since, in this regime, $D(a,b;\delta)= \varepsilon^2 A$. 

Note that in the particular regime $\Delta \ll \delta \ll 1$, the quantity $B$ is proportional to $\Delta/\delta^3$ and is therefore much smaller than $A^2 \propto 1/\delta^2$, meaning that 
$r(s)$ becomes concentrated around $s=A$, with fluctuations of order $\sqrt{\Delta/\delta^3}$.  
This result is also consistent with the direct calculation of the root-mean squared fluctuations of $s$, as obtained in Appendix B, see equation \eqref{small_fluc_sd}.

\begin{figure}[h!btp] 
		\center
		\includegraphics[scale=0.5]{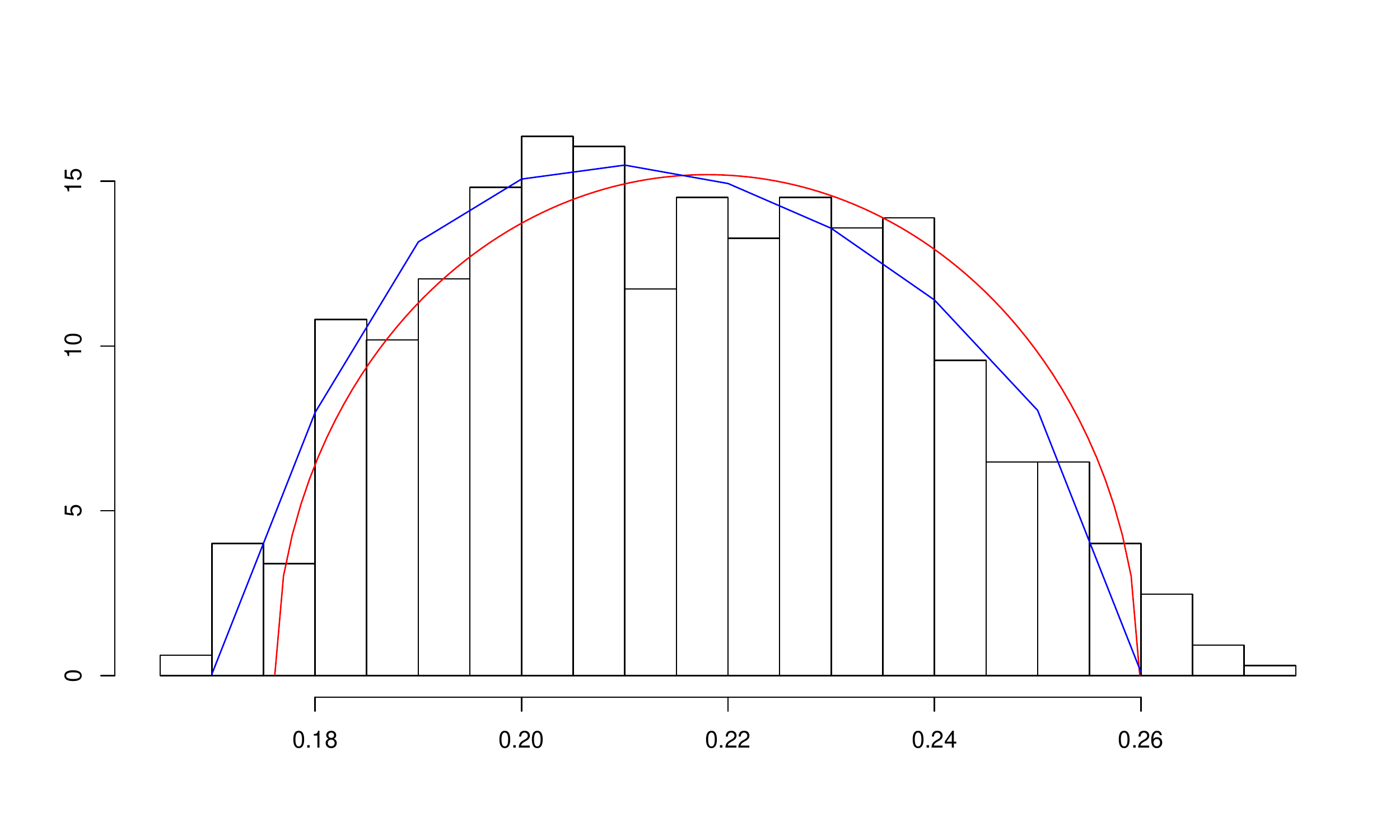}
                \caption{\small{The histogram represents a numerical simulation of the density of states of the matrix $\bf \Sigma$ (computed with $100$ independent samples).
                The red curve is the theoretical corresponding density for $r(s)$ in small fluctuations regime given by \eqref{eq_density_r_small_fluct}. The blue curve 
                represents also the theoretical density $r(s)$ but computed numerically by solving directly the system \eqref{eq_g} and \eqref{eq_r=0}.  
                For this figure, we chose $a=0,b=0.01,\Delta = 0.01,\delta=1$.}}        \label{density_r_smallfluct.pdf}
\end{figure}

\subsubsection{Strong fluctuations regime $\delta \ll \Delta \ll 1$}

To simplify notations, we will suppose in the following that $a$ and $b$ are such that $\rho(a)\geq\rho(b)$.
Let us first consider the right edge $s_{max}$ as given by \eqref{smax}. We need to find an asymptotic expansion in this regime of the $\bar{g} \in (0;1/m_0)$ which verifies \eqref{eq_barg}.  
So we start by defining $\alpha:= \bar{g} m_0 \in (0;1)$ and investigate its behavior when $\delta \ll \Delta \ll 1$.
Since $m_0\sim_{\delta\to0} \rho(a)/\delta$, equation \eqref{eq_barg} now rewrites as 
\begin{equation}\label{eq_alpha}
\frac{\alpha^2 \delta^2}{\rho(a)^2 N_a^b}  \int\limits_{[-2;2]\setminus[a-\delta;b+\delta]} dx \, \frac{\rho(x)\sigma(x)^2}{(1-\alpha \frac{\sigma(x)}{m_0})^2} \sim 1\,.
\end{equation}
In the limit $\Delta \ll 1$, it is easy to see that the function $\sigma$ can be written for $x< a$ as 
\begin{equation}\label{forme_sigma}
\sigma(x) = \frac{\rho(a)}{a-x} f\left(\frac{a-x}{\Delta}\right)\,, 
\end{equation}
where the function $f$ verifies $f(u)\sim_{u\rightarrow 0} 1$ and $f(u)\sim_{u\to \infty} 1/u$.
Using \eqref{forme_sigma}, we can write 
\begin{align*}
\int_{-2}^{a-\delta} dx \, \frac{\rho(x)\sigma(x)^2}{(1-\alpha \sigma(x) \frac{\delta}{\rho(a)})^2} & = 
\rho(a)^2 \int_{-2}^{a-\delta} dx \, \frac{\rho(x)f^2(\frac{a-x}{\Delta})}{(a-x-\alpha f(\frac{a-x}{\Delta}) \delta)^2} \\ 
&= \frac{\rho(a)^2}{\Delta} \int_{\frac{\delta}{\Delta}}^{\frac{a+2}{\Delta}} du\, \frac{\rho(a-u\Delta) f^2(u)}{(u-\alpha \frac{\delta}{\Delta} f(u))^2}\,,
\end{align*} 
where we did the change of variables $u=(a-x)/\Delta$ for the last line. In the limit $\delta \ll \Delta \ll 1$, this last integral is dominated by the region 
where $u$ is small and $f(u)\sim 1$. We thus have 
\begin{align*}
\int_{-2}^{a-\delta} dx \, \frac{\rho(x)\sigma(x)^2}{(1-\alpha \sigma(x) \frac{\delta}{\rho(a)})^2} &\sim \frac{\rho(a)^3}{\Delta} \int_0^{+\infty} \frac{du}{(u-\alpha
\frac{\delta}{\Delta})^2} \\
&\sim \frac{\rho(a)^3}{\delta} \frac{1}{1-\alpha}\,.
\end{align*}  
Then, using \eqref{eq_alpha} and with the same argument now for $x>b$, we get 
\begin{equation*}
\alpha = 1 -  \frac{2\delta}{\Delta}\,.
\end{equation*}
The corresponding $\bar{g}$ is $\bar{g}= \delta/\rho(a) (1-2\delta/\Delta)$ and plugging this value of $\bar{g}$ in \eqref{smax} gives
\begin{equation*}
s_{max} \sim \frac{\rho(a)}{\delta} + \frac{1}{N_a^b} \int\limits_{[-2;2]\setminus[a-\delta;b+\delta]} dx \, \frac{\rho(x)\sigma(x)}{1-\sigma(x)
\bar{g}}\,.
\end{equation*}
But, it is plain to check that the second term is of order at most $\ln(\delta/\Delta)/\Delta \ll 1/\delta$ in the limit $\delta \ll \Delta \ll 1$. 

\begin{figure}[h!btp] 
		\center
		\includegraphics[scale=0.75]{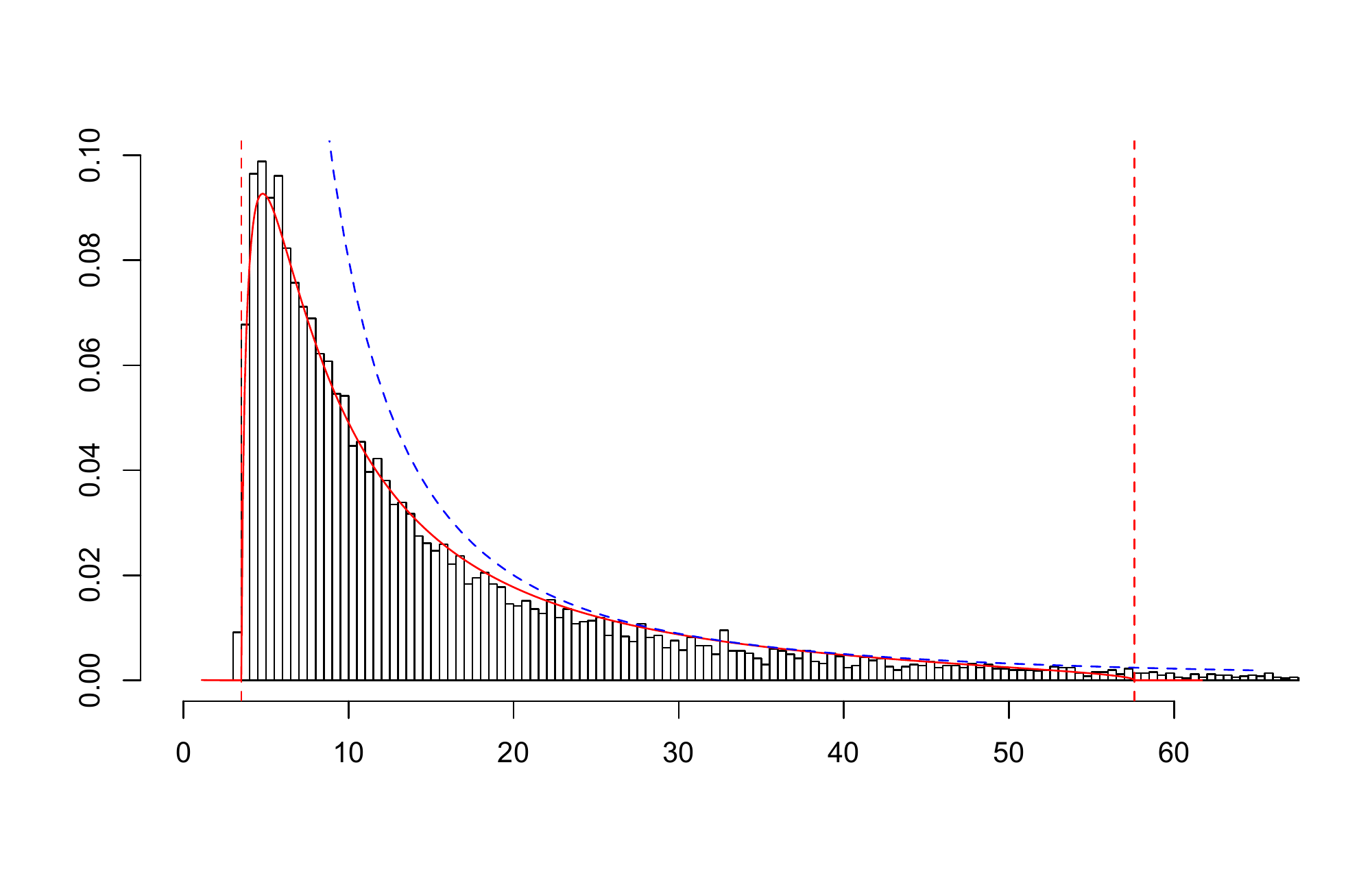}
                \caption{\small{The histogram represents a numerical simulation of the density of states of the matrix $\bf \Sigma$ (computed with $20$ independent samples).
                The red curve is the theoretical corresponding density for $r(s)$, it is computed numerically by solving the system \eqref{eq_g} and \eqref{eq_r=0}. The red dotted vertical lines 
                show the left and right edges of the density $r(s)$.
                The blue dotted curve is the graph of the function $8/x^2$. 
                For this figure, we chose $a=0,b=0.1,\Delta=0.1,\delta=0.01$.}}         \label{density_r_largefluct.pdf}
\end{figure}

\begin{figure}[h!btp] 
		\center
		\includegraphics[scale=0.8]{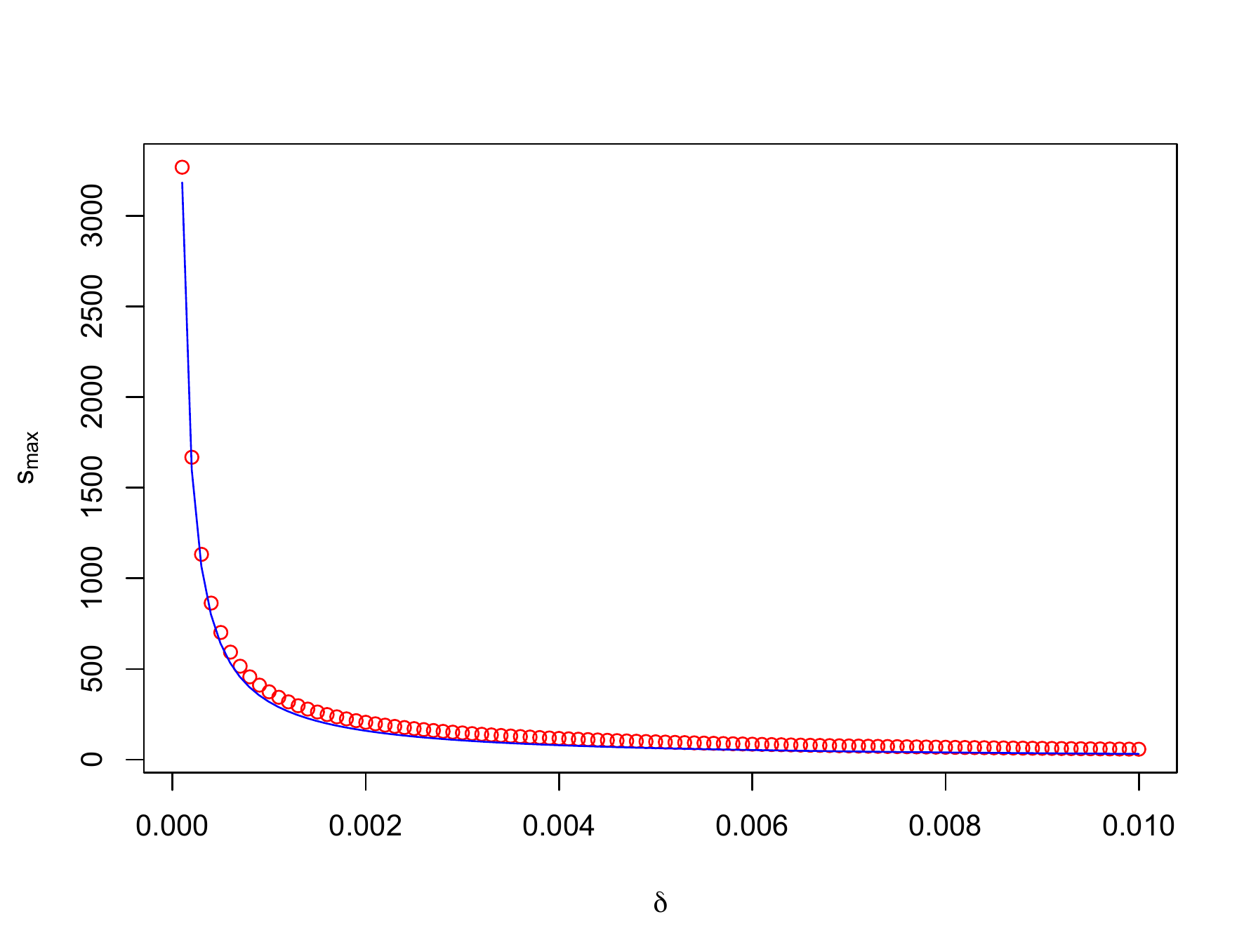}
                \caption{\small{The points represent the function $s_{max}(\delta)$ as a function of $\delta$. They are computed numerically through equations \eqref{eq_barg} and \eqref{smax}. 
                The blue dotted line is the function 
                $\delta \rightarrow \rho(a)/\delta$. In this figure, we chose $a=0,b=0.1,\Delta=0.1$.}}        \label{fig_s_max}
\end{figure}

The determination of $s_{min}$ proceeds similarly, and the calculations are detailed in Appendix C. The final result is that
\begin{equation*}
s_{min} = \frac{c\rho(a)}{\Delta},
\end{equation*}
where $c>0$ is a number of order unity that can be determined if needed. 

To summarize, in this regime, the minimum and maximum eigenvalues $s_{min}$ and $s_{max}$ of the random matrix $\Sigma$ are asymptotically given by:
\begin{align*}
s_{min} \sim \frac{c \rho(a)}{\Delta}\, \ll \, s_{max} \sim \frac{\rho(a)}{\delta}.
\end{align*} 
We verified the result for $s_{max}(\delta)$ with numerical simulations (see Fig. \ref{fig_s_max}).

Together with the exact result on the average value of $r(s)$ in this regime (given by $D(a,a+\Delta;\delta)/\varepsilon^2$) and its variance computed in 
Appendix B, we conjecture that the asymptotic behaviour of $r(s)$ in the region $s_{min} \ll s \ll s_{max}$ is given by:
\begin{equation}\label{r(s)}
r(s) \propto \frac{s_{min}}{s^2}\,. 
\end{equation}
Since the integral of $s r(s)$ is logarithmically divergent (but cut-off at  $s_{min}$ and $s_{max}$), it is easy to see that this form reproduces exactly 
the logarithmic behavior of $D(a,a+\Delta;\delta)$ in this regime, see equation \eqref{cases}. On the other hand, the integral of $s^2 r(s)$ is dominated by its upper bound, 
leading to a variance of the spectrum given by $s_{min} \times s_{max}$, in agreement with the exact result obtained in Appendix B, see equation \ref{strong_fluc_sd}. Therefore, in this regime, the situation is particularly interesting: while most eigenvalues are close to $s_{min}$, there is a slow power-law tail in $r(s)$ that makes the 
average of $s$ logarithmically divergent when $\delta \to 0$. This is why we call this a strong fluctuation regime: the `overlap' distance $D$ between 
the initial and the target spaces is large because a relatively small number of directions are completely lost.

\section{Eigenvector stability for covariance matrices}\label{section_cov_matrix}
\subsection{Eigenvectors of Spiked matrices}

In this subsection, we will assume that $({\bf C}_N)$ is a sequence of positive definite matrices. 
We will denote by $\lambda_1^N,\dots,\lambda_N^N$ the eigenvalues of $({\bf C}_N)$ in decreasing order and we will suppose that
\begin{itemize}
\item there exists a fixed number $k< N$, $q\in (0;1)$ and $(\lambda_1>\dots>\lambda_k>(1+\sqrt{q})^2)$ such that 
$$(\lambda_1^N,\dots,\lambda_k^N) \rightarrow_{N\to \infty} (\lambda_1,\dots,\lambda_k).$$
\item the empirical measure $\mu_N\equiv \frac{1}{N} \sum_{i=k+1}^N \delta_{\lambda_i}$ converges in the limit 
of large $N,T$ with $N/T=q$ to the Marchenko-Pastur distribution whose density with respect to Lebesgue measure is given by 
$$ \rho(x) \equiv \frac{1}{2\pi q x} \sqrt{(\gamma_+-x)(x-\gamma_-)},\quad a<x<b,$$
where $\gamma_- \equiv(1-\sqrt{q})^2$ and $\gamma_+ \equiv (1+\sqrt{q})^2$.   
\end{itemize} 
For each $N$, ${\bf C}_N$ is the true covariance matrix (also called ``population covariance matrix"). This particular choice
for the shape of the matrices ${\bf C}_N$ is rather natural in view of applications. For example, in financial market, the
correlation (or covariance) matrix has $k$ isolated eigenvalues well separated from the other eigenvalues which form the noisy part 
of the spectrum ({\it Marchenko-Pastur sea} or the {\it bulk}).      

We now
consider the associated empirical covariance matrix ${\bf E}_N$ defined as: 
\begin{equation*}
{\bf E}_{N,ij} \equiv \frac1T\sum_{t=1}^{T} r_i^t r_j^t
\end{equation*}
where the $(r_1^t,\dots,r_N^t), 1\leq t \leq T$ are i.i.d. Gaussian vectors of covariance ${\bf C}_N$.

The question we ask in this subsection is: how close are the eigenvectors of ${\bf E}_N$ to those of the matrix 
 ${\bf C}_N$ ? In the following two paragraphs, we treat the two cases of the eigenvectors associated to eigenvalues in the Marchenko-Pastur sea  
and of those associated to the isolated eigenvalues $\lambda_1,\dots,\lambda_k$.

This question falls under the scope of section \ref{stat_tools} since
the matrix ${\bf E}_N$ can be written as a perturbation of the matrix ${\bf C}_N$. Indeed we have:
\begin{equation}\label{E=C+eps}
{\bf E}_N = {\bf C}_N + {\bf \mathcal{E}}_N, \quad {\rm {with}} \quad \mathcal{E}_{N,ij} = \frac{1}{T} \sum_{t=1}^{T} r_i^t r_j^t - C_{N,ij}.
\end{equation}
and the matrix elements of $\mathcal{E}$ are (because of the Central Limit Theorem) of order $1/\sqrt{T}$ which is 
much smaller than $1$ as $T$ is large. 
However, this problem is of different nature than the one treated in section \ref{GOE} because of the non-trivial dependance structure for the matrix elements of the perturbation matrix $\mathcal{E}$. It is given by 
\begin{equation}\label{cov_struc_mcorr}
\overline{\mathcal{E}_{N,ij} \mathcal{E}_{N,k\ell}} =  (C_{N,ik} C_{N,j\ell} + C_{N,i\ell} C_{N,jk})/T.
\end{equation}

In the whole current section, $\overline{\cdot\cdot\cdot}$ denotes an averaging over the $r_i^t$.
 
\subsubsection{Eigenvectors in the Marchenko-Pastur sea}

The results of subsections \ref{Distance_eigenspaces}, \ref{full_distrib} and \ref{qual_prop_spectrum} can be extended to this context. 
We consider the subspace of eigenvectors of ${\bf C}_N$ corresponding to all the eigenvalues $\lambda$ contained in a certain finite interval $[a,b]$ included 
in the Marchenko-Pastur sea $[\gamma_-,\gamma_+]$. 
We want to compute the distance $D$ between this subspace and the subspace spanned by 
the perturbed eigenvectors of ${\bf E}_N$ corresponding to all eigenvalues of ${\bf E}_N$ contained in $[a-\delta,b+\delta]$, where $\delta$ is a positive parameter. 
Using formula \eqref{expr_D_dimension_finie} as before, we find that in 
the limit of large $N,T$ (with $N/T= q$), as soon as $\delta>0$, the mean overlap distance $\overline{D}$ is given (using \eqref{cov_struc_mcorr} for the averaging) 
by:
\begin{equation}\label{det(G^TG)cov}
\overline{D}(a,b;\delta) \sim \frac{1}{2 T N_a^b} \int_a^b d\lambda \int\limits_{[-2;2]\setminus[a-\delta;b+\delta]} 
d\lambda' \,\, \frac{\lambda \lambda' \rho(\lambda)\rho(\lambda')}{(\lambda - \lambda')^2},
\end{equation}
where $\rho(\lambda)$ is now the Marchenko-Pastur distribution of parameter $q=N/T$. Obviously, when $T$ is infinite, $D=0$ since ${\bf E}_N={\bf C}_N$.

For the singular value density of states $r(s)$, the resolvent of the matrix ${\bf \Sigma}$ defined as ${\bf \Sigma}\equiv T ({\bf I} - {\bf G G^\dagger})$  
now verifies:
\begin{equation}
z = \frac{1}{Z(z)} + \frac{1}{N_a^b} \int\limits_{[-2;2]\setminus[a-\delta;b+\delta]} d\lambda \frac{ \rho(\lambda)\nu(\lambda)}{1-\nu(\lambda) Z(z)}\,
\end{equation}
where $\nu$ is defined as $\nu(\lambda) \equiv \lambda \int_a^b dx \frac{\rho(x)}{(x-\lambda)^2}.$
As before, it is easy to show that the density of states of $\bf \Sigma$ is compactly supported and to find numerical evaluations of the left and right edges. One can 
also study the limit shape of 
the density of states in the two regimes $\Delta\ll\delta$ and $\delta\ll\Delta\ll 1$, with results very similar to the GOE ones above. 

The matrix ${\bf GG^\dagger}$ in this case gives a precise information on the relationship between the eigenvectors of the {\it population covariance matrix} (or true covariance matrix) 
$\bf C_N$ and the eigenvectors of the sample covariance matrix $\bf E_N$. Previous works along these lines can be found in \cite{Peche,Bai}.   

\subsubsection{Isolated eigenvectors}

In this paragraph, we now consider the case of eigenvectors associated to isolated eigenvalues $\lambda_1,\dots,\lambda_k$. We denote by 
$|\phi_1\rangle,|\phi_2\rangle,\dots,|\phi_k\rangle$ the corresponding eigenvectors of ${\bf C}_N$ and by $|\phi_1'\rangle,|\phi_2'\rangle,\dots,|\phi_k'\rangle$
\footnote{The vectors $|\phi_i\rangle$ and $|\phi_i'\rangle$ depend on $N$ but to simplify notations, we drop the subscript.}
the corresponding eigenvectors of ${\bf E}_N$.

To understand precisely how the $|\phi_i'\rangle$ decompose on the basis of the $|\phi_j\rangle$ in the limit of large $N$,
we want to compute the limit of the average local density of states for each state $|\phi_i'\rangle$ ($1\leq i\leq k$), that is the probability measure 
\begin{equation*}
\nu_N^{(i)}(\lambda) \equiv \frac1N \sum_{j=1}^N\overline{ \langle \phi_i' | \phi_j \rangle^2} \,\, \delta(\lambda -\lambda_j)
\end{equation*} 
where $\overline{\cdot\cdot\cdot}$
denotes an average over ${\bf E}_N$. This expresses the way $|\phi_i'\rangle$ is scattered over the unperturbed eigenvectors.

Perturbation theory again allows to compute the quantities $\langle \phi_i' | \phi_j \rangle^2$ for $i\not=j$:
\begin{align*}
\overline{\langle \phi_i' | \phi_j \rangle^2} = \overline{\frac{\langle\phi_i|{\bf \mathcal{E}}|\phi_j\rangle^2}{(\lambda_i^N-\lambda_j^N)^2}} 
= \frac{1}{T} \frac{\lambda_i^N \lambda_j^N}{(\lambda_i^N-\lambda_j^N)^2} \sim_{N\to\infty} \frac{1}{T} \frac{\lambda_i \lambda_j}{(\lambda_i-\lambda_j)^2},
\end{align*}
and for $i=j$, \[\overline{\langle \phi_i' | \phi_i \rangle^2} = 1 - \sum_{k\not=i} \overline{{\frac{\langle\phi_i|{\bf \mathcal{E}}|\phi_k\rangle^2}{(\lambda_i^N-\lambda_k^N)^2}}}
\sim_{N\to \infty}1- \frac{1}{T} \sum_{k \not=i}\frac{\lambda_i \lambda_k}{(\lambda_i-\lambda_k)^2}.\]
Note that the random variables $\langle\phi_i|{\bf \mathcal{E}}|\phi_j\rangle, i \not= j$ are uncorrelated. 
Thus, the local density of states $\nu_N^{(i)}$ has $k$ atoms and (for large $N,T$ with $N/T=q$) admits a continuous density in the Marchenko-Pastur sea. 
The atom are localized on the $\lambda_j, j=1,\dots,k$ and have weights $\frac{1}{T} \frac{\lambda_i \lambda_j}{(\lambda_i-\lambda_j)^2}$ for $j\not=i$. The continuous density in the Marchenko-Pastur sea $[\gamma_-,\gamma_+]$ is given by: 
\begin{equation}
\frac1T \frac{\lambda_i \lambda}{(\lambda_i-\lambda)^2} \rho(\lambda) d\lambda, \quad \gamma_- < \lambda < \gamma_+.
\end{equation}
This asymptotic for the probability measure $\nu_N^{(i)}$ has been verified with numerical simulations. 

\subsection{Stability of eigenspaces}\label{eigenspace}

We now want to characterize the stability of the subspace spanned by the eigenvectors associated to the (largest) isolated eigenvalues. 
The theory we develop here provides a precise estimate of the amount of eigenspace instability induced by measurement noise. 
This sets a benchmark that will allow us to detect any extra dynamics of the eigenvectors of the correlation matrix of stock returns in financial markets 
not explained by measurement noise and therefore attributable to a genuine evolution of the market (see section \ref{empirical_results}).

As shown by Eq. (\ref{E=C+eps}) above, the sample covariance matrix ${\bf E}$\footnote{As $N$ does not have to be necessarily large in this subsection 
and in the next section, we drop the subscript $N$ for 
the matrices ${\bf C}$ and $\bf E$.} is a perturbed version of ${\bf C}$.   
Using again the framework of section \ref{stat_tools}, one can calculate the distance (or overlap) between 
the top $P$ eigenvectors of the true correlation matrix ${\bf C}$ and the top $Q$\footnote{We take $Q\geq P$ as before in section \ref{stat_tools} .} 
eigenvectors of the empirical correlation matrix ${\bf E}$.  

Provided $T$ is large enough for the above perturbation theory to be valid, and upon averaging over the measurement noise, one gets the 
following expression for the overlap distance $D$:
\begin{equation} \label{noiseVecteur}
\overline{D}(P,Q) = \frac{1}{2TP} \sum_{i=1}^{P} \sum_{j=Q+1}^{N} \frac{\lambda_i \lambda_j}{(\lambda_i - \lambda_j)^2}, 
\end{equation}  
where the $\lambda_i$s are the eigenvalues of ${\bf C}$, in decreasing order. 

Note that one can extend the previous result \eqref{noiseVecteur} to the case where the vectors $(r_1^t,\dots,r_N^t),t\geq 0$ are distributed according to a  
multivariate Student distribution with $\nu$-degrees of freedom 
and covariance matrix $\bf C$. In this case\footnote{see e.g. Eq. (9.28) p. 154 of \cite{bookBP} that replaces Eq. \eqref{cov_struc_mcorr} above.}, Eq. \eqref{noiseVecteur} becomes 
\begin{equation} 
\overline{D}(P,Q) = \left(\frac{\nu-2}{\nu-4}\right) \frac{1}{2TP} \sum_{i=1}^{P} \sum_{j=Q+1}^{N} \frac{\lambda_i \lambda_j}{(\lambda_i - \lambda_j)^2}\,.
\end{equation}  
Note that the Gaussian case corresponds to $\nu \to \infty$. For $\nu \to 4^+$, on the other hand, fluctuations become divergent.

In practice for applications (see section \ref{empirical_results}), one does not know the true correlation matrix ${\bf C}$ and thus it is in fact not possible 
to compute empirically the overlap distance between the eigenvectors of ${\bf C}$ and the eigenvectors of the empirical correlation matrix ${\bf E}$. 
However, if one is given a time series
of empirical correlation matrix $({\bf E}^t)_{t\geq 0}$ defined for all $t$ as 
\be
E_{ij}^t = \frac1T \sum_{u=1}^T r_i^{t+u} r_j^{t+u}\,,
\ee
where, $(r_1^v,\dots,r_N^v),v\geq 0$ are independent Gaussian vectors of covariance matrix $\bf C$,
one can similarly define the distance between the eigenspaces 
of two independent sample covariance matrices 
${\bf E}^s$ and ${\bf E}^t$ (determined on two non overlapping time periods, i.e. such that $|t-s| > T$).
In this case, the above formula Eq. (\ref{noiseVecteur}) is simply multiplied by a factor $2$. 

For the comparison between the eigenvalues of ${\bf E}^s$ and ${\bf E}^t$, one can show using 
perturbation theory (see equation \eqref{pt_eigenvalues} and also equation \eqref{cov_struc_mcorr} for the averaging) 
that the measurement noise is, for $T$ large enough, given by: 
\begin{equation}\label{noise_value}
\overline{\left(\lambda_i^{s} - \lambda_i^{t}\right)^2}_{|t-s|>T} \approx \frac{4 \lambda_i^2}{T}. 
\end{equation} 
where the $\lambda_i$ are the eigenvalues of the matrix ${\bf C}$ measured empirically using the whole period of time and 
where $\overline{\cdot\cdot\cdot}_{|t-s|>T}$ denotes an empirical average over all 
$s,t$ such that ${|t-s|>T}$. As before, if the vectors $(r_1^v,\dots,r_N^v),v\geq 0$ are distributed according to a  
multivariate Student distribution with $\nu$-degrees of freedom and covariance matrix $\bf C$, one finds an extra multiplicative term $(\nu-2)/(\nu-4)$ in \eqref{noise_value}.  

Another characterization of the stability of eigenspaces was proposed by Zumbach \cite{Zumbach}. 
The idea here is to study the stability of the {\it spectral projectors} associated to the top $k$ eigenvalues. The spectral projector of rank $k$ associated to the top $k$ eigenvalues is defined as follows:
\begin{equation*}
\chi_k = \sum_{i=1}^k |\phi_i\rangle\langle\phi_i|\,,
\end{equation*} 
where the $|\phi_i\rangle, i \in \{1,\dots,k\}$ are the eigenvectors of ${\bf C}$. As before the {\it true} spectral projector $\chi_k$ is measured through an empirical 
covariance matrix $\bf E$ and the resulting spectral projector $\chi_k'$ will be affected by measurement noise. The aim is again to compute  
properties of this spectral projector $\chi_k$, so as to be able to separate the measurement noise effect from a true temporal evolution 
of the matrix $\bf C$. 

Using perturbation theory in Eq. (\ref{E=C+eps}), we have:
\begin{align*}
\chi_k' &= \sum_{i=1}^k |\phi_i'\rangle\langle\phi_i'| 
= \sum_{i=1}^k \left(1- \sum_{j \neq i} \frac{\langle \phi_i|{\bf{\mathcal{E}}}|\phi_j \rangle^2}{(\lambda_i - \lambda_j)^2}  \right)|\phi_i\rangle\langle\phi_i|\\
&+ \sum_{i=1}^{k} \sum_{j \neq i} \frac{\langle \phi_i|\mathcal{E}|\phi_j \rangle}{\lambda_i - \lambda_j} \left(|\phi_i\rangle\langle\phi_j| + |\phi_j\rangle\langle\phi_i| \right) \\ 
&+   \sum_{i=1}^{k} \sum_{j \neq i} \alpha_{i,j} (|\phi_i\rangle\langle\phi_j|+|\phi_j\rangle\langle\phi_j|) 
+   \sum_{i=1}^{k} \sum_{j \neq i} \sum_{\ell \neq i}\frac{\langle \phi_i|{\bf{\mathcal{E}}}|\phi_j \rangle \langle \phi_i|{\bf{\mathcal{E}}}|\phi_\ell \rangle }{(\lambda_i - \lambda_j)(\lambda_i - \lambda_\ell)} 
|\phi_j\rangle\langle\phi_\ell |
\end{align*}
where $$\alpha_{i,j} = \frac{1}{\lambda_i - \lambda_j} \left( \sum_{\ell \neq i} \frac{\langle \phi_j|{\bf{\mathcal{E}}}|\phi_\ell \rangle \langle 
\phi_\ell|{\bf{\mathcal{E}}}|\phi_i \rangle }{\lambda_i - \lambda_\ell} - \frac{\langle \phi_i|{\bf{\mathcal{E}}}|\phi_i \rangle \langle \phi_i|{\bf{\mathcal{E}}}
|\phi_j \rangle}{\lambda_i - \lambda_j} \right)\,.$$

Using again equation \eqref{cov_struc_mcorr}, 
\begin{equation*}
\overline{\langle \phi_j|{\bf{\mathcal{E}}}|\phi_i \rangle \langle \phi_\ell |{\bf{\mathcal{E}}}|\phi_i \rangle} = 
\begin{cases}
0 & \quad \text{if $\ell \neq j$},\\
\lambda_j \lambda_i/T &\quad  \text{if $j = \ell, j \neq i$},\\  
2 \lambda_i^2/T & \quad \text{otherwise},
\end{cases}
\end{equation*}
we get:
\begin{equation*}
\overline{\chi_k'} = \sum_{i=1}^k \left(1- \frac{1}{T} \sum_{j \neq i} \frac{\lambda_i \lambda_j }{(\lambda_i - \lambda_j)^2}  \right)|\phi_i\rangle\langle\phi_i| 
+ \frac{1}{T} \sum_{i=1}^{k} \sum_{j \neq i} \frac{\lambda_i \lambda_j }{(\lambda_i - \lambda_j)^2} 
|\phi_j\rangle\langle\phi_j |\,.
\end{equation*}

We see that the vectors $\phi_i, i\in\{1,\dots,N\}$ are also eigenvectors of $\overline{\chi_k'}$, but with shifted eigenvalues. 
More precisely, we have, for $i\leq k$
\begin{align}\label{benchmark_spectrum_1}
\overline{\chi_k'}|\phi_i\rangle =\left( 1- \frac{1}{T} \sum_{j=k+1}^N \frac{\lambda_i\lambda_j}{(\lambda_i-\lambda_j)^2}\right) |\phi_i\rangle\,,
\end{align}
and, for $i >k$,
\begin{align}\label{benchmark_spectrum_2}
\overline{\chi_k'}|\phi_i\rangle =\frac{1}{T} \sum_{j=1}^k \frac{\lambda_i\lambda_j}{(\lambda_i-\lambda_j)^2}\, |\phi_i\rangle \,.
\end{align}
Therefore, in the absence of measurement noise (i.e. for $T \to \infty$), $\overline{\chi_k'}$ has $k$ eigenvalues exactly equal to unity, and $N-k$ eigenvalues equal to zero, as expected since in this case 
$\overline{\chi_k'}=\chi_k$. All the above results will be compared with empirical data
(for the case of financial markets) in section \ref{sect_emp_eigenspace_stability} below.

\section{The case of an isolated top eigenvalue}\label{isolated_top_eigenvalue}

\subsection{A Langevin equation for the top eigenvalue and eigenvector}

A more detailed characterization of the dynamics of the top eigenvalue and eigenvector can be given in the case where this top eigenvalue is 
well separated from all the others, as is well known to be the case for financial covariance matrices. 
The financial interpretation of this large eigenvalue is the so-called `market mode': in a first approximation, all stocks move together, up or down. 
In this subsection, we assume that the {\it true} covariance matrix ${\bf C}$ has one large eigenvalue $\lambda_1$ of order $N$ well separated 
from the other ones, which are {\it all} equal to $\lambda_2$. We suppose that $\lambda_1 \gg \lambda_2$. 

Let $\left(r_i^t\right)_{1\leq i \leq N}, 1\leq t \leq T$ be i.i.d. Gaussian vectors of covariance ${\bf C}$. Both for technical convenience and to 
follow market practice, we suppose that the covariance matrix is now measured through an exponential moving average of the $r_i^t$.  
This means that the matrix ${\mathbf E}$ evolves in time as:
\begin{equation}\label{EMA}
{\bf E}_{ij,t} = (1 - \varepsilon) {\bf E}_{ij,t-1} + \varepsilon r_i^t r_j^t.
\end{equation}

We address the following question: what is the dynamics of the top eigenvalue $\lambda_1(t)$ and of the top eigenvector $\phi_1^t$ of 
the empirical covariance matrix ${\bf E}_t$? Of course, the largest eigenvalue and eigenvector of the
empirical covariance matrix will be, as discussed at length above, affected by measurement noise. Can one make 
predictions about the fluctuations of both the largest eigenvalue and the corresponding eigenvector induced by 
measurement noise?  We shall see that such a decomposition is indeed possible in the limit where $\lambda_1 \gg \lambda_2$. 
The calculations in this section and in Appendix D follow closely those made in \cite{Krakow} which were slightly incorrect (see below).

We keep the same notations as in the previous section for the eigenvalues of ${\bf C}$. The eigenvalues and eigenvectors of 
${\bf E}_t$ will be respectively denoted as $\lambda_1^t,\dots,\lambda_N^t$ and $\phi_1^t,\dots,\phi_N^t$.

Standard perturbation theory, valid for $\varepsilon \ll 1$, gives:
\begin{equation*}
\lambda_{1}^t = (1 - \varepsilon) \lambda_{1}^{t-1} + \varepsilon \langle \phi_{1}^{t-1} | C | \phi_{1}^{t-1} \rangle 
+ \varepsilon \langle \phi_{1}^{t-1} | \eta_{t} | \phi_{1}^{t-1} \rangle,
\end{equation*}
with $\eta_{ij} = r_i r_j - {\bf C}_{ij}$. Because the returns are Gaussian, we have:
\begin{equation*} \label{etavar}
\overline{ \eta_{ij}\eta_{k\ell} } = C_{ik}C_{j\ell} + C_{i\ell}C_{jk}.
\end{equation*}
In the limit where $\lambda_1$ becomes much larger than all 
other eigenvalues, the above equation simplifies to:
\begin{equation}\label{OU}
\lambda_{1}^{t} \approx (1 - \varepsilon) \lambda_{1}^{t-1} 
+ \varepsilon \cos^2(\theta_{t-1}) \lambda_1 \left[ 1 + \xi_t \right],  
\end{equation}
where $\cos(\theta_t) \equiv \langle \phi_{1}^{t} | \phi_1 \rangle$ and $\xi_t$ is a random noise term of mean zero and 
variance equal to $2$. In the limit of large matrices and $\varepsilon \to 0$, the above difference equation can be written as
a Langevin (or stochastic differential) equation, in the It\^o sense: 
\begin{equation}\label{Orn1}
{\rm d}\lambda_1^t = \varepsilon\left[(\lambda_1 - \cos^2(\theta_t) \lambda_1^t) {\rm d}t + \sqrt{2} \lambda_1 \cos^2(\theta_t) \, {\rm d}B_t \right].
\end{equation}
where $B_t$ is a standard Brownian motion. We have neglected in the above equation a deterministic term equal to 
$\varepsilon \sin^2(\theta_t) \lambda_2$, which will turn out to be a factor $\lambda_2/\lambda_1$ smaller than the terms retained in Eq. (\ref{Orn1}).
As we shall show below, the angle $\theta_t$ turns out to be small, so that one can replace $\cos(\theta_t)$ by unity in the above equation, which becomes a simple 
Ornstein-Uhlenbeck process. We therefore find for the variogram of $\lambda_1$:
\be
\left\langle\left(\lambda_1^{s} - \lambda_1^{t}\right)^2\right\rangle \approx 2 \varepsilon \lambda_1^2 \left(1 - \exp(-\varepsilon|t-s|)\right),
\ee
a result that we mentioned in the above section \ref{eigenspace}.

A similar SDE can be written for the projection of the instantaneous eigenvector 
$|\phi_{1}^{t}\rangle$ on the true eigenvector $|\phi_1\rangle$. This can again be done using perturbation theory, as is detailed in Appendix D. 
The quantity $\cos(\theta_t)$ is found to be close to $1$ when $\varepsilon$ is small, so we set $x_t\equiv1-\cos(\theta_t)$.

Keeping only the leading term in the three small parameters $\varepsilon, \lambda_2/\lambda_1$ and $x_t$, we finally find the following 
Langevin equation for $x_t$ (in the It\^o sense):
\begin{equation}\label{xt}
{\rm d}x_t = 2\varepsilon \left(\mu  - x_t\right)  {\rm d}t + \varepsilon \sqrt{2 x_t(4 x_t +\frac{\lambda_2}{\lambda_1})} \, \, {\rm d}B_t
\end{equation}
with, for $N \to \infty$, $\varepsilon \to 0$,
\begin{equation*}
\mu:= \frac{q}{4} \frac{\lambda_2}{\lambda_1}\,, \quad {\mbox{with}} \quad q \equiv \varepsilon N.
\end{equation*}
Equation (\ref{xt}) defines a very interesting class of random processes, that we call ``P\"oschl-Teller" processes, on which we say more in Appendix E.

\begin{figure}[h!btp] 
	\center
		\includegraphics[scale=0.6]{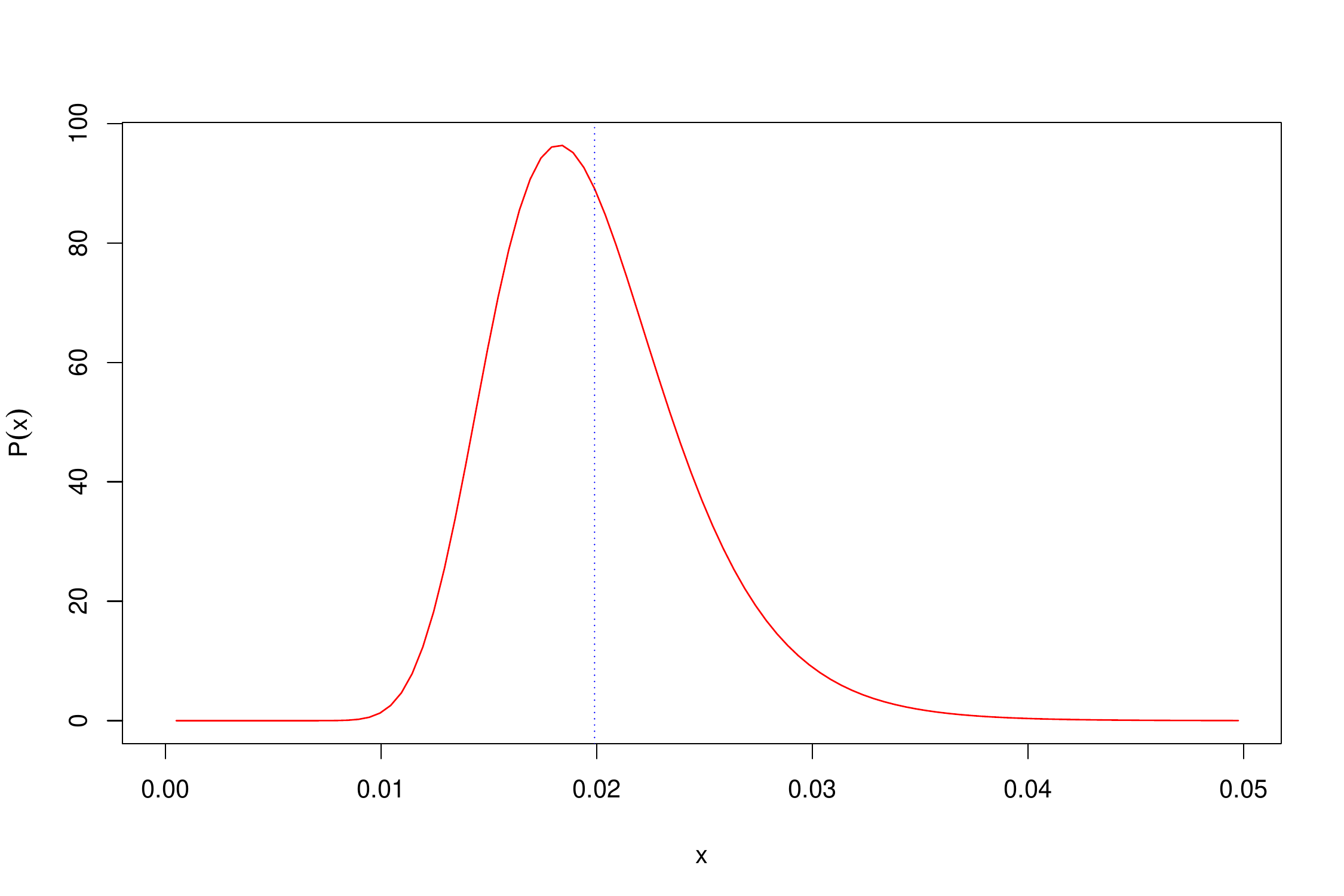}
            \caption{\small{A picture of the stationary probability density $P(x)$ of the process $x_t$ verifying \eqref{xt}. 
            The parameters are: $\varepsilon = 1/50$, $N = 200$ (corresponding to $q=4$) and $\lambda_2/\lambda_1=0.02$.
            The vertical blue dotted line shows the position of $\mu \approx 0.02$ for this choice of parameters.}} 
             \label{statio_density}      
\end{figure}

In the continuous time limit, we have therefore established two coupled Langevin equations (SDEs) for the top eigenvalue $\lambda_{1}^{t}$
and $x_t$. To leading order and for $N \to \infty$, $\varepsilon \to 0$, the stationary solution for the ``angle" $x_t$ can be computed to be:
\begin{equation*}
P(x) \propto \left(\frac{4x}{4x+\frac{\lambda_2}{\lambda_1}}\right)^{\frac{N}{2}} 
\left(\frac{1}{4x+\frac{\lambda_2}{\lambda_1}}\right)^{\frac{1}{2\varepsilon}},
\end{equation*}
which corrects the result obtained in \cite{Krakow}, and is plotted in Fig. \ref{statio_density}.
From the above Langevin equation, it is immediate to see that the average value of $x$ is given by $\overline{ x } = \mu$. It is 
nicer to rewrite the stationary distribution in terms of $\hat x=x/\mu$. The interesting regime is when $q$ remains of order unity when $N \to \infty$ and
$\varepsilon \to 0$, in which case:
\begin{equation*}
P(\hat x) \approx Z e^{-\frac{Nf(\hat x)}{2}}, \qquad f(\hat x)= \frac{\ln(1 + q\hat x)}{q} + \ln \left(1 + \frac{1}{q\hat x}\right)\,,
\end{equation*} 
where $Z$ is a normalisation. It is easy to see that $f(\hat x)$ has a minimum for $\hat x=1$, or $x=\mu$ (corresponding to the most probable value), and that 
$f''(1)=1/(1+q)$. This shows that the fluctuations of $\hat x$ around $\hat x=1$ are of order $\sqrt{(1+q)/N}$ and thus very small in the large $N$ limit.

Note finally that according to Eq. (\ref{Orn1}), the largest eigenvalue is on average {\it shifted upwards} compared to the true value $\lambda_1$, by
a factor $\approx (1 + 2 \mu) = (1 + \frac{q}{2} \frac{\lambda_2}{\lambda_1})$. This is the analogue of a similar well-known result for flat-window
averages of empirical covariance matrices -- see \cite{BaiSilverstein,BenArous}.

\subsection{Variograms}\label{variogram}

\begin{figure}[h!btp] 
	\center
		\includegraphics[scale=0.7]{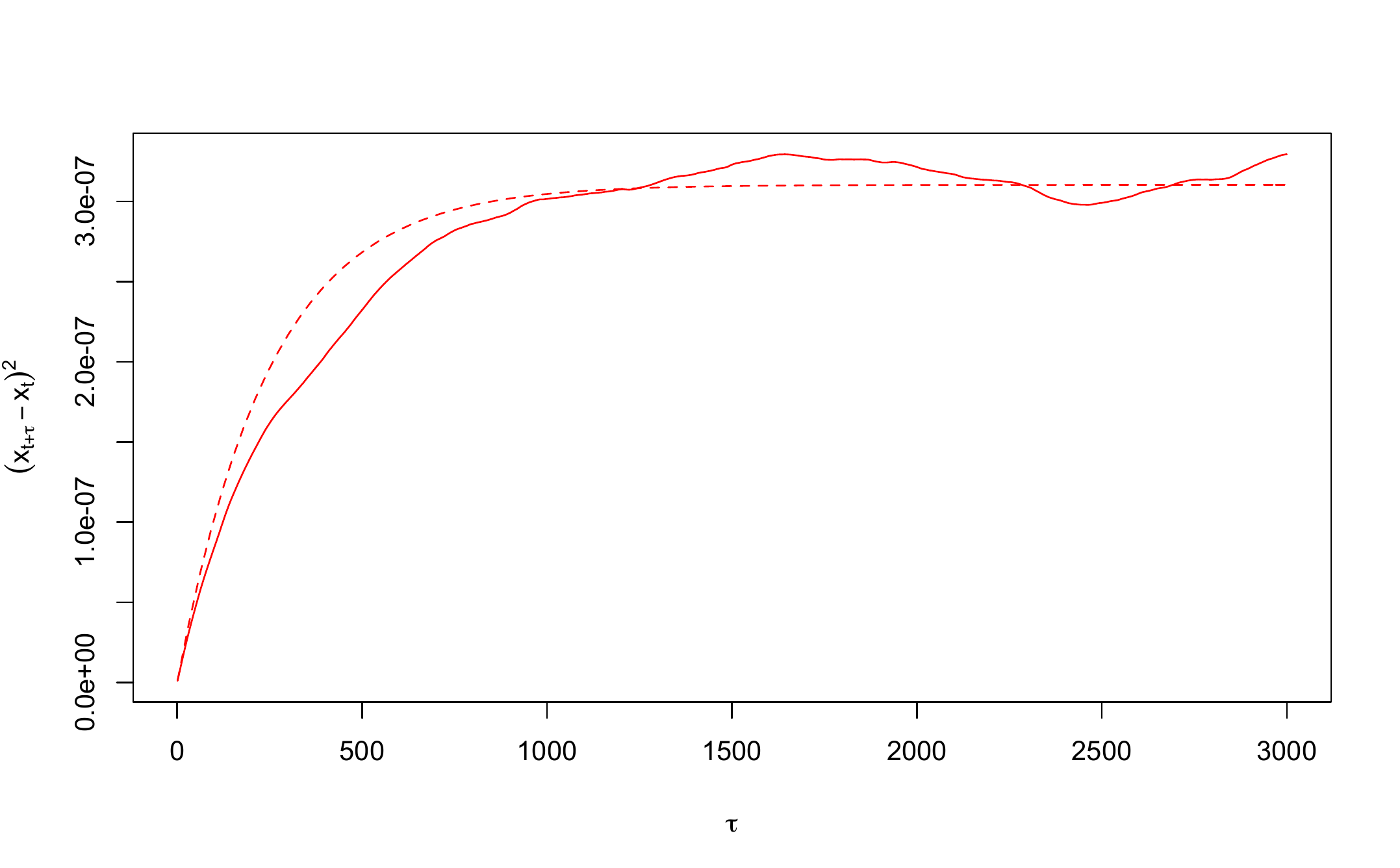}
            \caption{\small{The plain line represents the function $\upsilon(\tau)$ as a function of $\tau$ for $\varepsilon=0.002, N=200$ ($q=0.4$) and $\lambda_2/\lambda_1=0.033$ = 30. 
            The dotted line is a numerical simulation of the semivariogram of $x_t$ in the benchmark case where there is a constant in time correlation matrix $\bf C$.} } 
             \label{fig_variogram}      
\end{figure}

From the Langevin equation one can easily compute the second moment $\overline{ x_t^2}$ with as initial condition $x_0=0$. Indeed, using It\^o's formula and taking expectations, 
we get: 
\begin{equation*}
\overline{ x_t^2}  = x_0^2 +4\varepsilon  \left(  \mu + \varepsilon \frac{\lambda_2}{2\lambda_1} \right) \int_0^t  
\overline{ x_s }{\rm d}s - 4 \varepsilon(1-2\varepsilon) \int_0^t \overline{ x_s^2} \, {\rm d}s\,.
\end{equation*}
Computing $\overline{ x_t}$ with the same technique, we can solve this ordinary differential equation to obtain that 
\begin{align*}
\overline{x_t^2} &= x_0^2 e^{-4\varepsilon (1-2\varepsilon)t} + \frac{\mu (\mu+\varepsilon \frac{\lambda_2}{2\lambda_1})}{1-2\varepsilon} 
\left( 1 - e^{-4\varepsilon(1-2\varepsilon)t} \right)\\
&+ \frac{(2\mu+\varepsilon \frac{\lambda_2}{\lambda_1})(x_0-\mu)}{1-4\varepsilon} \left(e^{-2\varepsilon t} - e^{-4\varepsilon (1-2\varepsilon) t}\right)\,.
\end{align*}
In order to characterize the dynamics of the angle fluctuations, we want to compute the variogram of $x_t$, defined as $\upsilon(\tau) := \overline{ (x_{t+\tau} - x_t )^2 }$ for $\tau\geq 0$, and in the limit $t \to \infty$. Using the previous computations, we obtain, in the scaling limit:
\begin{equation*}
\upsilon(\tau) \approx \frac{q^2(1+q)}{4N} \left(\frac{\lambda_2}{\lambda_1}\right)^2 \left(1-e^{-2\varepsilon \tau}\right).
\end{equation*}
We show in Fig. \ref{fig_variogram} a numerical simulation of the dynamics of the top eigenvector of a fixed matrix ${\bf C}$ such that $\lambda_2/\lambda_1=0.033$. The 
resulting variogram compares very well with the above prediction.

However, the above calculation is not particularly useful for financial applications, since the ``true'' top eigenvector $|\phi_1\rangle$, needed to define the angle $\theta_t$, is in general not known. A more appropriate quantity to describe the dynamical fluctuations of $|\phi_1\rangle$ is, as suggested in \cite{Krakow}, the function 
$\tau \rightarrow \overline{\langle \phi_1^t | \phi_1^{t+\tau}\rangle}$, which we now study analytically. Let us write $|\phi_1^t\rangle$ as 
\begin{equation}\label{phi_1^t}
|\phi_1^t \rangle = \cos(\theta_t) |\phi_1 \rangle  + | \varphi_\perp^t \rangle,
\end{equation}
where $\varphi_2^t$ is a vector in the eigenspace corresponding to the small eigenvalues $\lambda_2$. Therefore: 
\begin{equation*}
\langle \phi_1^t | \phi_1^{t+\tau}\rangle  = \cos(\theta_t) \cos(\theta_{t+\tau}) + \langle \varphi_\perp^t | \varphi_\perp^{t+\tau}\rangle\,.
\end{equation*}
Now, it is easy to have an explicit expression for $\varphi_2^t$ by considering the empirical covariance (or correlation) matrix ${\bf E}^t$
as a perturbation of the true covariance matrix ${\bf C}$, as we did above. Standard perturbation theory then gives 
\begin{align*}
|\phi_1^t \rangle = \left(1-\frac{1}{2}\sum_{i\neq 1} \frac{\langle \phi_1 | {\bf \mathcal{E}}^t | \phi_i \rangle^2 }{(\lambda_1-\lambda_2)^2} \right) | \phi_1 \rangle + 
\sum_{i\neq 1} \frac{\langle \phi_1 | {\bf \mathcal{E}}^t | \phi_i \rangle}{\lambda_1-\lambda_2} | \phi_i \rangle
\end{align*}
where 
\begin{equation*}
\mathcal{E}^t_{ij} = \varepsilon \sum_{s=0}^{+\infty} (1-\varepsilon)^s \left(r_i^{t-s} r_j^{t-s} - C_{ij}\right)\,. 
\end{equation*}
It is clear that the last term of the above expression is exactly $| \varphi_2^t \rangle$, which enables us to obtain:
\begin{equation*}
{\langle \varphi_\perp^t | \varphi_\perp^{t+\tau} \rangle} = \frac{1}{(\lambda_1-\lambda_2)^2} \sum_{i\neq 1} \langle \phi_1 | {\bf \mathcal{E}}^t | \phi_i \rangle
\langle \phi_1 | {\bf \mathcal{E}}^{t+\tau} | \phi_i \rangle\,.
\end{equation*}
But, by noting that: 
\begin{equation*}
 \mathcal{E}^{t+\tau}_{ij} = (1-\varepsilon)^\tau  \mathcal{E}^{t}_{ij} + \varepsilon \sum_{s=0}^{\tau-1} (1-\varepsilon)^s \left(r_i^{t+\tau-s} r_j^{t+\tau-s} - C_{ij}\right)
\end{equation*}
and with the fact that $\overline{ \langle \phi_1 | {\bf \mathcal{E}}^t | \phi_i \rangle^2 } = \varepsilon \lambda_1\lambda_2 /2$, we get that:
\begin{equation*}
\overline{ \langle \varphi_\perp^t | \varphi_\perp^{t+\tau} \rangle} \approx 2\,\mu\, e^{-\varepsilon \tau}\,,
\end{equation*}
and hence, our final result, to lowest order in $\mu$:
\begin{align}
\overline{\langle \phi_1^t | \phi_1^{t+\tau} \rangle}  &= \overline{ (1-x_t)(1-x_{t+\tau}) } + \overline{ \langle \varphi_\perp^t | \varphi_\perp^{t+\tau} \rangle } \\
& \approx 1-2 \mu \left(1-e^{-\varepsilon \tau} \right) \,.  
\end{align}
which is similar to the result obtained in \cite{Krakow}, except that the coefficient $\mu$ was a factor $N$ too small in that paper. This result
will be compared with empirical data in section \ref{emp_dyn_top_eigenvector}.

\subsection{Transverse fluctuations of the top eigenvector}\label{transvese_top_ev}

In order to go further and describe the evolution of the top eigenvector of ${\bf E}$ (the so-called ``market mode'' in the context of financial
markets), we need to study the statistics of the transverse component $| \varphi_\perp^t \rangle$. 
In order to make sense of the pattern created by these transverse fluctuations, we propose to introduce the correlation matrix of the components of 
$| \varphi_\perp^t \rangle$ in the eigen-basis of the true correlation matrix. We therefore define the following $N-1 \times N-1$ matrix:
\begin{equation*}
{\bf F}_{ij} = \frac1T \sum_{t=1}^T \langle \varphi_\perp^t | \phi_i \rangle \langle \varphi_\perp^t | \phi_j   \rangle \, (i,j \geq 2)
\end{equation*} 
The eigenvalues and eigenvectors of this new correlation matrix (not to be confused with the empirical correlation matrix ${\bf E}$ needed to 
define $| \varphi_2^t \rangle$!) will entirely characterize the transverse fluctuations of the ``market mode''.

In the benchmark case where there is a true correlation matrix $\bf C$ stable in time, one can check that: 
\begin{equation*}
{\bf F}_{ij} = \frac1T \sum_{t=1}^T \frac{\langle \phi_1 | {\bf \mathcal{E}}^t | \phi_i \rangle}{\lambda_1-\lambda_i} \frac{\langle \phi_1 | {\bf \mathcal{E}}^t | \phi_j \rangle}{\lambda_1-\lambda_j}\,.
\end{equation*}
What is the eigenvalue spectrum of ${\bf F}$ for this benchmark case? In our case where for all $i\neq 1$, $\lambda_i=\lambda_2$, 
the density of states of this type of random matrix has been studied before in the literature (see \cite{Schlemm}). 
Indeed the random variables $\langle \phi_1 | {\bf \mathcal{E}}^t | \phi_i \rangle$ are uncorrelated for $i\neq j$, 
their mean is $0$ and their variance is given by: 
\begin{equation}
\overline{ \langle \phi_1 | {\bf \mathcal{E}}^t | \phi_i \rangle^2 } = \frac{\varepsilon\lambda_1\lambda_2}{2}\,.
\end{equation}
However, the random variables $\langle \phi_1 | {\bf \mathcal{E}}^t | \phi_i \rangle$ {\it are correlated in time} and thus the density of states in the limit of large matrices 
will not be given by the usual Marchenko-Pastur law. Rather, $\langle \phi_1 | {\bf \mathcal{E}}^t | \phi_i \rangle$ follows an auto-regressive linear process, for which the 
authors of \cite{Schlemm}, give a precise way to compute the density of states in the limit of large matrices by mean of its Stieltjes transform. This probability density 
depends as expected of the parameter $N/T$ but also of the parameter $\varepsilon$ of the auto-regression. In the case where  $\lambda_{i>1}=\lambda_2$, one furthermore expects that the eigenvectors of ${\bf F}$ are isotropically distributed in the $N-1$ dimensional subspace spanned by 
$|\phi_2 \rangle, \dots |\phi_N \rangle$. This means that the transverse fluctuations $|\varphi_\perp \rangle$ of the top eigenvector have no particular structure. 

In the more general context where the $\lambda_i$ for $i\neq 1$ are not all equal to $\lambda_2$, the eigenvalue spectrum of ${\bf F}$ must be characterized numerically, see below.

\section{Empirical results}\label{empirical_results}

For the following analysis, we have used the daily returns of several pools of stocks belonging to $4$ major indices: SP500, Nikkei, DAX \& CAC 40. The number of stocks are respectively 
$N=500, 204,30,39$ and the period of interest is $2000-2010$ (11 years of data, corresponding to $\approx 2750$ days).
The main issue, as alluded to above, is that the empirical determination of correlation\footnote{Our results in the previous sections hold for empirical covariance matrices. Hence 
we centered and normalized our empirical time series of returns so as to use them. } matrices requires some measurement time $T$. If this time is 
too short, the empirical correlation matrix will appear to evolve with time, but this may just be due to the measurement noise that one would like to 
distinguish from a genuine evolution of the underlying structure of correlation. If the measurement time is too long, on the other hand, one may miss important correlation shifts and 
get exposed to unwanted sources of risk.

\subsection{Stability of eigenspaces}\label{sect_emp_eigenspace_stability}

\begin{figure}[h!btp] 
	\center
		\includegraphics[scale=0.40]{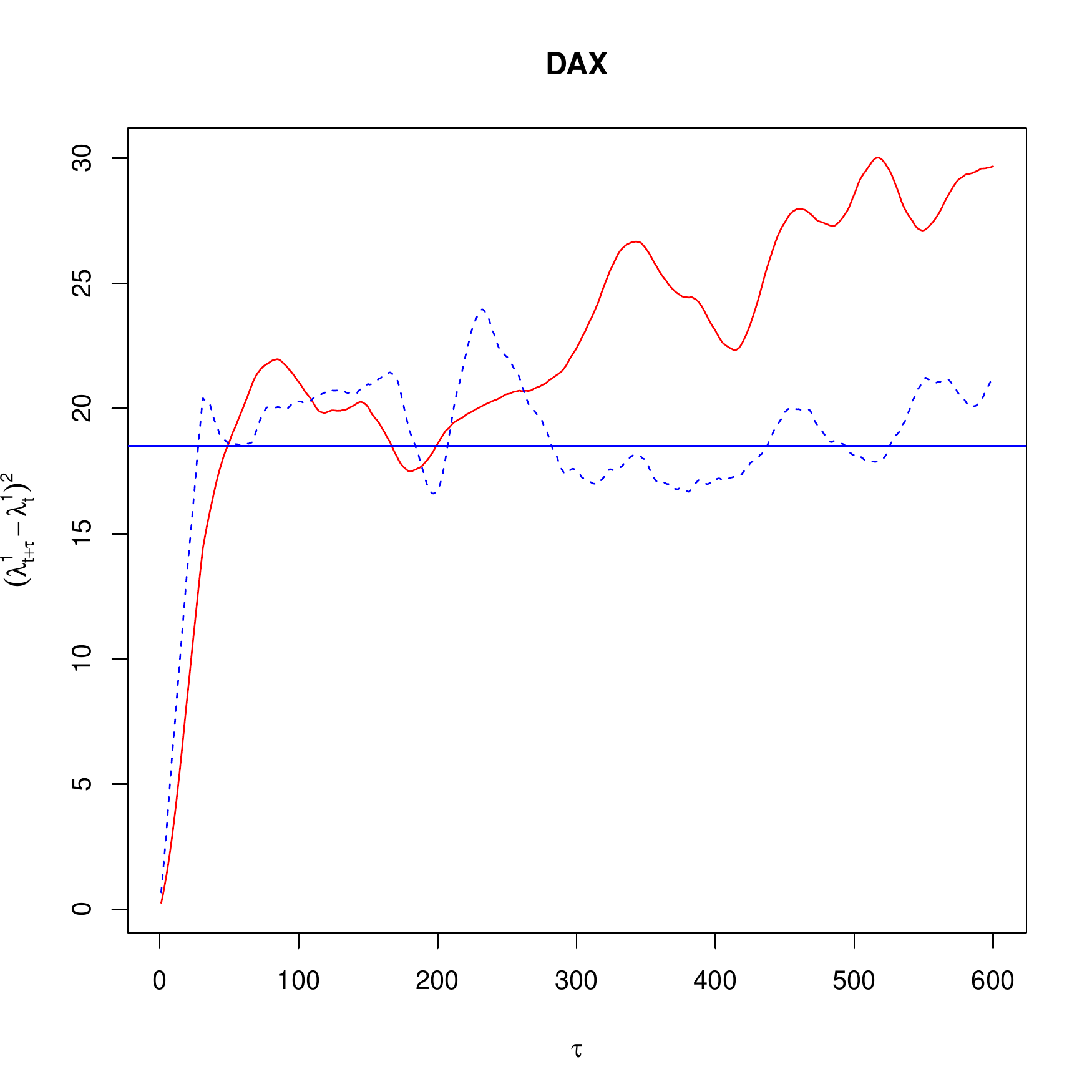}
        \includegraphics[scale=0.40]{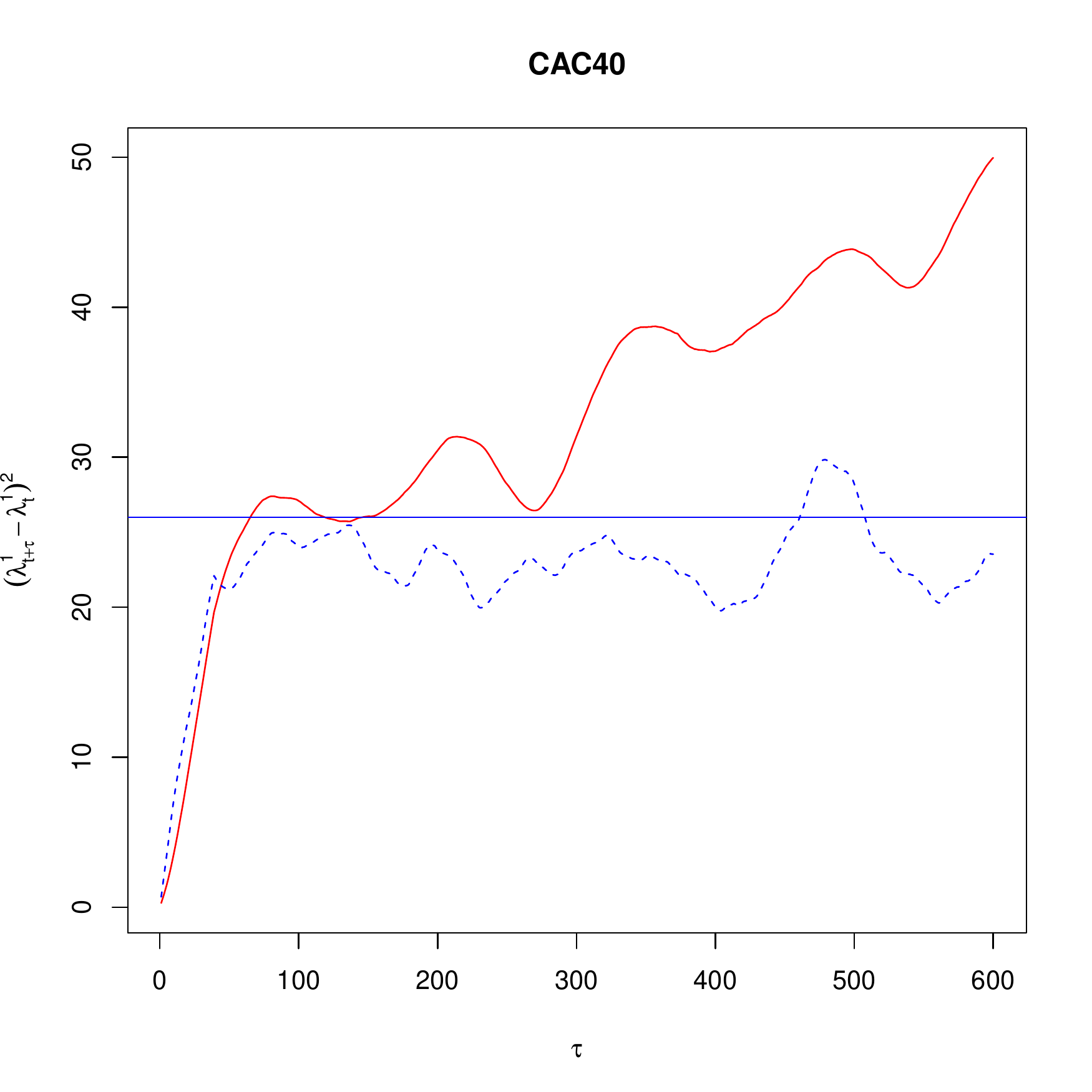}
        \includegraphics[scale=0.40]{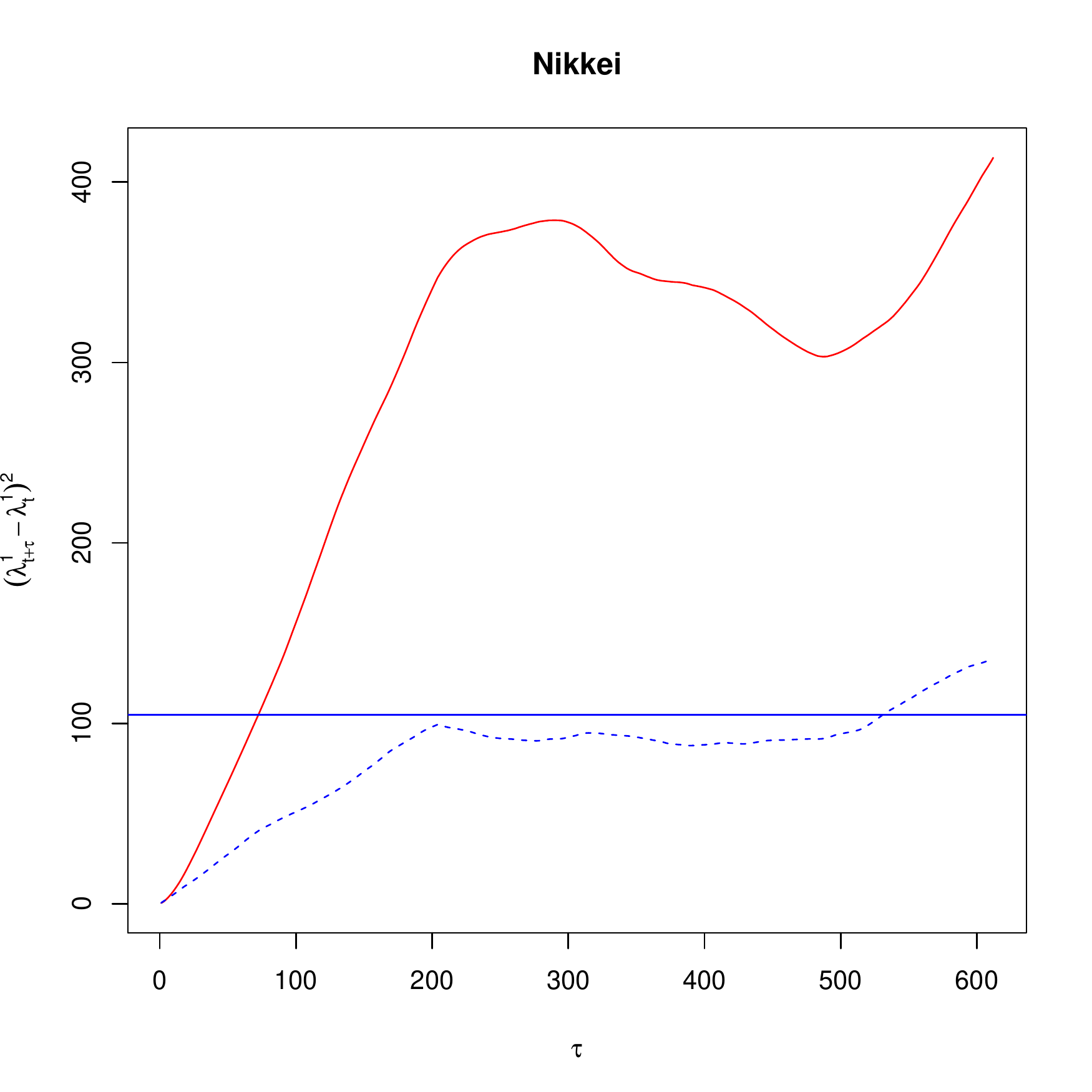}
        \includegraphics[scale=0.40]{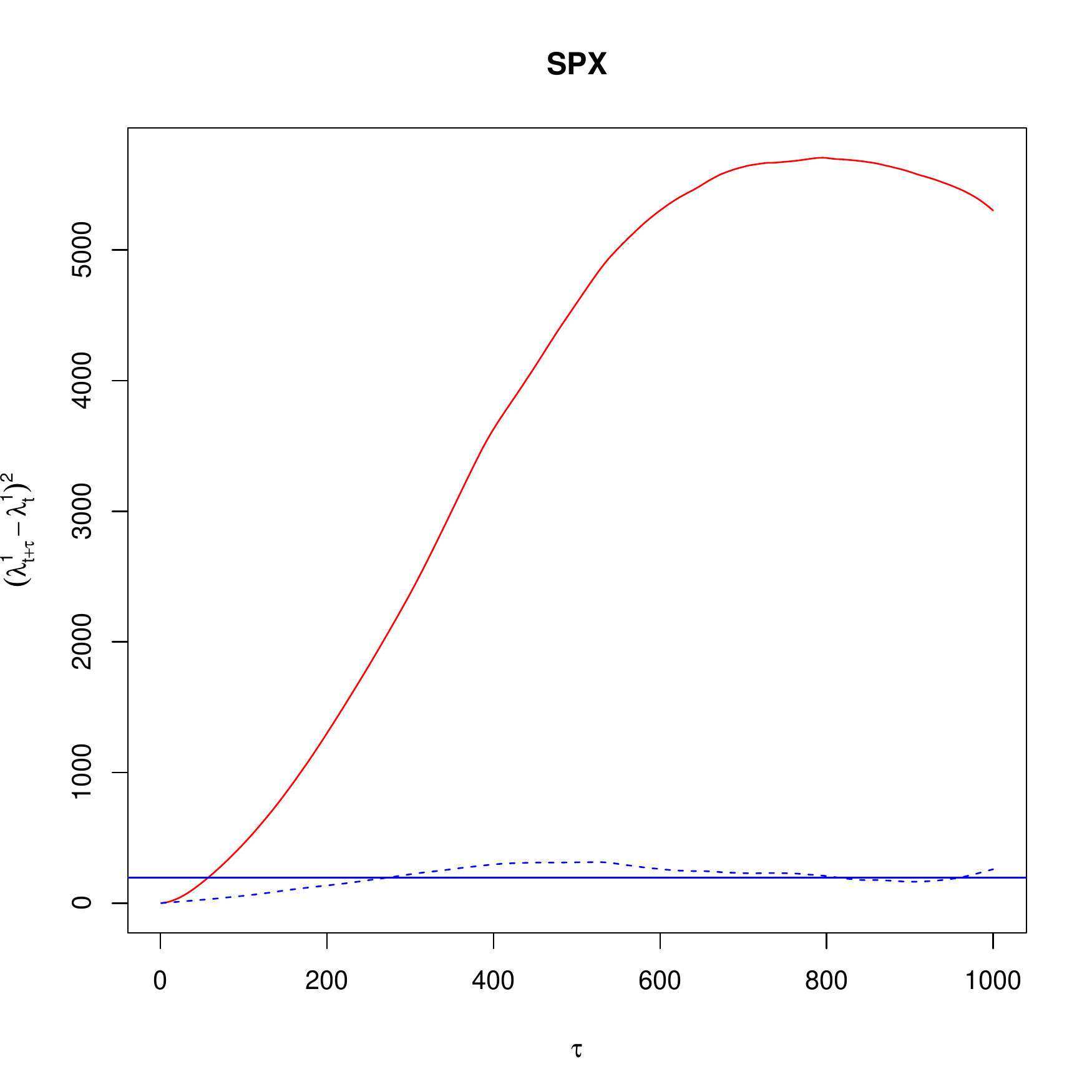}
        \caption{\small{Plot of $\langle(\lambda_1^{s} - \lambda_1^{t})^2\rangle_{|t-s|=\tau}$ as a function of $\tau$ for the four different indexes of our sample. 
        The empirical correlation matrices are computed on a sliding window of 
        size $T=N$. The red line corresponds to the empirical datas from our pools of stocks, the plain blue line is the theoretical prediction $4\lambda_1^2/T$ (valid in the limit of large $T$) and 
        the dotted blue line represents a numerical simulation of the benchmark case. Very similar curves hold for the second and third eigenvalues as well.} }       \label{fig_V_lambda_1}      
\end{figure}

We first determined the empirical variograms $\langle(\lambda_i^{s} - \lambda_i^{t})^2\rangle_{|t-s|=\tau}$ for $i=1,2$, the result (for $i=1$) is shown in Figure \ref{fig_V_lambda_1} 
and is found to be much larger 
than the above theoretical prediction, i.e. $4 \lambda_i^2/T$, shown as a horizontal
plain line. The fact that the empirical (red) curve starts from $0$ for $\tau=0$ and increases to reach the stationary noise level at time $\tau=T$ is simply due to the overlapping between the sliding periods.  
For those figures, we computed the time series of correlation matrices using a sliding window of size $T=N$ (recall $N$ is the number of stocks). Thus, for small markets like DAX and CAC40, 
this value is quite small (respectively $30$ and $40$) and we find that the first eigenvalue of the correlation matrix does not evolve too much during the following (non overlapping) period 
$\tau \in [T;250]$ days. After this time period, 
the evolution appears and from this point, the difference between the two non overlapping periods increases significantly with the time lag. 
For larger markets such as SPX and Nikkei, the value of $T$ is quite large as $N$ is respectively equal to $500$ and $200$. So it is not very surprising the temporal evolution shows up 
immediately.       
This clearly shows that there is a genuine evolution of the 
eigenvalues of ${\bf C}$ with time. For the top eigenvalue, this is a well known effect (see \cite{Krakow} and section \ref{variogram}, 
Fig. \ref{fig_variogram} below): both the volatility of individual stocks 
and the average correlation between stocks are indeed time dependent, and tend to
increase in crisis periods \cite{pa_leverage,Ingve}. We see that the same is true for smaller eigenvalues too, reflecting the instability of intra-sector correlations (data not shown). 

But what about the eigenvectors? One could be in a ``mixed'' situation where the eigenvectors of the true underlying covariance $\bf C$
keep a fixed direction through time\footnote{Here we mean that the non-perturbed (or population) eigenvectors do not evolve with time; obviously 
we do not talk about the sample eigenvectors of the empirical 
covariance matrix $\bf E$ which will be affected by measurement noise, evolving around the population eigenvectors.}  
while its eigenvalues are moving around. 
But if the eigenvalues of the matrix $\bf C$ (that was always supposed not to depend of time in the previous sections) themselves are evolving with time, 
the formulas derived in the theoretical section above need to be upgraded. Let us assume that the true covariance matrix ${\bf{C}}_t$ has time dependent eigenvalues 
$\lambda_1^t,\dots,\lambda_N^t$ but with constant eigenvectors that will be denoted $|\phi_1\rangle, \dots, |\phi_N\rangle$ as above. For times $s<t$ with $|t-s|\geq T$, we 
define the overlap matrix 
${\bf{G}}^{s,t}$ as: $G_{ij}^{s,t} = \langle \phi_i^{s} | \phi_j^{t} \rangle$. Under the assumption that the eigenvalues are varying sufficiently slowly with time, one now finds that:
\begin{align}
D&(P,Q;s,t) = - \frac{1}{2P}  \left\langle \ln |\det({{\bf G}^{s,t}}^\dagger{\bf G}^{s,t})| \right\rangle\nonumber \\ 
&\approx  \frac{1}{2TP} \sum_{i=1}^{P} \sum_{j=Q+1}^{N} \left(\frac{\lambda_i^s \lambda_j^s}{(\lambda_i^s - \lambda_j^s)^2}  
+ \frac{\lambda_i^t \lambda_j^t}{(\lambda_i^t - \lambda_j^t)^2} \right). \label{timeDep}
\end{align}
Up to corrections of order $T^{-3/2}$, one can replace in the above formulas the $\lambda^{s,t}$ by their empirical estimates. 
We finally compute the theoretical distance $D_{th}(P,Q,\tau)$ as an average over all $s,t$ such that $|t-s|=\tau$ of the above quantity. 

\begin{figure}[h!btp] 
	\center
		\includegraphics[scale=0.40]{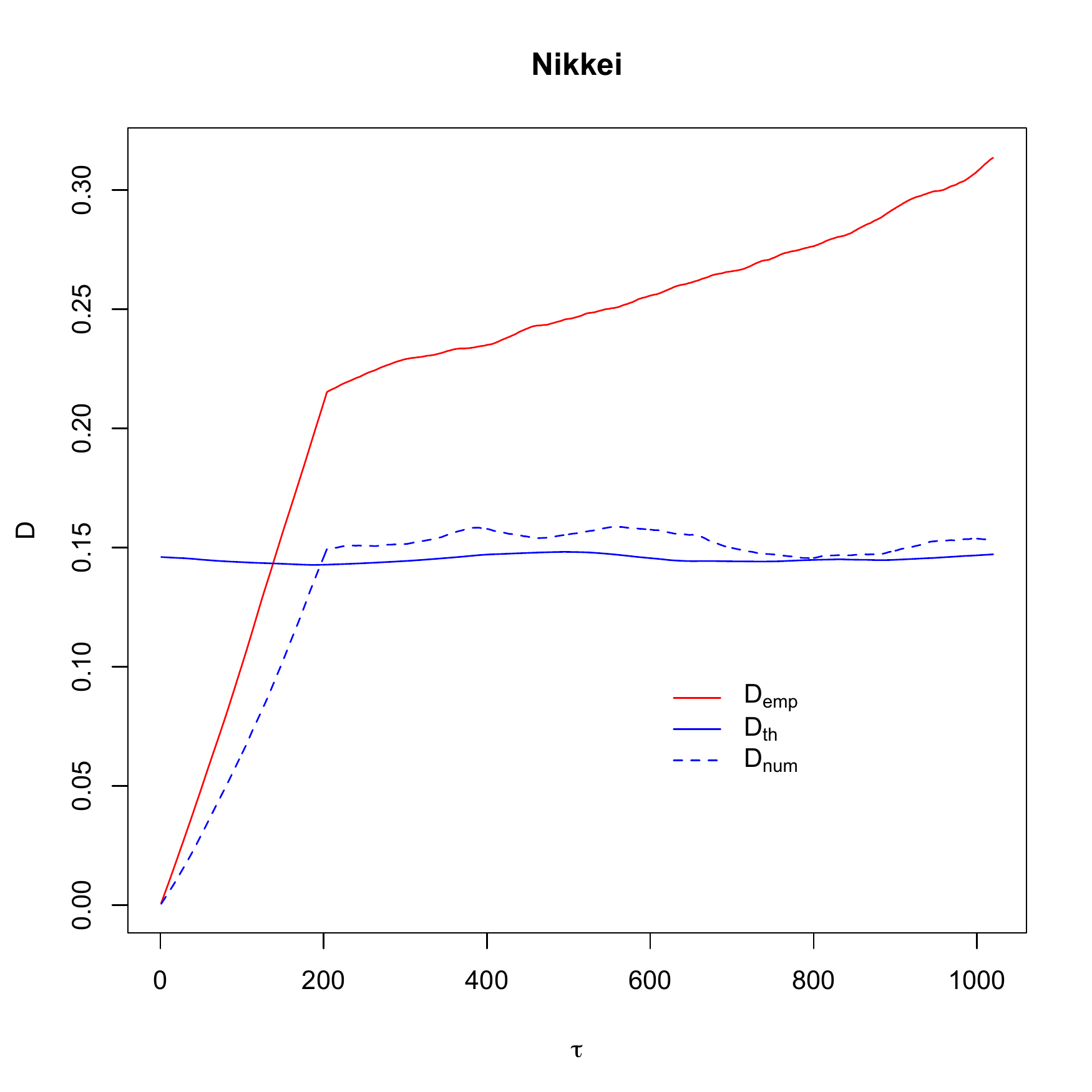}
        \includegraphics[scale=0.40]{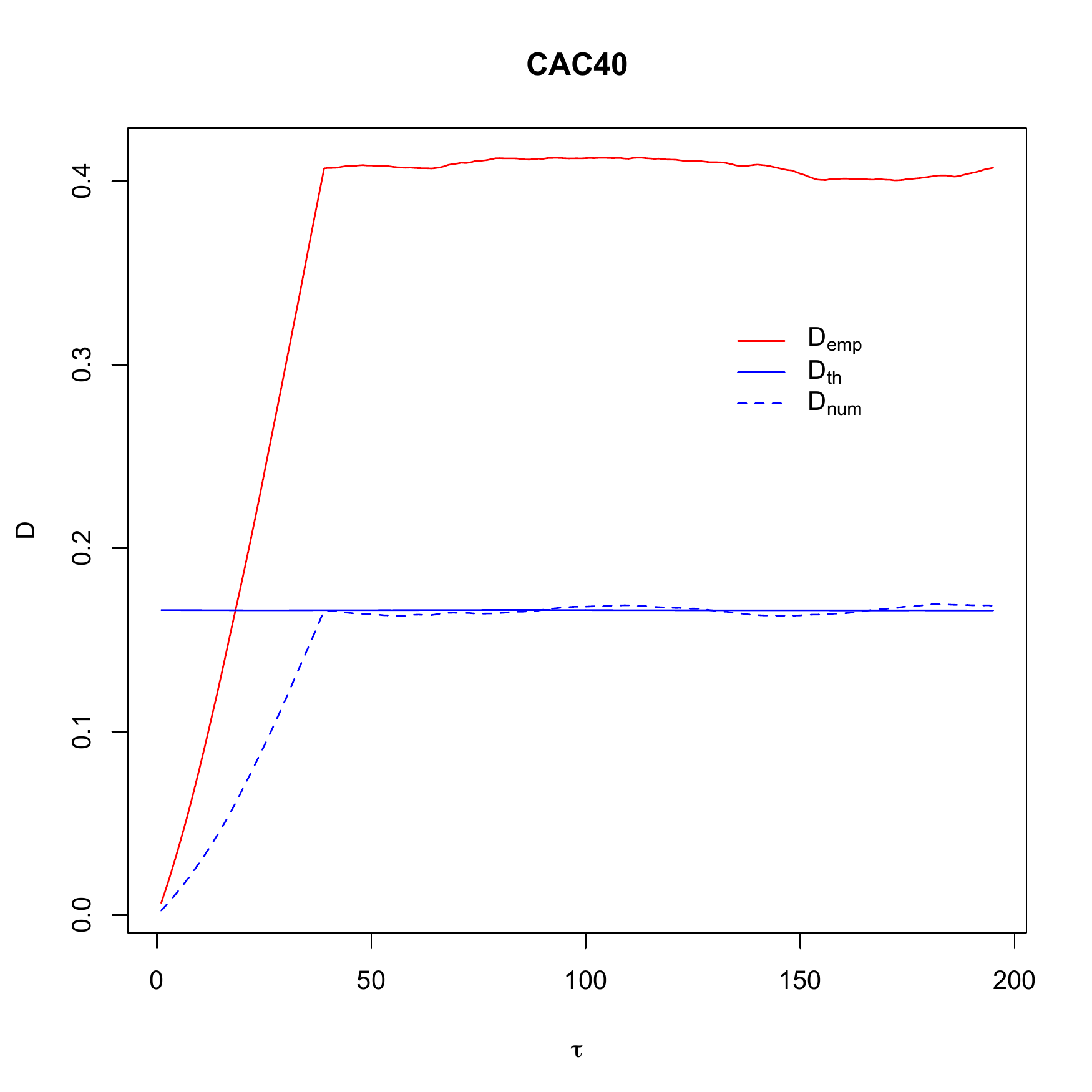}
        \includegraphics[scale=0.40]{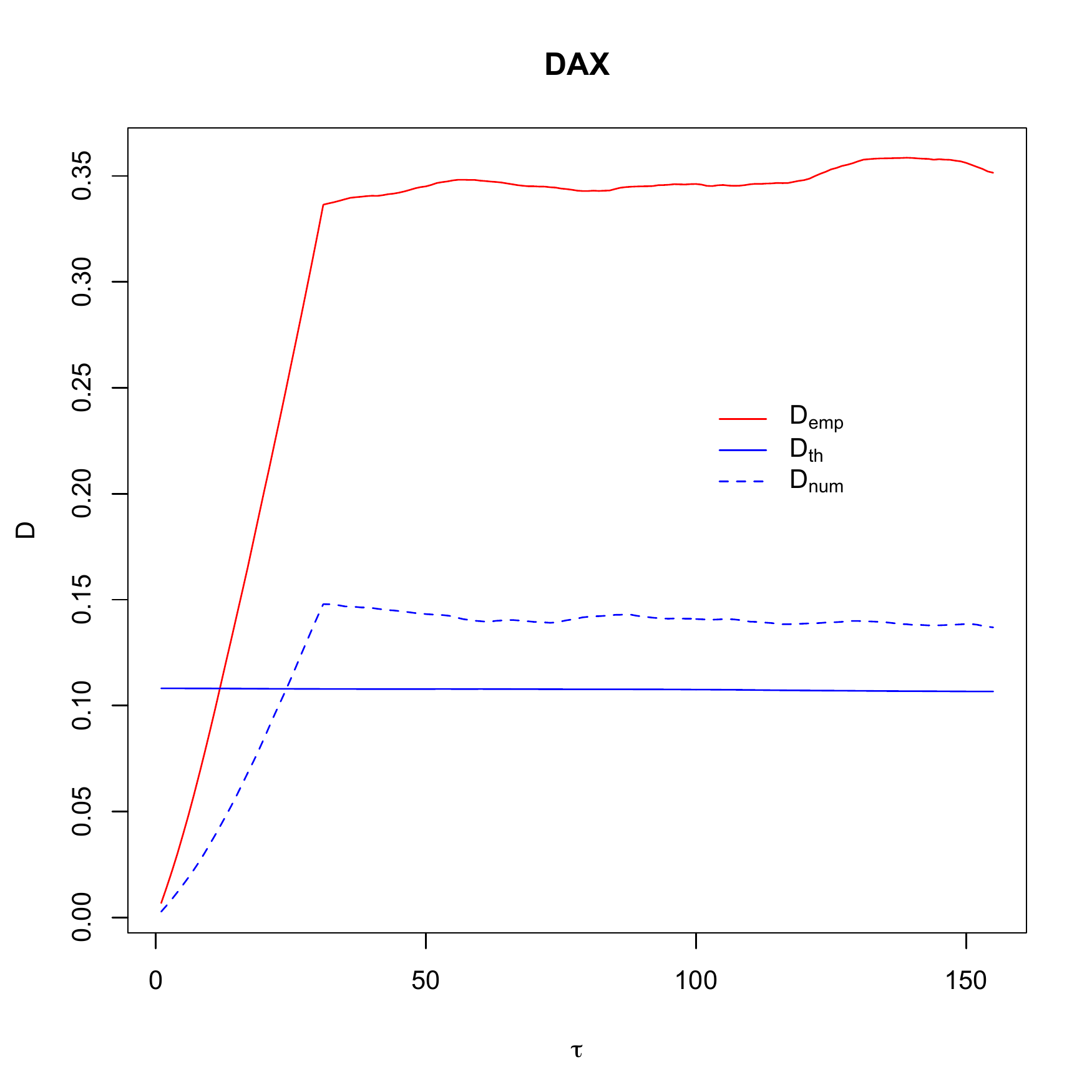}
        \includegraphics[scale=0.40]{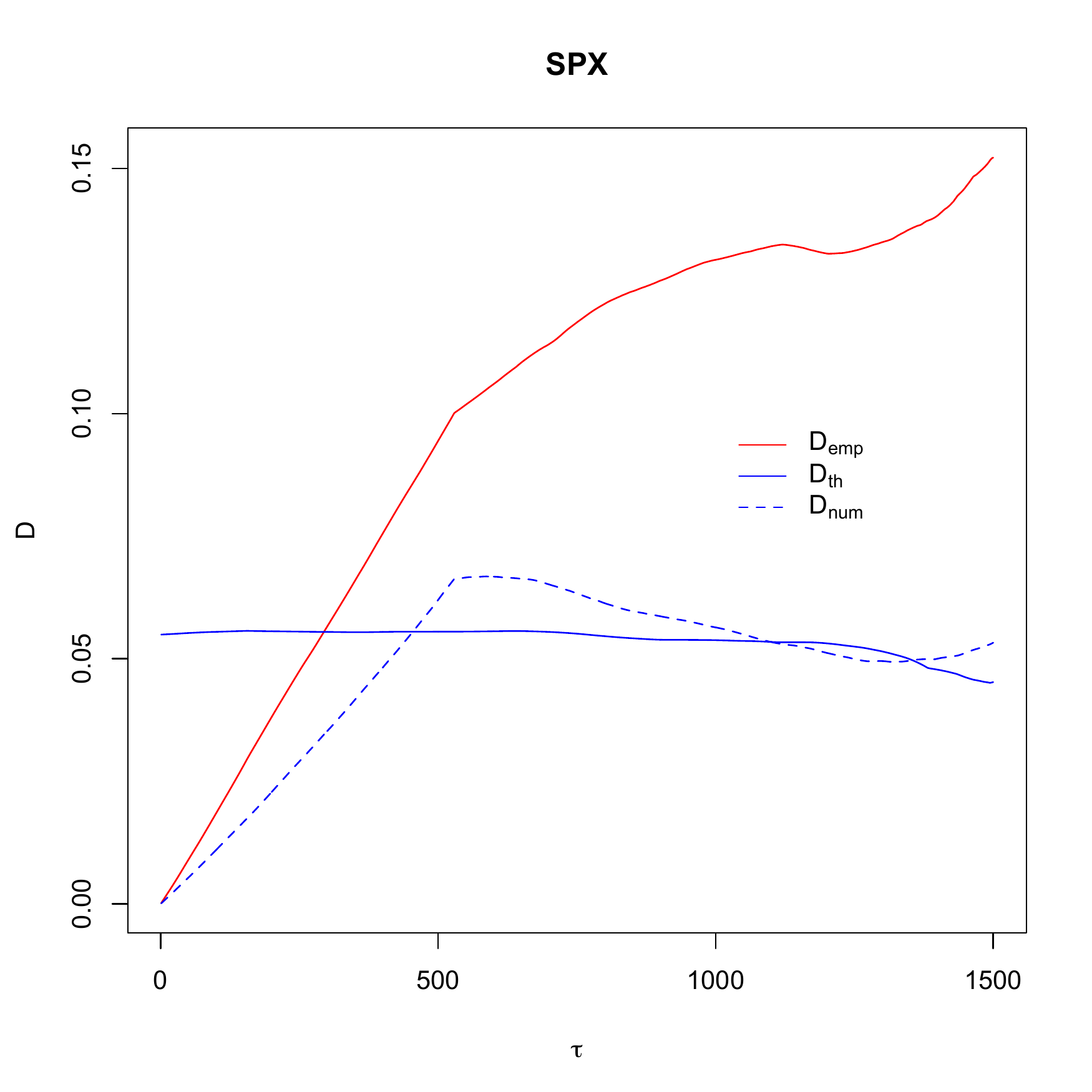}
        \caption{\small{Plot of $D_{th}, D_{num}, D_{emp}$  for $T=N$, $P=5,Q=10$ for the four indices considered here. The blue lines are theoretical benchmark results for fixed eigenvector directions 
        (plain line: analytical result, dotted line:
        numerical simulations, while the red line is the empirical result). These plots clearly show that the subspace spanned by the $5$ top eigenvectors evolve with time.}}   \label{Les_3_D}      
\end{figure}

We now compare our null hypothesis formula, Eq. (\ref{timeDep}) with (a) an empirical determination of $D_{emp}(P,Q,\tau)$ using financial data and (b) 
a numerical determination of $D_{num}(P,Q,\tau)$ using synthetic 
time series of returns that abide to the hypothesis of a covariance matrix ${\bf C}_t$ with {\it fixed} eigenvectors, but time dependent eigenvalues.
To achieve this, we choose an arbitrary (but fixed) set 
of orthonormal vectors $|\psi_1\rangle, \dots, |\psi_N\rangle$ and define ${\bf C}_t$ as ${\bf C}_t = \sum_{i=1}^{N} \lambda_i^{t} |\psi_i\rangle \langle \psi_i|$,
where the $\lambda^{t}$ are the empirical eigenvalues obtained on the financial return time series. We then use ${\bf C}_t$ to generate synthetic Gaussian 
multivariate returns $\left\{r_i(u)\right\}$. We show the corresponding results in Fig. \ref{Les_3_D}, with the choice $P=5,Q=10$, as a function of $\tau$ and for $T=N$ days.
As above, the study concerns the same $4$ different pools of stocks corresponding to $4$ major indices: SP500, Nikkei, DAX, CAC 40. 
We conclude that (i) the theoretical formula Eq. (\ref{timeDep}) is indeed in very good agreement with the numerical results obtained with synthetic data: 
$D_{num} \approx D_{th}$; whereas (ii) the financial data clearly 
departs from the null hypothesis of constant eigenvectors, since $D_{emp} > D_{th}$. The same conclusion holds for different values of $P,Q$. 

\begin{figure}[h!btp] 
	\center
		\includegraphics[scale=0.40]{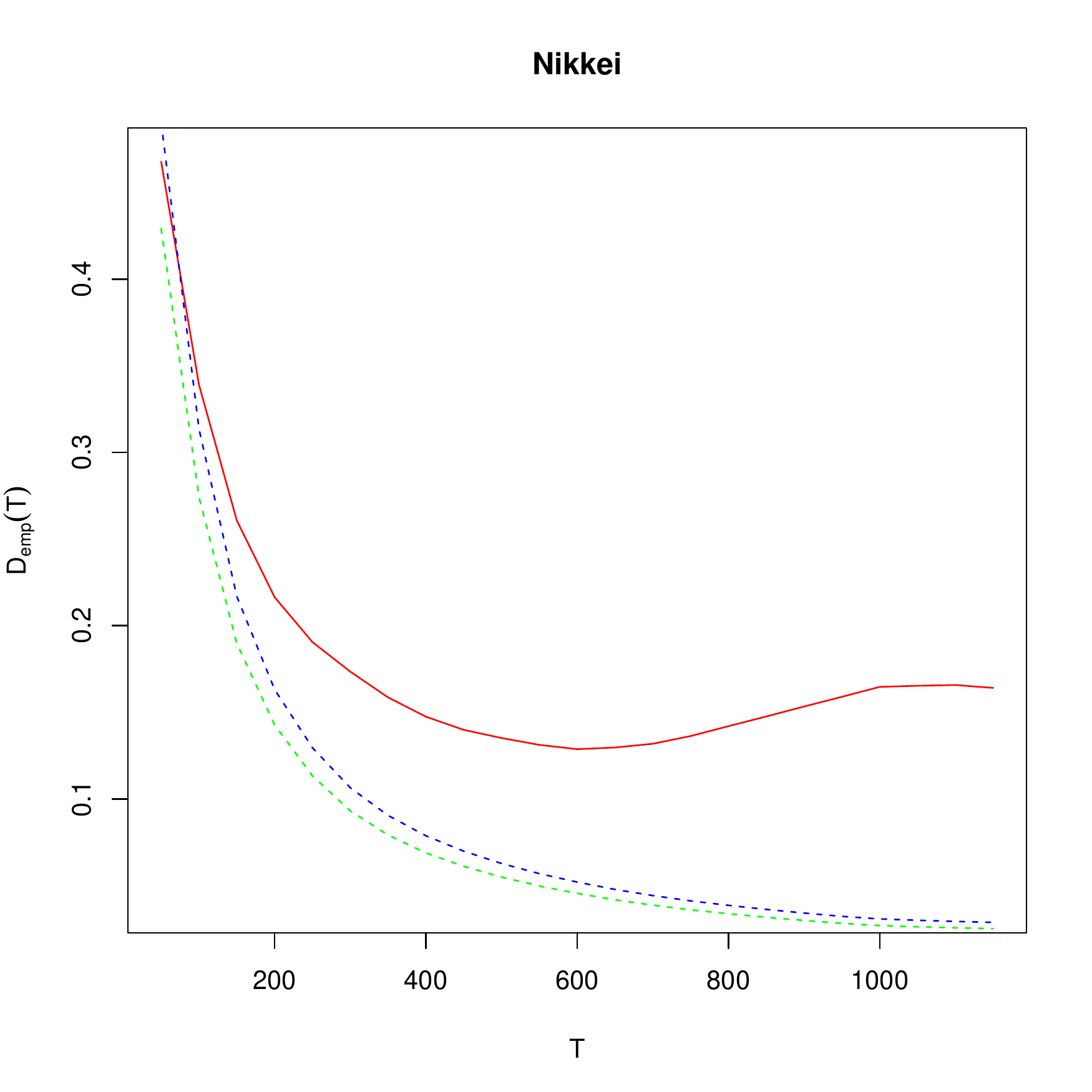}
        \includegraphics[scale=0.40]{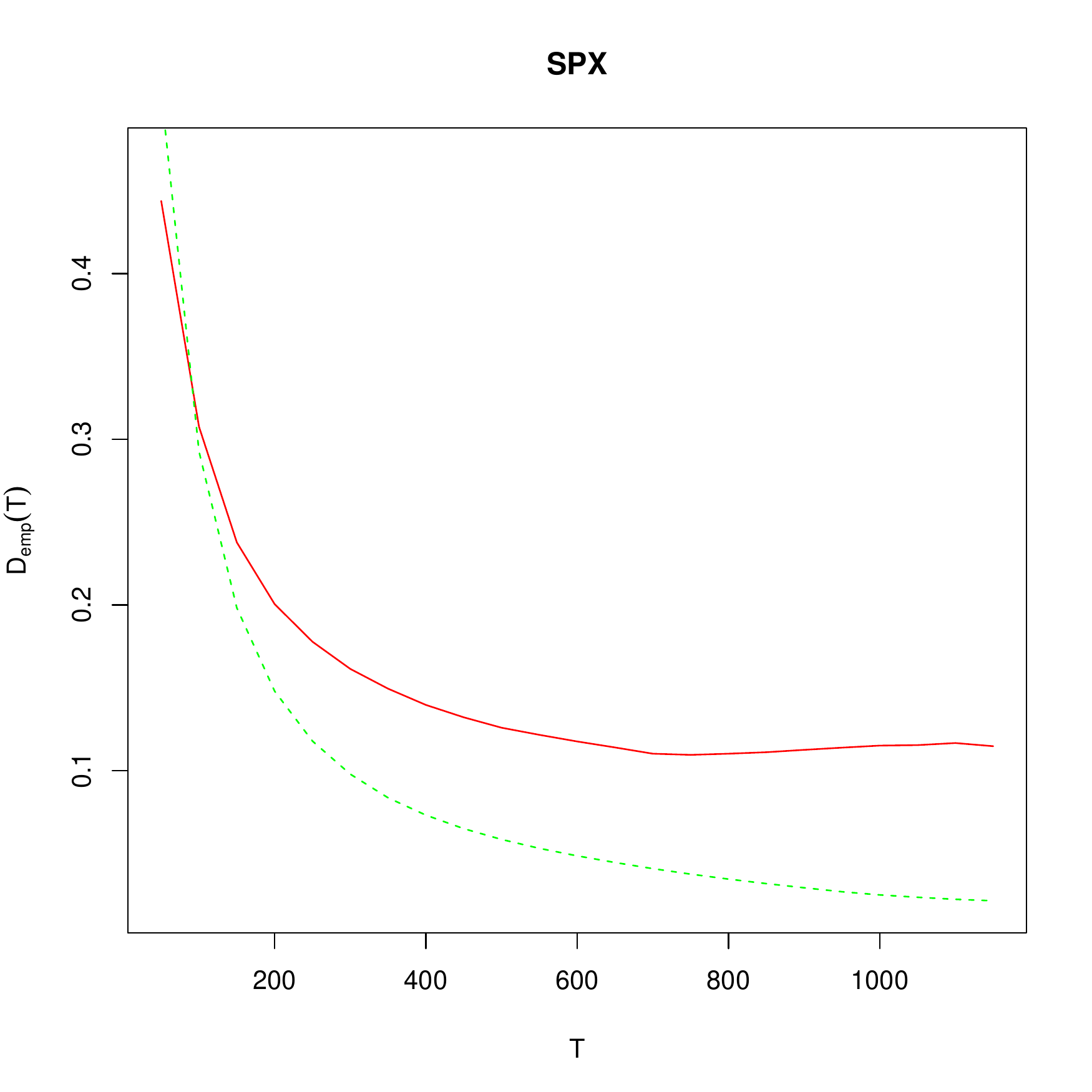}
        \includegraphics[scale=0.40]{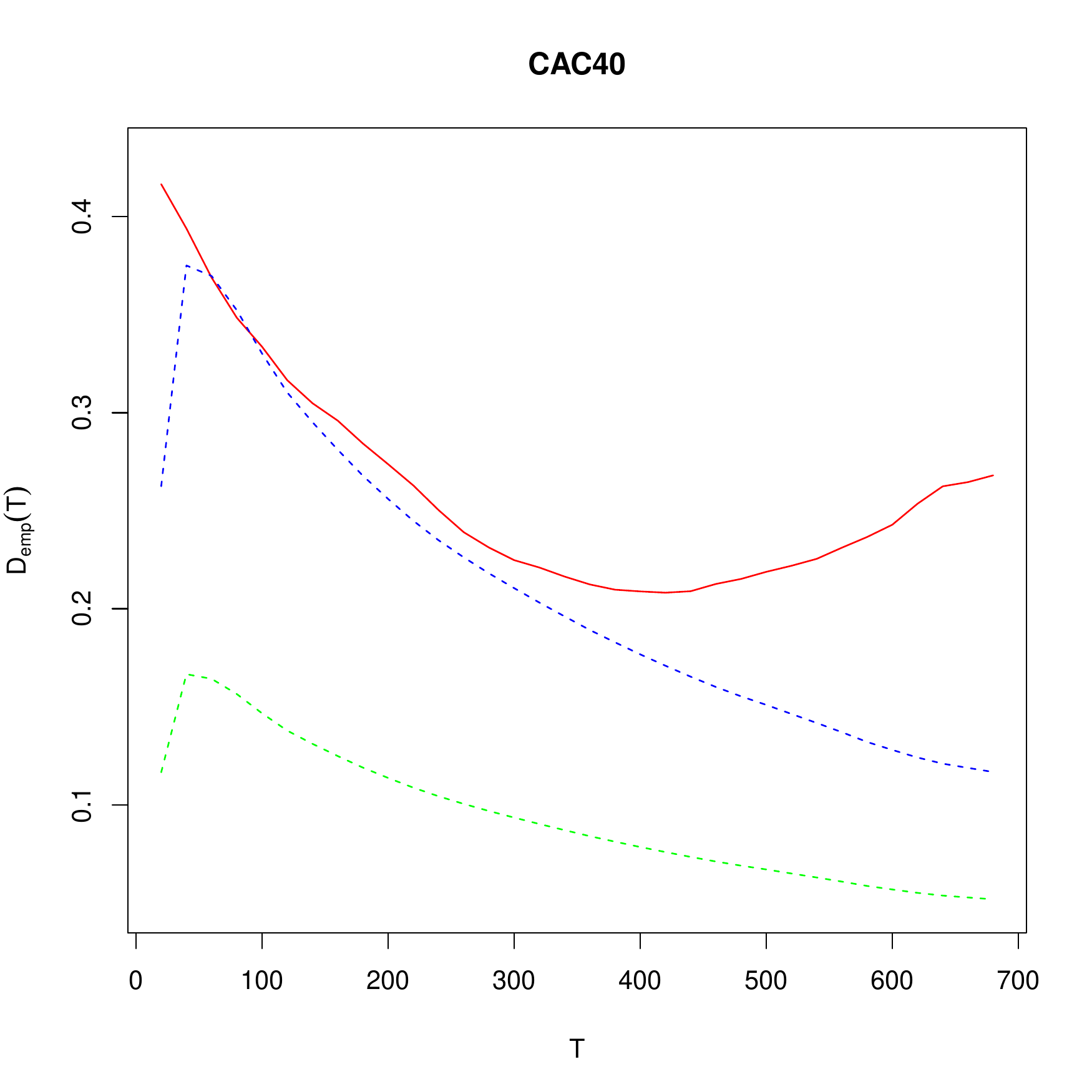}
        \includegraphics[scale=0.40]{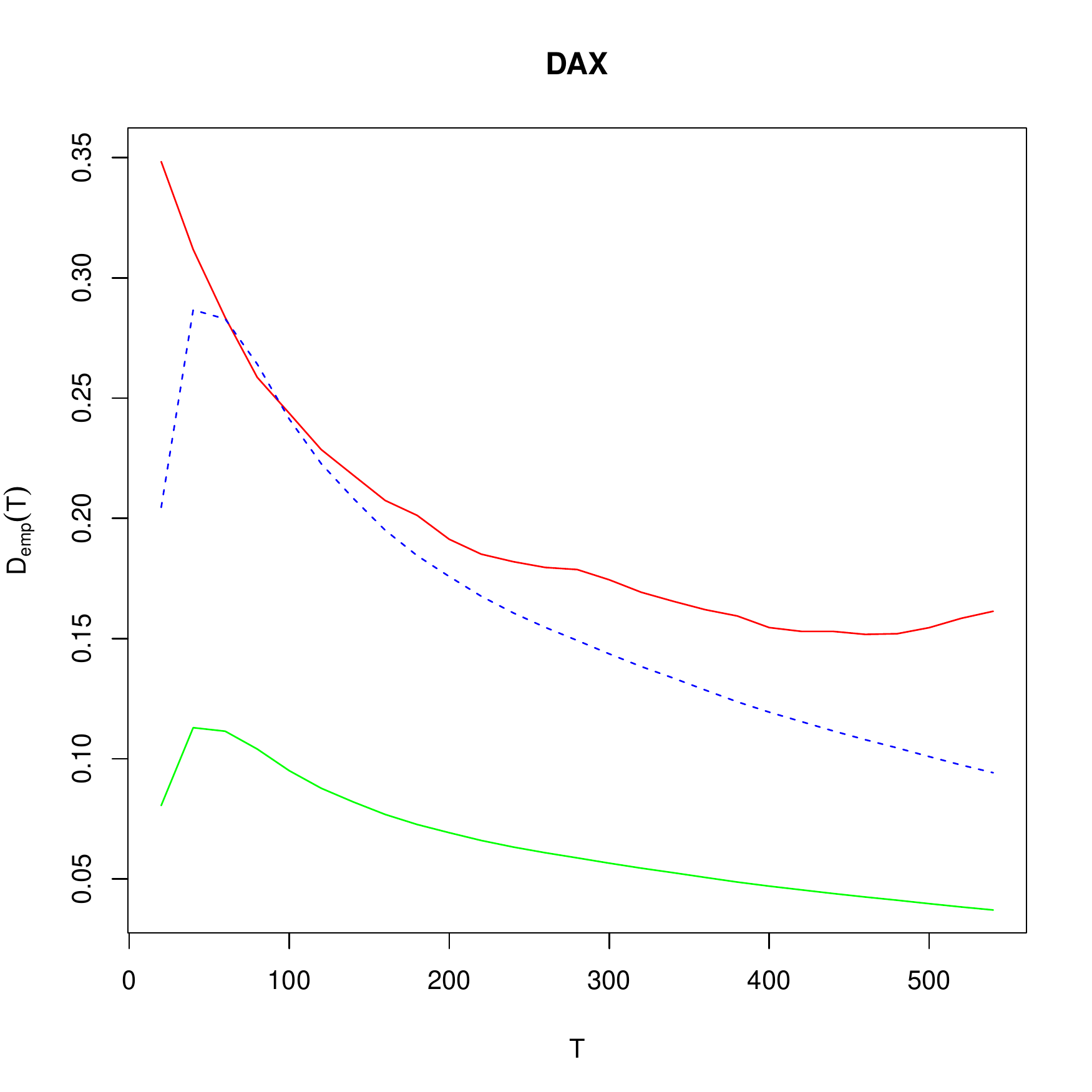}
        \caption{\small{Plot of $D_{emp}(\tau=T)$ (red line) and $D_{th}(\tau=T)$ as a function of $T$, $P=5,Q=10$ 
        for the four indices considered here. The dotted green line represents $D_{th}(\tau=T)$ in the benchmark case where the returns are Gaussian with constant covariance matrix $\bf C$ and
        the dotted blue line represents $D_{th}(\tau=T)$ in the benchmark case where the returns are distributed with a multivariate Student distribution with $\nu$-degrees of freedom and with 
        a constant covariance matrix $\bf C$. The constant $\nu$ is chosen equal to $5.5$ for the CAC40 and DAX indexes and to $18$ for the Nikkei index. 
        The initial decline as $T$ increases follows 
        from reducing the
        measurement noise. However, when $T$ becomes very large, the ``true'' evolution of the eigenvectors is being felt, and leads to an increase of $D_{emp}$. This plot suggests that the 
        optimal time scale to measure the empirical eigenspaces is around two years ($T^*=500$ days).} }  \label{fig_D_emp}      
\end{figure}

We have also computed the value $D_{emp}(\tau=T)$ for different values of $T$ for every pool of stocks, the result is shown in Fig. \ref{fig_D_emp}. 
We compare the empirical function $T \rightarrow D_{emp}(T)$ with the theoretical value $D_{th}(T)$ in the benchmark case where the stock returns 
are distributed as Gaussian vectors of constant covariance matrix $\bf C$. At first sight, the noise contribution appears to be too small to 
explain the value of $D_{emp}(T)$ at small $T$s, at least for the pool of the CAC40 and DAX indices. Nevertheless, if we now compare 
the value of $D_{emp}(\tau=T)$ for small value of $T$ with the value of $D_{th}(\tau=T)$ in the benchmark case 
where the stock returns are distributed with a multivariate Student distribution with $\nu$-degrees of freedom and with a constant covariance matrix $\bf C$,  
we see that we can make the two curves coincide for small values of $T$. Therefore, the initial decline as $T$ increases indeed follows from a reduction of the measurement noise. 
However, when $T$ becomes very large, the ``true'' evolution of the eigenvectors starts being visible, and leads 
to an increase of $D_{emp}$. This plot suggests that the optimal time scale to measure the empirical eigenspaces is around $T^*=600$ days for the stocks from the 
Nikkei index, $T^*=400$ days for the ones from CAC40, $T^*=450$ days for the ones from DAX and $T^*=700$ days for the ones from the SP500 index.


\begin{figure}[h!btp] 
	\center
		\includegraphics[scale=0.6]{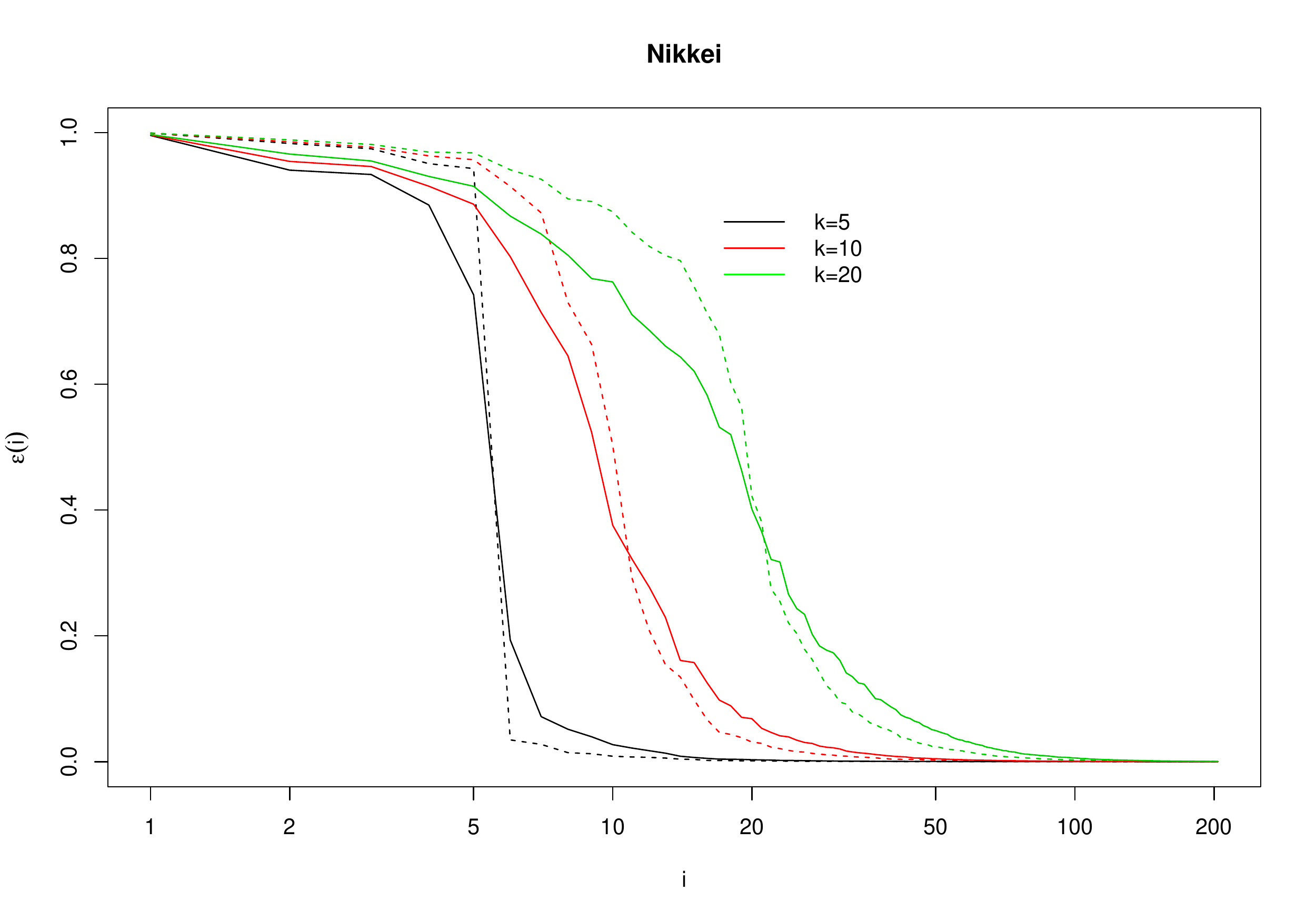}
            \caption{\small{The dotted lines represent the eigenvalues $\epsilon_i$ of the spectral projector of rank $k$ 
            (for $k=5,10,20$ see the legend) as a function of  $\log(i), i=1,\dots,M$ in 
            the benchmark case where the
            true correlation matrix $\bf C$ is not evolving but dressed by measurement noise. The plain lines represent the same function for the empirical data 
            from the Nikkei index ($204$ stocks between $2000-2010$). Here $T=600$.
            In the ideal case (constant correlation matrix, $T \to \infty$), these functions should be step functions: $\epsilon_{i \leq k}=1$ and $\epsilon_{i > k}=0$.}}   
            \label{fig_spectral_projectors}      
\end{figure}

The above results are fully confirmed, and made more precise, by the spectral projector analysis proposed by Zumbach.  
In Fig. \ref{fig_spectral_projectors} we plot, as in \cite{Zumbach}, the eigenvalues of the average spectral projector $\overline{\chi_k'}$ as a function of its theoretical rank $k$, for several values 
of $k$. We show in plain lines the eigenvalues of the empirically determined
$\overline{\chi_k'}$ for the Nikkei idex, where the averaging is made over (overlapping) periods of length $T=600$ days, and in dotted lines the corresponding theoretical 
predictions Eqs. \eqref{benchmark_spectrum_1} and \eqref{benchmark_spectrum_2} for the benchmark case where the eigenspaces are fixed in time, but are blurred by measurement noise. Here again we find clear signals 
of a true evolution of the eigenspaces. 
The results for other stock indices are very similar.

\subsection{The dynamics of the top eigenvector}\label{emp_dyn_top_eigenvector}

As explained above, one expects in general the top eigenvector to wobble around its ``true'' direction $|\phi_1 \rangle$. The fluctuations around 
$|\phi_1 \rangle$ have two possible origins: one is measurement noise, the other is the presence of a systematic rotation of the top eigenvector due
to some financial mechanism. 

\begin{figure}[h!btp] 
	\center
		\includegraphics[scale=0.6]{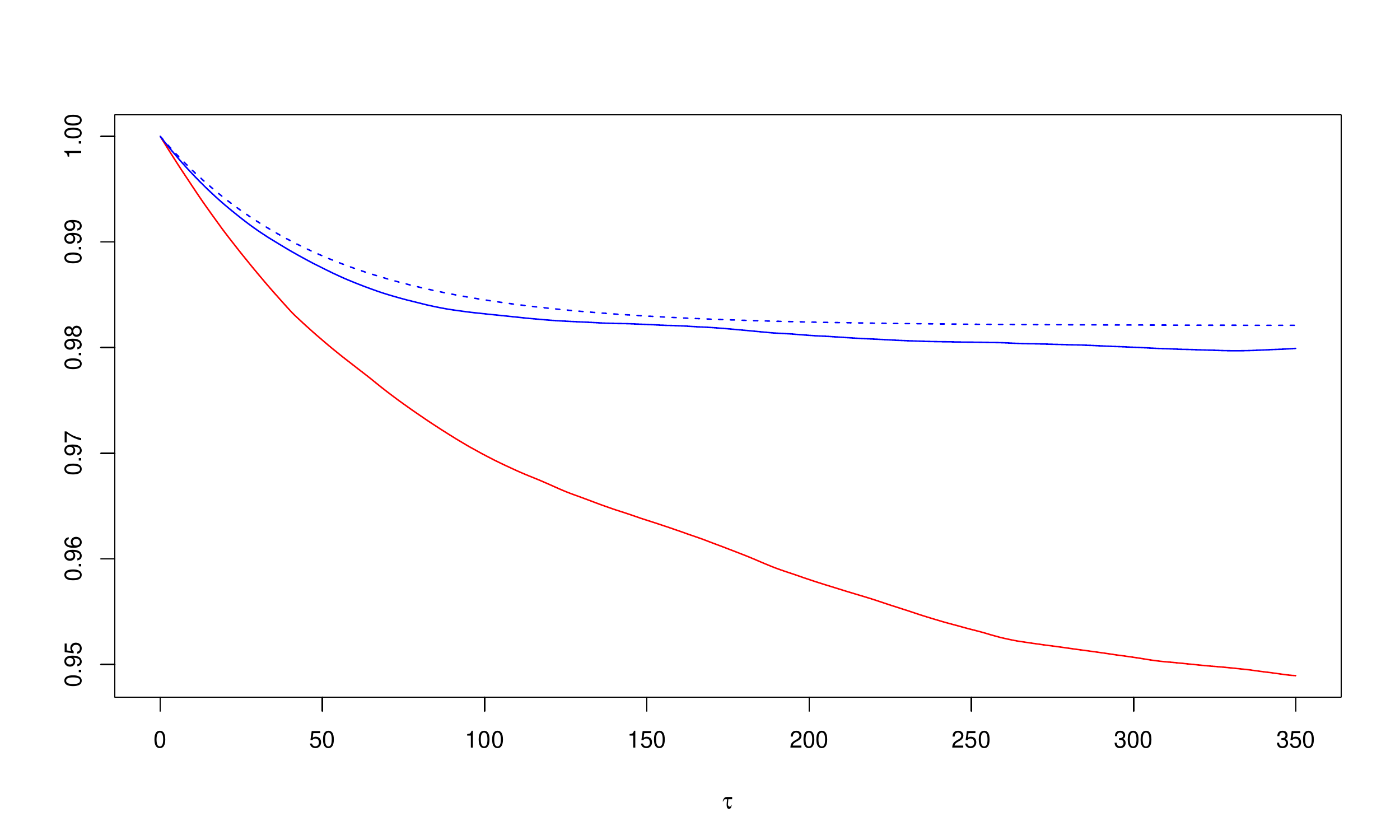}
            \caption{\small{The plain line represents the empirical function $\langle \phi_1^t | \phi_1^{t+\tau} \rangle$ as a function of $\tau$. The period on which the average is 
            performed has $2336$ days starting 01/01/2000. 
            There are $N=204$ stocks from the Nikkei index. The exponential moving average is made with a parameter $\epsilon=1/50$. The true empirical correlation matrix ${\bf C}$ is 
            chosen to be the empirical correlation matrix computed using the data on the whole period. For this ${\bf C}$, we have $\lambda_1 \approx 73$ and $\lambda_2 \approx 0.7$.
            The beginning of the period is used to initialize the exponential moving average. The plain blue is a numerical simulation in the benchmark case. 
            The dotted line represents the function $\tau \rightarrow 1 - 2 \mu(1-\exp(-\varepsilon \tau))$ which corresponds to the benchmark case when there is a constant in time correlation matrix.
            }} 
             \label{varioangle}      
\end{figure}

As a further check that measurement noise is not enough to explain the observed dynamics of $|\phi_1^t \rangle$, we have studied numerically the
average overlap of the top eigenvector measured a time $\tau$ apart: $\overline{\langle \phi_1^t | \phi_1^{t+\tau}\rangle}$. This is an interesting 
quantity because it does not require the knowledge of the true direction $|\phi_1 \rangle$. As shown above, this quantity should be approximately given 
by $1-2 \mu \left(1-e^{-\varepsilon \tau} \right)$ if measurement noise is the only source of fluctuations. We show in Fig. \ref{varioangle} a 
comparison between this prediction and empirical data on the market mode of the Nikkei index. Here again, we find that the decorrelation of the top eigenvector is much stronger than the benchmark. 
The deviation from unity is, for $\tau=350$, more than three times larger than the benchmark case, 
with no signs of saturation. 
 
\begin{figure}[h!btp] 
	\center
		\includegraphics[scale=0.6]{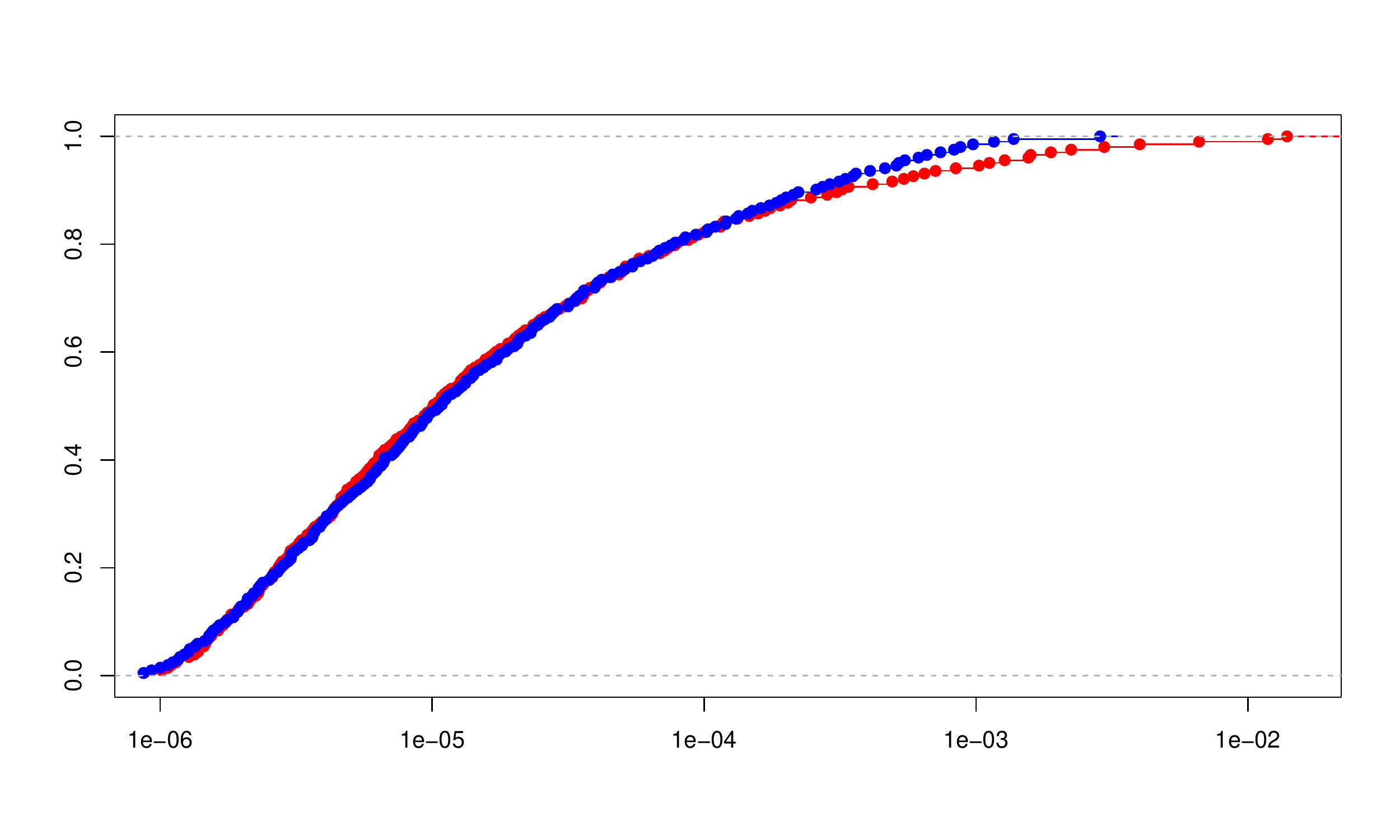}
            \caption{\small{The red curve represents the cumulative distribution of the density of states of the matrix $\bf F$ for the Nikkei index 
            with $N=204$ stocks, in the period $2000-2010$, with $\varepsilon = 1/50$. 
            The blue curve is a numerical simulation for the benchmark case with the true correlation matrix $\bf C$ chosen to be 
            the empirical correlation matrix using the whole period. 
            For this period and pool of stocks, we have $\lambda_1\approx 73$ and $\lambda_2 \approx 0.7$. }}  
             \label{spectrum_D}      
\end{figure}

So there is a genuine motion of the top eigenvector in time. This was already pointed out in \cite{pa_leverage}, where we established empirically that
the top eigenvector rotates towards the uniform vector $|e \rangle = (1,1,\dots,1)/\sqrt{N}$ when the market goes down, and away from $|e \rangle$ 
when the market goes up. In order to be more comprehensive and understand in details the dominant transverse fluctuations of the top eigenvector, 
we have studied the correlation matrix ${\bf F}$ defined in subsection \ref{transvese_top_ev} above. We first determined the eigenvalue spectrum of ${\bf F}$ numerically, both for the benchmark 
case (with only measurement noise) and for real empirical data, see Fig. \ref{spectrum_D}. From this figure, we
conclude that, for the Nikkei index during the period $2000-2010$, there are $3$ (maybe $4$) eigenvalues of the empirical matrix $\bf F$ that 
reside outside the spectrum of the corresponding benchmark matrix. This suggests that these $3$ or $4$ modes are real and correspond to true fluctuations 
of the market mode, that contribute 
to the discrepancy displayed in Fig. \ref{varioangle} above. We are now in a position
to identify the corresponding eigenvectors, i.e. the directions in which the market mode most likely to tilt. 

It is natural to think that these directions should themselves correspond to large eigenvectors of the correlation matrix ${\bf C}$. Therefore we look for the decomposition of the top three 
eigenvectors of ${\bf F}$ (that we call $|\omega_1\rangle, |\omega_2\rangle, |\omega_3\rangle$) in terms of 
$|\phi_i \rangle, i \in \{2,3,4,5\}$. A  singular value analysis of the $3 \times 4$ overlap matrix shows that that one can indeed explain $\approx 85 \%$ of these three eigenvectors in this way, with:
\begin{align}\label{omega}
| \omega_1 \rangle & \approx  -0.34 \,|\phi_2\rangle + 0.29\, |\phi_3\rangle + 0.30 \,|\phi_4 \rangle + 0.84 \,|\phi_5 \rangle \notag \\
| \omega_2 \rangle & \approx  0.53 \,|\phi_2\rangle + 0.45\, |\phi_3\rangle + 0.47 \,|\phi_4 \rangle - 0.54 \,|\phi_5 \rangle \\ 
| \omega_3 \rangle & \approx  0.77 \,|\phi_2\rangle + 0.40\, |\phi_3\rangle + 0.48 \,|\phi_4 \rangle \notag \,.
\end{align}
This means that all the four top eigenvectors of ${\bf C}$ contribute to the ``tilt motion'' of the market mode. 
To check that this result is significant, we ran numerical simulations for this singular value decomposition in the benchmark case with a constant 
correlation matrix $\bf C$ chosen as before to be the empirical correlation matrix computed using the whole period of time (here the decade $2000-2010$).
The $3 \times 4$ singular values analysis now give an explanatory power of $\approx 70 \%$, which is clearly less than the $85 \%$ obtained above. 
Still, a large part of this explanatory power seems to trivially come from the non random structure of ${\bf C}$ itself. 

\begin{figure}[h!btp] 
    \center 
         \includegraphics[scale=0.6]{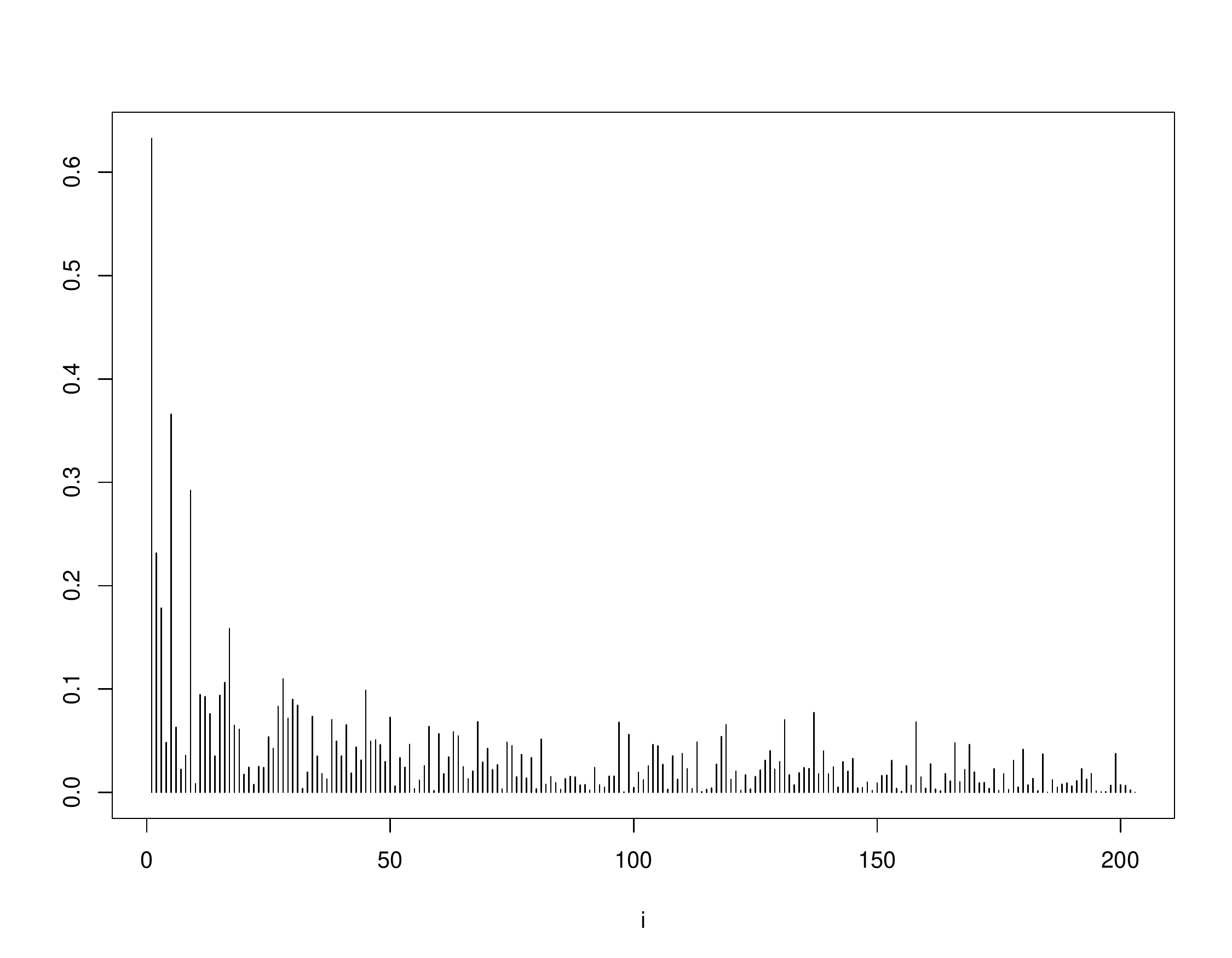}
             \caption{\small{Plot of the overlap $|\langle e_\perp|\phi_i \rangle|$ as a function of $i$ for $i\geq 2$. This graph shows that the main contribution to $| e_\perp \rangle$ comes
             from the top eigenvectors of the correlation matrix ${\bf C}$. }}\label{overlap_e_perp}
\end{figure}

In order to revisit the result found in \cite{pa_leverage}, we need to understand the link between the uniform vector $|e\rangle$ and the 
eigenvectors $|\phi_2\rangle, \dots, |\phi_5\rangle$ of the correlation matrix $\bf C$. Thus, we look at the orthogonal projection $| e_\perp \rangle
:= (|e\rangle-\langle e | \phi_1 \rangle |\phi_1\rangle)/\mathcal{N}$ ($\mathcal{N}$ is chosen such that $\langle e_\perp | e_\perp \rangle^2=1$ )
of the uniform vector $|e\rangle$ in the space generated by the $|\phi_i\rangle, i\geq 2$. The overlap $\langle e_\perp | \phi_i\rangle$ for all $i>2$ are
shown in fig. \ref{overlap_e_perp} for the Nikkei index during the period $2000-2010$. We see that $|e_\perp\rangle$ has indeed very strong overlap 
with $|\phi_2\rangle, |\phi_3\rangle, |\phi_4\rangle, |\phi_5\rangle$, and hence, from the above results, also with $|\omega_1\rangle, |\omega_2\rangle, |\omega_3\rangle$.
Therefore, the fact that the main fluctuation modes of $|\phi_1\rangle$ are along these three $\omega$ directions is compatible with the tilt motion towards 
$|e \rangle$. However, other modes, not mentionned in \cite{pa_leverage}, are detected by the present analysis.

\section{Conclusion \& Open problems}

Let us try to summarize what we have achieved in this paper. We have developed general tools to describe the dynamics of eigenvectors under the influence 
of small random perturbations and to study the stability of the subspace spanned by $P$ consecutive eigenvectors of a generic symmetric matrix. This problem 
is relevant in various contexts, including 
quantum dissipation and financial risk control, but hopefully the ideas and methods introduced here can be used in a much broader context. 


We argue that the problem can be formulated in terms of the singular values of the overlap matrix between the initial eigenspace and the target eigenspace, that allows one to define an 
overlap distance, which is small if most of the initial information is conserved. We first specialize our results for the case of a Gaussian Orthogonal Ensemble, for which 
the full spectrum of singular values can be explicitly computed in the limit of large matrices under the regime where the entries of the perturbation are very small 
compared to the mean level spacing of the non-perturbed matrix. We argue that our setting with rectangular $Q\times P$ overlap matrices $\bf G$ allows 
to extend our results to perturbations with entries larger than the mean level spacing. We provide some numerical evidences that it is indeed true. 
We find two regimes, depending on the dimension of the target space $Q$ compared to that of the initial space $P$. If $Q \gg P$, all singular values 
are close to one another, and their distribution is given by Wigner's semi-circle. If on the other hand $(Q-P)/P \ll 1$, the singular values $s$ are distributed 
according to a very broad law that decays as $s^{-2}$. These results are actually universal, and apply for other matrix 
ensemble as well -- for example the case of empirical covariance matrices -- provided one is interested in eigenspaces deep in the bulk.

We have also studied the case of isolated eigenvalues, that are usually very important for applications, for example in finance. In most cases, empirical 
correlation matrices are noisy measurements of the true covariance matrix, that can lead to an apparent evolution of the top eigenspace, whereas in reality the underlying process is stationary. 
We have derived exact expressions both for the overlap distance and for the average spectral projectors (introduced by Zumbach \cite{Zumbach}) that can be directly 
compared to empirical results. The special case where the top eigenvalue is much larger than all the other ones can be investigated in full detail. In particular, the 
dynamics of the angle made by the top eigenvector and its true direction defines an interesting new class of random processes, for which we have provided explicit analytical results. 

When compared to empirical correlation matrices of several major stock markets, our results allow us to unambiguously conclude that there is a genuine 
evolution in time of the true underlying correlation matrix: measurement noise in itself is unable to explain the observed variability (in time) of the top eigenspaces. We have found that 
the overlap distance is minimized when the measurement time is on the order of two to three years. Both for shorter and 
longer averaging times, measurement noise and the genuine evolution of the market leads to an instability of the correlation matrix, and to exposure to 
unwanted sources of risk. 

The case of the top eigenvector of the correlation matrix, usually called the market mode, is particularly interesting. We have suggested a characterization of the evolution of its direction through a new correlation matrix, which measures the amplitude of its fluctuations transverse to its average direction. We found that the dominant modes are in the space spanned by the largest eigenvectors of the correlation matrix itself. 

Now the genuine evolution of the correlation structure of stock returns is well characterized, one should aim at devising quantitative models for this
evolution. As usual, there are two ways to do this. One is to postulate an econometric model and try to calibrate it on data. In this line of thought, 
extensions of the GARCH framework have been proposed: multivariate GARCH, BEKK model, etc.\cite{BEKK}, but they often lack intuition (to say the least) and are very hard to calibrate (the a priori number of parameters 
is of order $N^4$!). 

The second approach is to think about mechanisms that can lead to changes of the correlation structure. For example, market drops may lead to panic sell-offs, that increase the top eigenvalue of the correlation matrix and tilt the top eigenvector towards uniformity, as reported in \cite{Ingve,pa_leverage}.
The impact of rebalancing or deleveraging complex portfolios can also lead to substantial changes in the correlation matrix -- see the insightful work of 
Cont and Wagalath \cite{Cont} in this direction. We hope that the tools provided in this paper will help building financially motivated, more efficient models of dynamical correlations and, correspondingly, second 
generation risk models where impact and feedback effects are accounted for \cite{Arthur}. 

\appendix

\section{Proof of the formula \eqref{delta=0}}

We need to introduce the two level density of states 
$$ \rho_2^N(\lambda,\lambda') = \frac{1}{N^2}\sum_{i,j=1}^{N}\delta(\lambda-\lambda_i,\lambda'-\lambda_j),$$
and to note from equation \eqref{expr_D_dimension_finie} that
\begin{equation}
D(V_0;V_1) = \frac{N \varepsilon^2}{2P} \int_a^b \int_{[-2;2]\setminus[a;b]} \frac{ \rho_2^{N}(\lambda,\lambda')}{(\lambda-\lambda')^2}.
\end{equation}
From \cite{Mehta}, we know the asymptotic behavior of the two level density of states in the limit of large matrices; 
more precisely, there exists a function $g$ such that, in the limit of large $N$,
\begin{equation}
 \rho_2^N(\lambda,\lambda') = g(N\rho(\lambda)|\lambda-\lambda'|)\rho(\lambda) \rho(\lambda') d\lambda d\lambda'
\end{equation} 
which is defined as $g(r) = 1 - \left(\frac{1}{2}- \int_0^r s(t)dt \right) s'(r) + s(r)^2$ with $s(r) = \frac{\sin(\pi r)}{\pi r}$. 
One can check that:
\begin{itemize}
\item $g(r)\leq 1$ for all $r$,
\item in the neighborhood of $0$, $g(r) \sim \frac{\pi^2}{2} r$, 
\item $g(r)$ tends to $1$ when $r$ goes to $\infty$,
\item $g'(r)=O(1/r^2)$ in the neighborhood of $\infty$.
\end{itemize}
We can write:
\begin{equation*}
D(a,b;\delta=0) = \frac{N\varepsilon^2}{2P} \int_a^b \int_{[-2;2]\setminus[a;b]} \frac{g(N\rho(x)|x-y|)}{(x-y)^2}
\rho(x) \rho(y) dx dy .
\end{equation*}
We want to do an asymptotic expansion of the right hand side when $N\to\infty$.
 
First, note that $N/P$ tends to $1/\int_a^b \rho$.

For the integral, we begin by doing an integration by part, we get 
for $x\in[-2;a]$:
\begin{align}\label{ipp}
 \int_a^b &\frac{g(N\rho(x)|x-y|)}{(x-y)^2}  \,dy = \frac{\rho(a) g(N\rho(x)(a-x))}{a-x} - \frac{\rho(b) g(N\rho(x)(b-x))}{b-x}&   \\
    &  + \int_a^b \frac{dy}{y-x} \left[ \rho'(y) g\left(N \rho(x)(y-x)\right) + N \rho(x) \rho(y) g'\left(N \rho(x)(y-x)\right) \right] \,.
\end{align}

We need to integrate equation \eqref{ipp} between $-2$ and $a$ and between $b$ and $2$ and to compute the asymptotic of 
every integrals of the right hand side. We will decompose each integral into two terms so as to take advantages of the asymptotic 
property of $g$ around $0$ and $\infty$. 

Set $\eta=N^{-1+\alpha}$ with $\alpha>0$.
First we consider the integral: 
\begin{align*}
 \int_{a-\eta}^{a} \rho(x)&\frac{g(N\rho(x)(a-x))}{a-x}dx = \int_0^{N^\alpha} \frac{dx}{x} \rho(a-\frac{x}{N})g(\rho(a-\frac{x}{N}) x)\\
 &\sim \rho(a)\int_0^{N^\alpha} \frac{dx}{x}  g(\rho(a) x) =  \rho(a)\left[\ln\left(\rho(a)N^{\alpha}\right) - 
  \int_0^\infty \ln(x) g'(x)dx\right]\,.
 \end{align*}
Using the fact that $g(r)$ tends to $1$ when $r$ goes to $\infty$, we easily get that, in the limit $N\to\infty$:
\begin{align*}
\int_{-2}^{a-\eta}  & \rho(x) \frac{g(N\rho(x)(a-x))}{a-x}dx \\ 
&\sim \int_{-2}^{a-\delta} \frac{\rho(x)}{a-x}dx = -\rho(a)\ln(N^{-1+\alpha}) + 
\int_{-2}^a \rho'(x) \ln(a-x) dx \,.
\end{align*}

Moreover, we easily find that, when $N$ tends to $\infty$: 
\begin{equation*}
\int_{-2}^{a-\eta} \rho(x) \frac{g(N\rho(x)(b-x))}{b-x}dx \rightarrow \int_{-2}^{a} \frac{\rho(x)}{b-x} dx \,,
\end{equation*} 
and 
\begin{equation*}
\int_{a-\eta}^{a} \rho(x) \frac{g(N\rho(x)(b-x))}{b-x}dx \leq \int_{a-\eta}^{a} \frac{\rho(x)}{b-x} dx \,,
\end{equation*}
which goes to $0$ as $\eta$ goes to $0$.

The next term is easy to control using the fact that $g(r)$ goes to $1$ when $r$ goes to $\infty$; as $N\to\infty$:
\begin{align*}
\int_{-2}^{a-\eta} dx \rho(x) \int_a^b \frac{\rho'(y)dy}{y-x} g\left(N \rho(x)(y-x)\right) \rightarrow \int_{-2}^{a} dx \rho(x) \int_a^b \frac{\rho'(y)dy}{y-x}\,.  
\end{align*}

Using the fact that $g'(r)$ is of order $1/r^2$ for large $r$, it is easy to check that 
\begin{equation*}
N \int_{-2}^{a-\eta} dx \rho(x) \int_a^b  \frac{dy}{y-x}  \rho(y) g'\left(N \rho(x)(y-x)\right)
\end{equation*}
is of order $N^{-\alpha}$.

The remaining term 
\begin{equation*}
\int_{a-\eta}^{a} dx \rho(x) \int_a^b \frac{dy}{y-x} \left[ \rho'(y) g\left(N \rho(x)(y-x)\right) + N \rho(x) \rho(y) g'\left(N \rho(x)(y-x)\right) \right] 
\end{equation*}
is of order $N^{-1+\alpha}$.

One has to go through the same steps to compute the asymptotic of the integrals between $b$ and $2$.

Finally, we get:  
\begin{equation}
D(a,b;\delta=0) \approx \ln N \,\, {\varepsilon^2} \,\, \frac{\rho(a)^2+\rho(b)^2}{2\int_a^b \rho(\lambda) d\lambda } + A(a,b)
\end{equation}
where 
\begin{align*}
A(a,b) &= \frac{\varepsilon^2}{2\int_a^b \rho}    \Bigg[ \left(\rho(a)^2+\rho(b)^2\right) \left(1- \int_0^\infty \ln(x) g'(x) dx\right)  \\ 
&+ \rho(a)^2 \ln(\rho(a)) +  \rho(b)^2 \ln(\rho(b))\\ 
& + \rho(a) \int_{-2}^a \rho'(x) \ln(a-x) dx - \rho(b) \int_{b}^2 \rho'(x) \ln(x-b) dx \\
&-\rho(b) \int_{-2}^a \frac{\rho(x)dx}{b-x}  -\rho(a) \int_{b}^2 \frac{\rho(x)dx}{x-a} \\
&+ \int_{-2}^{a} dx \rho(x) \int_a^b \frac{\rho'(y)dy}{y-x} - \int_{b}^2 dx \rho(x) \int_a^b \frac{\rho'(y) dy}{x-y}  \Bigg]. 
\end{align*}

\section{Derivation of the standard deviation $\sigma(r)$ of $r(s)$.}
We have 
\begin{align*}
\sigma^2(r) &= \langle r(s)^2 \rangle - \langle r(s) \rangle^2 \\
 &\approx \frac{1}{P} \langle \tr\left(\Sigma^2\right) \rangle - \langle \left(\frac{1}{P} \tr(\Sigma)\right)^2 \rangle \,.
\end{align*}
But the two quantities are computable easily in the limit of large matrices using the convergence of the density of states for ${\bf H}_0$; We obtain 
\begin{align*}
\frac{1}{P} \langle \tr\left(\Sigma^2\right) \rangle &\approx \frac{\varepsilon^4}{N_a^b} \int\limits_{[a;b]} d\lambda \int\limits_{[-2;2]\setminus[a-\delta;b+\delta]} d\lambda'
\int\limits_{[-2;2]\setminus[a-\delta;b+\delta]} d\lambda'' \frac{\rho(\lambda)\rho(\lambda')\rho(\lambda'')}{(\lambda-\lambda')^2(\lambda-\lambda'')^2}\\
&+ \frac{\varepsilon^4}{N_a^b} \int\limits_{[a;b]} d\lambda \int\limits_{[a;b]} d\lambda'  \int\limits_{[-2;2]\setminus[a-\delta;b+\delta]} d\lambda'' \frac{\rho(\lambda)\rho(\lambda')\rho(\lambda'')}{(\lambda-\lambda'')^2(\lambda'-\lambda'')^2}
\end{align*}
and 
\begin{align*}
\langle \left(\frac{1}{P} \tr(\Sigma)\right)^2 \rangle \approx  \frac{\varepsilon^4}{{N_a^b}^2} \int\limits_{[a;b]} d\lambda \int\limits_{[a;b]} d\lambda' 
\int\limits_{[-2;2]\setminus[a-\delta;b+\delta]} d\lambda'' 
\int\limits_{[-2;2]\setminus[a-\delta;b+\delta]} d\lambda''' \frac{\rho(\lambda)\rho(\lambda')\rho(\lambda'')\rho(\lambda''')}{(\lambda-\lambda'')^2(\lambda'-\lambda''')^2} \,.
\end{align*}
Those two expressions give in the regime $\delta \ll \Delta \ll 1$
\begin{equation}\label{strong_fluc_sd}
\sigma(r) \approx \frac{\rho(a)}{\sqrt{\delta\Delta}} \,, 
\end{equation}
and in the regime $\Delta \ll \delta \ll 1$,
\begin{equation}\label{small_fluc_sd}
\sigma(r) \approx \rho(a) \sqrt{\frac{2\Delta}{3\delta^3}} \,.
\end{equation}

\section{Determination of $s_{min}$ in the strong fluctuation limit}

It is given by \eqref{smin} and we have to compute the $\tilde{g}\in (-\infty;0)$ that verifies \eqref{eq_barg}.  
For simplicity, we set $\hat{g}=-\tilde{g}$ and we aim to compute $\hat{g}\geq 0$ such that 
\begin{equation*}
 \frac{1}{N_a^b}  \int\limits_{[-2;2]\setminus[a-\delta;b+\delta]}dx\frac{\rho(x)\sigma(x)^2}{(1+\sigma(x)\hat{g})^2} = \frac{1}{\hat{g}^2} \,,
\end{equation*}
As $\hat{g}$ is non-negative, the integral on the left hand side converges when $\delta$ goes to $0$ and hence $\hat{g}$ verifies in fact 
\begin{equation}\label{eq_hatg}
 \frac{1}{N_a^b}  \int\limits_{[-2;2]\setminus[a;b]}dx\frac{\rho(x)\sigma(x)^2}{(1+\sigma(x)\hat{g})^2} = \frac{1}{\hat{g}^2} \,,
\end{equation}
We now need to estimate the integral in the limit $\Delta \ll 1$. As before, we can write using \eqref{forme_sigma}
\begin{align*}
\int_{-2}^a dx \, \frac{\rho(x)\sigma(x)^2}{(1+\sigma(x)\hat{g})^2} =  \frac{\rho(a)^2}{\Delta} \int_0^{\frac{a+2}{\Delta}} du \, 
\frac{\rho(a-u \Delta) f^2(u)}{(u+ \frac{\rho(a)\hat{g}}{\Delta} f(u))}\,.
\end{align*}
In the limit $\Delta\ll1$, this integral is dominated by the region where $u$ is small and $f(u)\sim 1$ and hence we have the following estimate 
\begin{align*}
\int_{-2}^a dx \, \frac{\rho(x)\sigma(x)^2}{(1+\sigma(x)\hat{g})^2} &\sim \frac{\rho(a)^3}{\Delta} \int_0^{+\infty} \frac{du}{(u+\frac{\rho(a)\hat{g}}{\Delta})^2}\\
&\sim\frac{\rho(a)^2}{\hat{g}}\,.
\end{align*}
Then we deduce from \eqref{eq_hatg} and with the same argument for the integral between $b$ and $2$ that 
\begin{equation*}
\hat{g} = \frac{\Delta}{2\rho(a)}\,.
\end{equation*}
Now we have to plug this $\hat{g}$ into equation \eqref{smin} to obtain
\begin{equation*}
s_{min} = -\frac{2\rho(a)}{\Delta} + \frac{1}{N_a^b} \int\limits_{[-2;2]\setminus[a;b]} dx \, \frac{\rho(x)\sigma(x)}{1+\sigma(x) \frac{\Delta}{2\rho(a)}} \,.
\end{equation*}
To evaluate the integral, we need to cut it into two parts. The first part is handled by
\begin{align*}
\int_{-2}^a  \frac{\rho(x)\sigma(x)}{1+\sigma(x) \frac{\Delta}{2\rho(a)}} &= \rho(a) \int_0^{\frac{a+2}{\Delta}}du\, \frac{\rho(a-u\Delta)f(u)}{u+f(u)/2} \\
&\rightarrow \rho(a)^2 \int_0^{+\infty} du\, \frac{f(u)}{u+f(u)/2}\,.
\end{align*}
Finally, we can deduce that
\begin{equation*}
s_{min} = \frac{2\rho(a)}{\Delta} \left(\int_0^{+\infty} du\,\frac{ f(u)}{u+f(u)/2} -1  \right) \,.
\end{equation*}

\section{Derivation of \eqref{xt} }

Using perturbation theory, one gets, in 
braket notation:
\begin{align*}
   | \phi_1^t \rangle &= \left(1-\frac{\varepsilon^2}{2} \sum_{i\not=1} \frac{\langle\phi_1^{t-1}|r^t{r^t}^*|\phi_i^{t-1}\rangle^2}{(\lambda_1^{t-1}-\lambda_i^{t-1})^2}\right) |\phi_1^{t-1}\rangle 
    + \varepsilon \sum_{i\not=1}    \frac{\langle\phi_1^{t-1}|r^t{r^t}^*|\phi_i^{t-1}\rangle}{\lambda_1^{t-1}-\lambda_i^{t-1}} |\phi_i^{t-1}\rangle\\
    &\approx    \left(1-\frac{\varepsilon^2}{2(\lambda_1^{t-1})^2} \sum_{i\not=1} \langle\phi_1^{t-1}|r^t{r^t}^*|\phi_i^{t-1}\rangle^2 \right) |\phi_1^{t-1}\rangle
    +\frac{\varepsilon}{\lambda_1^{t-1}} \sum_{i\not=1}    \langle\phi_1^{t-1}|r^t{r^t}^*|\phi_i^{t-1} \rangle |\phi_i^{t-1}\rangle\\ 
    &=  \left(1-\frac{\varepsilon^2}{2(\lambda_1^{t-1})^2} \left(\langle\phi_1^{t-1}|(r^t{r^t}^*)^2|\phi_1^{t-1}\rangle - \langle\phi_1^{t-1}|r^t{r^t}^*|\phi_1^{t-1}\rangle^2\right) \right) |\phi_1^{t-1}\rangle\\
    &+ \frac{\varepsilon}{\lambda_1^{t-1}}\left( r^t{r^t}^*|\phi_1^{t-1}\rangle - \langle \phi_1^{t-1} |r^t{r^t}^*|\phi_1^{t-1}\rangle|\phi_1^{t-1}\rangle \right)\,.
\end{align*}

Since $\cos(\theta_t)=\langle\phi_1^t|\phi_1\rangle$, we can write \[\phi_1^t=\cos(\theta_t) |\phi_1\rangle + \sin(\theta_t) |\varphi^t_\perp \rangle\]
where $|\varphi_\perp \rangle$ is a vector lying in the subspace spanned by the vectors $|\phi_2\rangle,\dots,|\phi_N\rangle$.
We want to describe the dynamic of $\cos(\theta_t)$; we deduce from the previous equation that
\begin{align}\label{langevin_cos_theta_d}
\cos(\theta_t) &=    \left(1-\frac{\varepsilon^2}{2(\lambda_1^{t-1})^2} \left(\langle\phi_1^{t-1}|(r^t{r^t}^*)^2|\phi_1^{t-1}\rangle - \langle\phi_1^{t-1}|r^t{r^t}^*|\phi_1^{t-1}\rangle^2\right) \right) 
\cos(\theta_{t-1})   \\ 
&+  \frac{\varepsilon}{\lambda_1^{t-1}}\left( \langle\phi_1|r^t{r^t}^*|\phi_1^{t-1}\rangle - \langle \phi_1^{t-1} |r^t{r^t}^*|\phi_1^{t-1}\rangle \cos(\theta_{t-1}) \right).
\end{align}
 
Since we have:
\begin{align*}
\langle\phi_1|{\bf C}|\phi_1^{t-1}\rangle&=\lambda_1 \cos(\theta_{t-1})\,,\\
\langle\phi_1^{t-1}|{\bf C}|\phi_1^{t-1}\rangle&=\lambda_1 \cos^2(\theta_{t-1})+\lambda_2 \sin^2(\theta_{t-1}) \,,\\
\overline{\langle\phi_1|\eta_t|\phi_1^{t-1}\rangle^2} &= 2 \cos^2(\theta_{t-1}) \lambda_1^2 + \sin^2(\theta_{t-1}) \lambda_1 \lambda_2\,, \\
\overline{\langle\phi_1^{t-1}|\eta_t|\phi_1^{t-1}\rangle^2} &= 2\left(\lambda_1 \cos^2(\theta_{t-1}) + \lambda_2 \sin^2(\theta_{t-1})\right)^2\,, \\
\overline{\langle\phi_1^{t-1}|r^t{r^t}^*|\phi_1^{t-1}\rangle^2}&=\lambda_1^2 \cos^4(\theta_{t-1}) + 2 \cos^2(\theta_{t-1}) \lambda_1^2 + \sin^2(\theta_{t-1}) \lambda_1 \lambda_2\,,\\
\overline{\langle\phi_1^{t-1}|(r^t{r^t}^*)^2|\phi_1^{t-1}\rangle}&= \cos^2(\theta_{t-1}) (3\lambda_1^2 +
(N-1)\lambda_1\lambda_2) +\sin^2(\theta_{t-1})((N+1)\lambda_2^2+\lambda_1\lambda_2)\,, 
\end{align*}
equation \eqref{langevin_cos_theta_d} can be rewritten, in the asymptotic regime where $\varepsilon \ll 1, N \gg 1$ with $q =\varepsilon N$ fixed and $\lambda_2 \ll \lambda_1$,
keeping up to terms of order $2$ for ``drift" terms and of order $1$ for noise terms in $\varepsilon$ and $\lambda_2/\lambda_1$\footnote{Note that $\sin^2(\theta_t)\approx 2 \mu$ is of order $\lambda_2/\lambda_1$
and that $1-\cos(\theta_t)\approx \mu$ is also of order $\lambda_2/\lambda_1$.}: 
\begin{align*}
d(\cos(\theta_t)) &= -\frac{\varepsilon^2}{2}\frac{1}{\lambda_1^2} \left[(\lambda_1^2+ N \lambda_1\lambda_2) \cos^2(\theta_t) -\lambda_1^2 \cos^4(\theta_t) \right]\cos(\theta_t) dt\\
&+ \varepsilon \cos(\theta_t)\sin^2(\theta_t) dt + \sigma_t dB_t
\end{align*}
where
\begin{equation} 
\overline{\sigma_t^2}=\frac{\varepsilon^2}{\lambda_1^2}\left[2\lambda_1^2\cos^2(\theta_t)\sin^2(\theta_t)+\lambda_1\lambda_2\cos^2(2\theta_t)\right]\sin^2(\theta_t).
\end{equation}
When $\theta_t \ll 1$, this leads to Eq. \ref{xt} given in the main text for $x_t=1-\cos(\theta_t)$.

\section{Transition probability of $x_t$}
In this appendix, we show that the function $P(x,t)$ giving the probability that the ``particle'' $x_t$ verifying \eqref{xt} is in $x$ at time $t$ 
can be computed explicitly. More generally, we will show that one can compute explicitly this transition density $P(x,t)$ for a process $x_t$ with initial condition in $t=0$ given by $x_0\geq 0$ verifying the Langevin equation
\begin{equation}\label{xt_general}
{\rm d}x_t = \theta (\mu-x_t) {\rm d}t + \sigma \sqrt{x_t(x_t+b)} {\rm d}B_t
\end{equation}
where $\theta,\mu, \sigma$ and $b$ are positive constants and $B_t$ a standard Brownian motion. 
One can proceed to the change of variables 
\begin{equation*}
y_t = \cosh^{-1}\left(\frac{2}{b} x_t +1\right) \quad \Leftrightarrow \quad x_t = \frac{b}{2} \left(\cosh(y_t)-1\right)\,,
\end{equation*}
and find that the process $y_t$ verifies 
\begin{equation}\label{yt}
{\rm d}y_t = \left(\theta (1+\frac{2\mu}{b}) \frac{1}{\sinh(y_t)} - (\theta+\frac{\sigma^2}{2}) \frac{\cosh(y_t)}{\sinh(y_t)}\right) {\rm d}t + \sigma \,{\rm d}B_t\,.
\end{equation} 
We will denote by $F(y)$ the drift coefficient of the previous stochastic differential equation \eqref{yt} and denote by $U$ its potential,
which verifies $U'= -F$. The transition density $\bar{P}(y,t)$ verifies the Fokker-Planck equation 
\begin{equation*}
\frac{\partial \bar{P}}{\partial t} = - \frac{\partial (F \bar{P})}{\partial y} + \frac{\sigma^2}{2} \frac{\partial^2 \bar{P}}{\partial y^2}  \,.
\end{equation*}
By setting $\bar{P}(y,t) := e^{-U(y)/\sigma^2} \psi(y,t)$, this equation becomes a Schrodinger equation:
\begin{equation*} 
\frac{\partial \psi }{\partial t} = \frac{\sigma^2}{2} \frac{\partial^2 \psi}{\partial y^2} -  V(y) \psi \,,
\end{equation*}
with the so-called P\"oschl-Teller potential $V(y)$:
\begin{align*} 
V(y) &= \frac{1}{2} \left( \frac{F^2(y)}{\sigma^2} + F'(y) \right)\\
&= \frac{1}{2} \left( \frac{\alpha}{\sinh^2(y)} -  \frac{\beta\,\cosh(y)}{\sinh^2(y)} + \gamma \right)\,
\end{align*}
with: 
\begin{align*}
\alpha &= \left(\theta + \frac{\sigma^2}{2} \right) \left(\frac{3}{2} + \frac{\theta}{\sigma^2} \right) + \frac{\theta^2}{\sigma^2} \left(1+\frac{2\mu}{b}\right)^2 \,,\\
\beta &= 2 \theta \left(1+\frac{2\mu}{b} \right) \left(1+\frac{\theta}{\sigma^2}\right)\,,\\
\gamma &= \frac{1}{\sigma^2} \left(\theta + \frac{\sigma^2}{2} \right)^2\,.
\end{align*}
Since the evolution of $\psi(y,t)$ is governed by a self adjoint operator 
\begin{equation*}
\mathcal{H}:=\frac{\sigma^2}{2} \frac{\partial^2 \cdot}{\partial y^2} -  V(y) \cdot
\end{equation*}
we can use its eigenfunctions to construct an orthonormal basis $(\psi_n)$ with corresponding eigenvalues $(-\lambda_n)$. The general solution 
$\psi(y,t)$ can thus be expanded in the following form
\begin{equation*}
\psi(y,t) = \sum_n c_n \psi_n(y) e^{-\lambda_n t}\,.
\end{equation*}
The general solution for $\bar{P}$ is thus given by 
\begin{equation*}
\bar{P}(y,t) = e^{-U(y)/\sigma^2} \sum_n c_n \psi_n(y) e^{-\lambda_n t}\,.
\end{equation*}
The initial conditions for $y_t$ determines the sequence $(c_n)$. In particular, if at time $t=0$, the probability $\bar{P}(y,0) = \delta(y-y_0)$ 
with $y_0 := \arg\cosh(\frac{2}{b} x_0 +1)$, then it is straightforward to see that 
\begin{equation*}
c_n = e^{U(y_0)/\sigma^2} \psi_n(y_0)\,.
\end{equation*}
The spectrum of $\mathcal{H}$ consists of a discrete and a continuous branch. 
The discrete energy levels (eigenvalues) are computed in, e.g. \cite{Dong} and are given for all $n \in \N, n \leq g/2$ with $g=1+2\theta\sigma^2$, 
by 
\begin{equation}\label{discrete_energy_levels}
\lambda_n = \frac{\sigma^2}{2} n \left(g - n \right)\,.
\end{equation}
The corresponding eigenvectors are also computed in \cite{Dong} and are expressed in terms of Jacobi polynomials.
To the best of our knowledge, the continuous branch of the spectrum has not been fully characterized in the literature. 
We should also mention that in the limit $b\rightarrow 0$ the corresponding process has been studied in details (see  \cite{Schenzle} and 
the appendix of \cite{Montus}). The problem can now be mapped into the Morse potential, which has exactly the same discrete spectrum as
above (as expected since $b$ does not appear), with eigenfunctions that can be expressed in terms of Laguerre polynomials. However, we have not been 
able to directly match the eigenfunctions in the two cases, and understand the $b \to 0$ limit in details. The limit $b \to \infty$ with $\sigma^2 b$ fixed, 
on the other hand, boils down to the standard Bessel process with mean-reversion.

\end{document}